\documentclass[11pt,a4paper]{article}

\usepackage[colorlinks=true, linkcolor=black!50!blue, urlcolor=blue, citecolor=blue, anchorcolor=blue]{hyperref}
\usepackage[font=small,labelfont=bf,margin=0mm,labelsep=period,tableposition=top]{caption}
\usepackage[a4paper,top=3cm,bottom=2.5cm,left=2.5cm,right=2.5cm,bindingoffset=0mm]{geometry}

\usepackage{graphicx}
\usepackage{float}
\usepackage{afterpage}
\usepackage{epsfig,cite}
\usepackage{amssymb}
\usepackage{amsmath}
\usepackage{dsfont}
\usepackage{multirow}
\usepackage{url}
\usepackage{xcolor}
\usepackage{float}
\usepackage{afterpage}

\usepackage{url}
\usepackage{hyperref}

\usepackage{multirow,booktabs,multirow}

\usepackage[T1]{fontenc} 

\newcommand{\be}{\begin{equation}}
\newcommand{\ee}{\end{equation}}
\newcommand{\bea}{\begin{eqnarray}}
\newcommand{\eea}{\end{eqnarray}}
\newcommand{\bi}{\begin{itemize}}
\newcommand{\ei}{\end{itemize}}
\newcommand{\ben}{\begin{enumerate}}
\newcommand{\een}{\end{enumerate}}

\newcommand{\as}{\alpha_s}

\def\frac#1#2{{{#1}\over {#2}}}
\def\gsim{\mathrel{\rlap{\lower4pt\hbox{\hskip1pt$\sim$}}
    \raise1pt\hbox{$>$}}}         
\def\lsim{\mathrel{\rlap{\lower4pt\hbox{\hskip1pt$\sim$}}
    \raise1pt\hbox{$<$}}}         

\def \n0{N_j^{(0)}}

\def \g{\gamma}

\def\gapprox{\lower .7ex\hbox{$\;\stackrel{\textstyle >}{\sim}\;$}}

\def\e{\epsilon}
\def\d{{\rm d}}

\def\qq{q \bar q}

\numberwithin{equation}{section}
\numberwithin{figure}{section}
\numberwithin{table}{section}

\begin{document}
\newgeometry{top=1.5cm,bottom=1.5cm,left=2.5cm,right=2.5cm,bindingoffset=0mm}
\vspace{-2.0cm}
\begin{flushright}
FERMILAB-PUB-18-018-T\\
Nikhef/2017-068\\
OUTP-17-16P
\end{flushright}
\vspace{2cm}

\begin{center}
  {\Large \bf
    Direct photon production and PDF fits reloaded
    }
\vspace{2.0cm}

John M. Campbell,$^{1}$ 
Juan Rojo,$^{2}$ 
Emma Slade,$^{3}$ 
and Ciaran Williams.$^{4}$

\vspace{1.0cm}
{\it \small
~$^1$Fermilab, P.O. Box 500, Batavia, IL 60510, USA.\\[0.1cm]
~$^2$Department of Physics and Astronomy, VU University, NL-1081 HV Amsterdam, \\
			and Nikhef Theory Group, Science Park 105, 1098 XG Amsterdam, The Netherlands.\\[0.1cm]
~$^3$Rudolf Peierls Centre for Theoretical Physics, 1 Keble Road, \\ University of Oxford, OX1 3NP Oxford, United Kingdom. \\[0.1cm]
~$^4$Department of Physics, University at Buffalo, \\The State University of New York, Buffalo 14260, USA. \\
}

\vspace{1.0cm}

{\bf \large Abstract}

\end{center}

Direct photon production in hadronic collisions provides
  a handle on the gluon PDF by means of the QCD Compton scattering process.
   In this work we revisit the impact of direct photon production on
  a global PDF analysis, motivated by the recent availability of the next-to-next-to-leading
  (NNLO) calculation for this process. 
  We demonstrate that the inclusion of NNLO QCD and leading-logarithmic electroweak corrections leads
  to a good quantitative agreement with the ATLAS measurements at 8 TeV
  and 13 TeV, except for the most forward rapidity region in the former case.
  By including
  the ATLAS 8 TeV direct photon production data in the NNPDF3.1 NNLO global
  analysis, we assess its impact on  the medium-$x$ gluon.
  We also study the constraining power of the direct photon  production
  measurements on PDF fits based on different
  datasets, in particular on the NNPDF3.1 no-LHC and collider-only fits.
  We also present updated NNLO theoretical predictions
  for direct photon production at 13 TeV that include the constraints
  from the 8 TeV measurements.

\clearpage

\tableofcontents

\section{Introduction}
\label{sec:intro}

The determination of parton distribution functions (PDFs) of
the proton~\cite{Forte:2013wc,Butterworth:2015oua,Rojo:2015acz,Gao:2017yyd} is an important component of the LHC program for many
analyses, from precision tests of the Standard Model to searches for new physics beyond it.
Within the global fitting framework, the gluon PDF has been traditionally constrained by the scaling violations
of deep-inelastic scattering (DIS) structure functions and from inclusive jet production~\cite{Rojo:2014kta}.
More recently, a number of additional collider observables have demonstrated their constraining
power on the gluon PDF, from differential distributions in top-quark pair production~\cite{Czakon:2016olj} to
the $Z$ boson transverse momentum~\cite{Boughezal:2017nla} and $D$ meson production in the
forward region~\cite{Gauld:2016kpd,Gauld:2017rbf}.
In this respect, the recent NNPDF3.1 global analysis~\cite{Ball:2017nwa} demonstrated how a robust determination of
the medium and large-$x$ NNLO gluon PDF can be achieved by the combination of LHC
measurements of top-quark pair, $Z$ $p_T$, and
inclusive jet production --- see also the discussion in~\cite{Nocera:2017zge}.

\begin{figure}[t]
\centering
\includegraphics[scale=0.34]{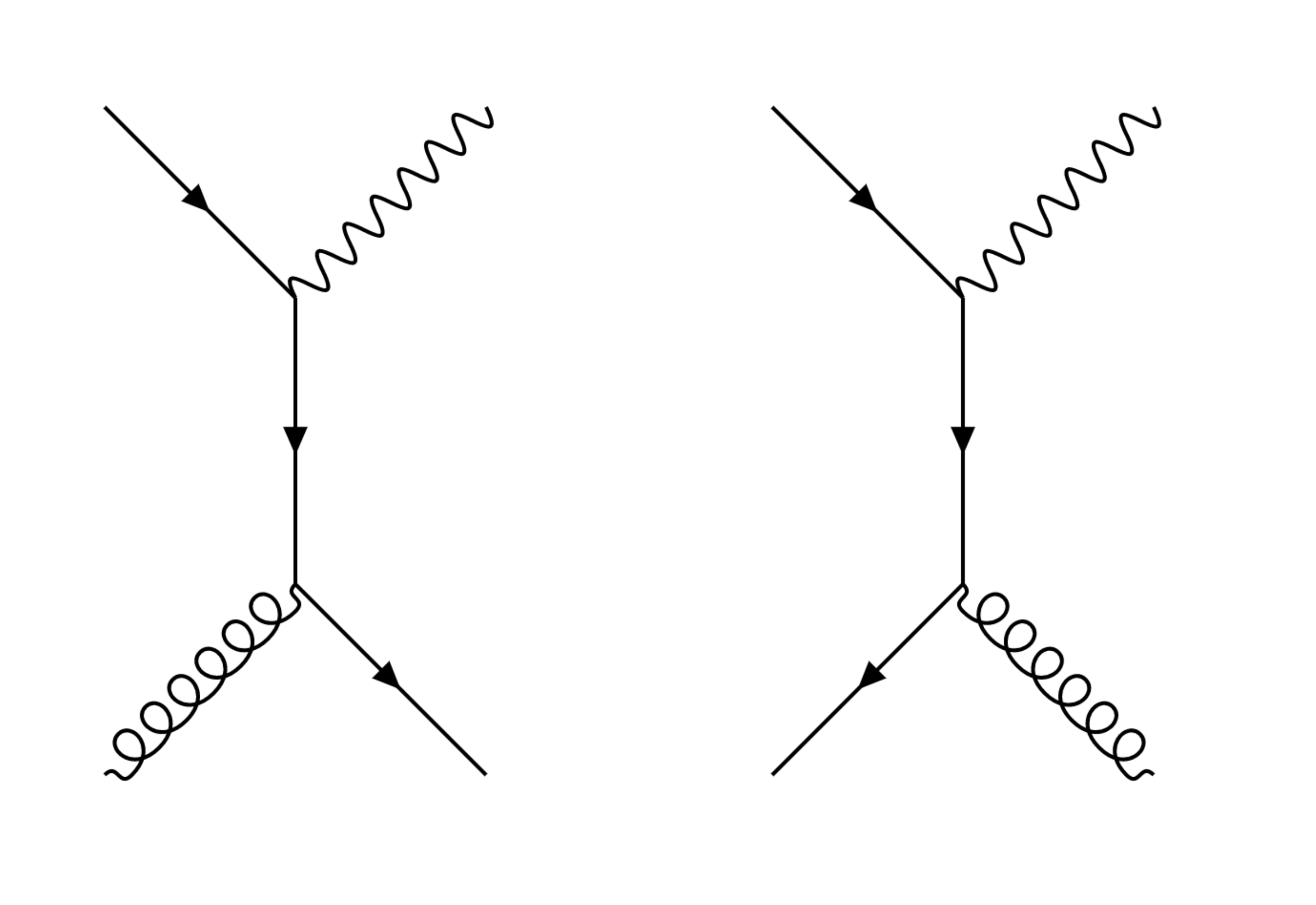}
\vspace{-0.8cm}
\caption{\small Feynman diagrams for direct photon production at leading order
  via the QCD Compton scattering
  process (left) and $\qq$ annihilation (right). \label{fig:direct_feynman}}
\end{figure}

Another process that has been advocated to constrain the gluon in a global PDF analysis is
direct (or ``prompt'') photon production at hadron colliders.
Indeed, direct photon production, $pp \rightarrow \g + X$, probes the gluon directly at leading order
through the QCD Compton scattering process $qg \rightarrow \g q$ shown in Fig.~\ref{fig:direct_feynman}.
Taking into account the kinematics of available LHC data, direct photon measurements provide
information on the gluon in the range between $x\simeq 10^{-3}$ and $x\simeq 0.1$~\cite{Ichou:2010wc,dEnterria:2012kvo}.
In addition, direct photons can also be produced via quark-antiquark annihilation (also shown in Fig.~\ref{fig:direct_feynman}),
such that this process also allows us to probe the contribution of different quark flavours in the same $x$ region.

However, exploiting collider measurements of direct photon production to constrain the gluon PDF is complicated by the fact that
high-$E_T$ photons can also be produced via the collinear splitting of a final-state quark.
These emissions have associated collinear singularities that are absorbed into non-perturbative quark-to-photon
and gluon-to-photon
fragmentation functions (FFs).
This fragmentation component is only loosely constrained by LEP data~\cite{Bourhis:1997yus}, therefore inducing
a potentially large source of theoretical uncertainty.

In spite of these complications, direct photon production data from fixed-target experiments
were used in early global PDF fits, such as those of~\cite{dflm,PhysRevD.42.798,Morfin:1990ck}.
However, the increased availability of  jet production data from the Tevatron,
together with the difficulties in reconciling NLO QCD
theory with some fixed-target measurements,
led to the abandonment of using photon data to constrain the large-$x$ gluon.
However, this general feeling that direct photon production data was not suitable for
PDF fits was demonstrated to be incorrect by the analysis of Ref.~\cite{dEnterria:2012kvo}.
There it was  shown that a good agreement between NLO theory and direct (isolated) photon
production measurements could be obtained
for a wide range of collider energies from RHIC and SPS to the Tevatron and the LHC
at $\sqrt{s}=7$ TeV.
This analysis also found that the LHC measurements lead to a moderate reduction
of the gluon PDF uncertainties in the region around  $x\simeq 0.02$.

Despite the results of this study,
none of the most recently-updated global PDF
fits~\cite{Harland-Lang:2014zoa,Ball:2017nwa,Dulat:2015mca,Alekhin:2017kpj} include collider direct photon measurements.
The reason for this is twofold.
On the one hand, the NLO QCD calculations are affected by large scale uncertainties,
thus making direct photon production inappropriate for NNLO global analyses that require
precision theoretical predictions.
On the other hand, in order to relate theory
calculations with experimental measurements
one needs to account for the poorly-understood fragmentation
component.

The first of these objections was removed by the availability of the NNLO QCD calculation~\cite{Campbell:2016lzl},
which together with the corresponding electroweak  corrections~\cite{Becher:2013zua}
was found to provide
a good quantitative description of the ATLAS measurements at $\sqrt{s}=8$
TeV~\cite{Aad:2016xcr}
at central photon rapidities~\cite{Campbell:2016lzl}.
The second objection can be somewhat alleviated by applying the smooth cone isolation prescription proposed by Frixione~\cite{Frixione:1998jh}.
This isolation condition removes the need for fragmentation functions from the theoretical calculation, at the cost of introducing a difference between 
the isolation definitions used in the theoretical and the experimental analyses.
However, as will be discussed later, this difference has been studied at NLO in great detail and found to be of limited practical consequence. 

In addition to their relevance for
PDF fits, photon production at the LHC is of great interest in searches for new physics beyond the standard model (BSM).
For instance, recent searches for BSM physics with photons in the final state from
ATLAS and CMS include searches for new particles by looking for high-mass resonances~\cite{Aaboud:2016tru,Aaboud:2017yyg,CMS_Zgamma}, anomalous couplings~\cite{Aaboud:2017tcq,Khachatryan:2016vif,Khachatryan:2016yro,Aad:2015uqa}, and
by measuring missing $E_T$ distributions~\cite{Sirunyan:2017ewk,Sirunyan:2017yse,Aaboud:2017uak,Sirunyan:2017hnk}.
These searches rely on a good understanding of the QCD background for photon production. It is therefore necessary
to account for higher order QCD and electroweak corrections and to use
recent global PDF fits that can properly model the background and signal events.

With this motivation, the goal of this paper is to revisit the impact of available
LHC direct photon production measurements on
  a global NNLO PDF analysis.
  Specifically, we will include the ATLAS 8 TeV measurements~\cite{Aad:2016xcr} into the NNPDF3.1 analysis,
  in order to  quantify the agreement between data and NNLO QCD theory and the corresponding
  impact on the gluon PDF.
  We find that a good description of this dataset is achieved,
  except for the most forward rapidity bin, and show that
  the inclusion of the photon data leads to a moderate reduction of the gluon PDF uncertainties at medium $x$.
  These fit results are cross-checked
  with those of the Monte Carlo Bayesian reweighting procedure~\cite{Ball:2010gb,Ball:2011gg}.
  In addition, we aim to study the constraining power of the direct photon  production
  measurements on PDF fits based on different
  datasets, in particular on the NNPDF3.1 no-LHC and collider-only sets.
  We also show that using state-of-the-art theory and including the constraints from the 8 TeV direct photon
  measurements leads to an excellent description of
  the recent ATLAS 13 TeV data~\cite{Aaboud:2017cbm}.
 
The paper is organised as follows.
In Sect.~\ref{sec:expts} we review existing
measurements of direct photon production, focusing on the ATLAS data used in the present study.
In Sect.~\ref{sec:theory} we discuss the theoretical setup for computing the
theoretical predictions for direct photon production.
The impact of the photon data upon the gluon PDF is
presented in Sect.~\ref{sec:results}, and in Sect.~\ref{sec:13tev} we provide
updated predictions for direct photon production at  13 TeV.
Finally,
in Sect.~\ref{sec:conc} we conclude with a summary of the results and outline possible future developments.
We assess the impact of the correlations among systematic uncertainties
in Appendix~\ref{sec:systematicbreakdown}, and
compare the results of the fits with those obtained with
the Bayesian reweighting method in Appendix~\ref{sec:Reweightingres}.

\section{Experimental data}
\label{sec:expts}
 
There exist many measurements of direct photon production
both at fixed-target and at collider experiments (we refer the reader to ref.~\cite{dEnterria:2012kvo}
for a detailed list).
The measurements performed at the highest
centre-of-mass energies are those from the LHC and from
the Tevatron.
At the Tevatron, the CDF and D\O~ experiments have measured direct photon cross-sections at $\sqrt{s} = 630$ GeV~\cite{Acosta:2002ya,Abazov:2001af}, 1.8 TeV~\cite{Abe:1994rra,Abe:1993qb,Abe:1992fd,Abbott:1999kd,Abachi:1996qz} and 1.96 TeV~\cite{Aaltonen:2013ama,Aaltonen:2009ty,Abazov:2005wc}.
More recently, the ATLAS and CMS experiments during Run I
have performed similar measurements at 7 TeV~\cite{Aad:2010sp,Aad:2011tw,Aad:2013zba,Chatrchyan:2011ue,Khachatryan:2010fm} and at 8 TeV~\cite{Aad:2016xcr}, while thus far during Run II only
ATLAS has measured direct photon production at 13 TeV~\cite{Aaboud:2017cbm}.

In this work, we will concentrate on the ATLAS
measurements at 8 TeV and 13 TeV.
The 8 TeV data exhibits reduced statistical
and systematic uncertainties as compared to their
7 TeV counterparts~\cite{Aad:2013zba,Khachatryan:2010fm},
and is thus suitable for inclusion in a global
PDF analysis.
The 13 TeV measurements will be used only to compare with our
theoretical predictions, but will not be included
in the fit since the experimental uncertainties
are larger than the 8 TeV data due to the limited
integrated luminosity, $\mathcal{L}_{\rm int}=3.2$ fb$^{-1}$.

The ATLAS 8 TeV direct photon production
measurement is presented as differential distributions in
the photon
transverse energy ($E_T ^\gamma$) in four photon pseudorapidity ($\eta^\gamma$)  bins:
\begin{align} \label{eq:rapidities}
&\text{region 1:}\,\, 0 < |\eta^\gamma| < 0.6 \,, \nonumber \\
&\text{region 2:}\,\, 0.6 \leq |\eta^\gamma| < 1.37 \,,  \\
&\text{region 3:}\,\, 1.56 \leq |\eta^\gamma| < 1.81\,, \nonumber\\
&\text{region 4:}\,\, 1.81 \leq |\eta^\gamma| < 2.37 \,. \nonumber
\end{align}
The measurements cover
the transverse energy range $25 < E_T ^\g < 1500$ GeV, though the
upper limit is reduced in the more forward bins. As we will discuss in Sect.~\ref{sec:QCD_EW}, 
the kinematic cuts applied constrain the number of points included in the fit to $N_{\rm dat}=49$.

For each of the experimental bins, the information
on the statistical, total systematic, and luminosity uncertainties
is provided by ATLAS.
The full breakdown of the experimental systematic uncertainties
including the information on cross-correlations
corresponding to this measurement
was only posted in HepData after the completion of the main
results of this work.
For this reason, the fits presented here are based
on a $\chi^2$ constructed by adding  the total systematic and statistical uncertainties in quadrature.
The luminosity uncertainty on the other hand is taken to
be fully correlated among all the bins, and correlated to other ATLAS measurements at 8 TeV included in the PDF fit.
In Appendix~\ref{sec:systematicbreakdown} we assess
the impact that including the correlation
between the experimental systematic uncertainties in
the $\chi^2$ definition
has at the level of both PDFs and at the level of fit quality.

As will be discussed in Sect.~\ref{sec:results},
the most forward rapidity bin, $1.81 \leq |\eta^\gamma| < 2.37$, is excluded from the fit due to the tensions between the experimental
data and the theoretical predictions.
In this respect, the fact that the covariance matrix was not available at the time of the completion of the main results of this work implies that we
cannot quantitatively study the origin of the tension in this
forward bin. We discuss in Appendix~\ref{sec:systematicbreakdown} the description of the 4th rapidity bin upon the inclusion of the covariance matrix.
Therefore, we have taken a conservative approach and excluded the anomalous bin; in Sect.~\ref{sec:data-theory} we discuss the impact in the fit of this bin and motivate in more detail
its exclusion from our analysis.

\begin{figure}[t]
\centering
  \includegraphics[scale=0.61]{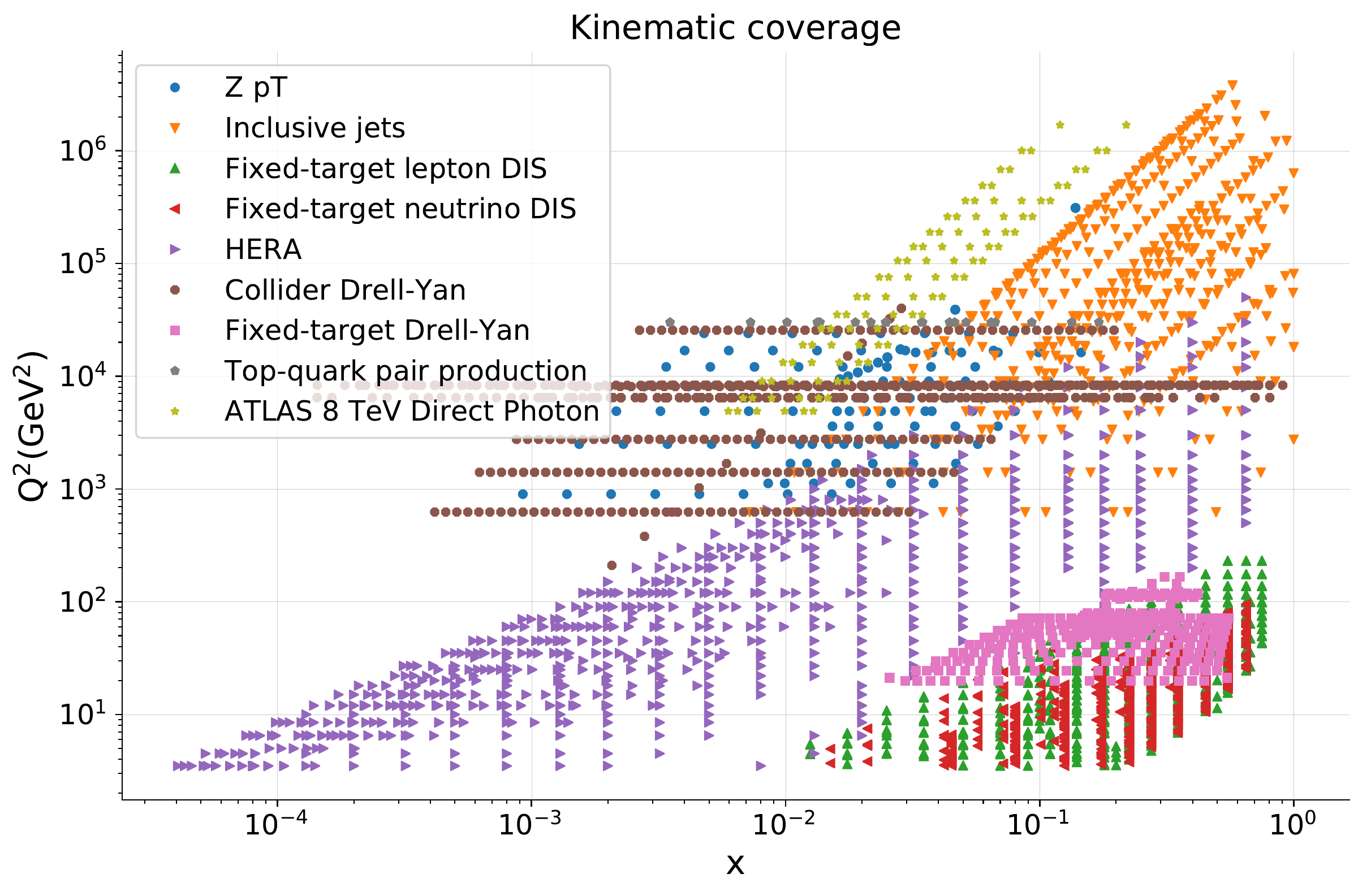} 
  \caption{\small The coverage in the $(x, Q^2)$ kinematic plane of the 8 TeV ATLAS photon measurements (using LO kinematics),
    compared to the dataset used in the global NNPDF3.1 fit.
  }
\label{fig:kinematic_range}
\end{figure}

In Fig.~\ref{fig:kinematic_range} we show the kinematic coverage of the ATLAS 8 TeV data in the $(x,Q^2)$ plane computed using LO kinematics, alongside with that
of the dataset used in the global NNPDF3.1 fit. At LO one may write $x_\pm = 2 E_T^\g \, \exp{(\pm \eta^\g)} / s$, so that each datapoint of the 8 TeV measurement corresponds to two points in the $(x,Q^2)$ plane.
From this comparison, we observe that
the photon data probes a $(x,Q^2)$ region only partly covered
by other experiments; specifically the medium-$x$ range for over two orders of magnitude in $Q^2$.
Therefore including the photon data allows one to constrain a new
kinematic region beyond the range of previous PDF fits. 

Concerning the ATLAS 13 TeV measurements, 
the data is presented in the same format as at 8 TeV
and covers a $E_T ^\gamma$ range between $125 < E_T ^\g < 1500$ GeV, for
a total of $N_{\rm dat}=53$ datapoints.
The covariance matrix is constructed in the same way as for the 8 TeV data,
namely by adding statistical and total systematic uncertainties in quadrature, and treating
the luminosity uncertainty as fully correlated among all the bins.

In order to distinguish prompt photons (produced in the hard-scattering process)
from secondary photons (which occur copiously in decays of hadrons) 
the experimental analyses apply isolation criteria to the measured photons.
Since secondary photons are predominantly associated with a large amount of hadronic activity,
the experiments restrict the hadronic radiation that is present in a cone around the photon candidate.
The isolation requirement used in the ATLAS analysis is $E_T^\g$-dependent, optimised to obtain the best signal-to-background ratio.
The additional advantage of the relatively tight isolation applied by ATLAS is that it significantly reduces the contribution from prompt photons which are produced in the fragmentation of a hard parton.
These contributions also have significant hadronic activity near the photon and are hence suppressed by the isolation condition. 

\section{Theoretical setup}
\label{sec:theory}
\label{sec:QCD_EW}
In this section we outline the NNLO QCD and
LL electroweak calculations that are used to compare with the
ATLAS direct photon production data.
We also describe the settings adopted in the calculation
of the NLO QCD predictions,
in particular the fast NLO {\tt APPLgrid} interpolation~\cite{Carli:2010rw} that is required
to include the photon data in the global analysis.

As discussed in the introduction,
direct photon production in hadronic collisions can proceed via two different types of processes.
The photon can either be directly emitted as part
of the hard-scattering interaction, as in Fig.~\ref{fig:direct_feynman},
or alternatively can be produced via the collinear fragmentation of a parton.
Taking into account these two contributions,
the differential cross-section as a function of $E_T ^\g$ can be written as
\begin{align}
\d\sigma &= \d\sigma_\text{dir} + \d\sigma_\text{frag} = \sum_{a,b = q, \bar{q}, g} \int \d x_a \d x_b f_a(x_a; \mu_F^2) f_b(x_b;\mu_F^2) \times \label{eq:fullcalc} \\
& \left[\d \hat{\sigma} ^\g _{ab} (p_\g, x_a, x_b; \mu_R, \mu_F, \mu_\text{ff}) + \sum_{c=q,\bar{q} ,g} \int_{z_\text{min}} ^1 \frac{\d z}{z^2} \d \hat{\sigma}^c _{ab}  (p_\g,x_a,x_b,z;\mu_R, \mu_F, \mu_\text{ff}) D_c ^\g( z; \mu_\text{ff}^2)\right]  \nonumber\,,
\end{align}
where $D_c ^\g(z,\mu_\text{ff}^2)$ is the fragmentation function of a parton $c$ to a photon carrying momentum fraction $z$ and $f_a(x_a; \mu_F^2)$ is the PDF of a parton $a$.
While $\mu_R$ and  $\mu_F$ are the standard renormalization
and factorization scales, note the appearance of a new scale
$\mu_\text{ff}$, known as the fragmentation scale.
The NNLO QCD corrections to the direct component
of the partonic cross-section $\hat{\sigma} ^\g _{ab}$ have
been computed in~\cite{Campbell:2016lzl}, while
the fragmentation component $\hat{\sigma}^c _{ab}$ is only known
at NLO.

The need to account for
the fragmentation functions $D^\g _c$ can be eliminated by adopting the smooth cone isolation criterion~\cite{Frixione:1998jh},
\begin{equation}
  \sum E_T^\text{had} (R) < \e_\g E_T ^\g \left( \frac{1- \cos R}{1 - \cos R_0} \right)^n \hspace{0.5cm} \forall R < R_0 \,,
\label{eq:smoothcone}
\end{equation}
with $E_T^{\rm had}$ being the hadronic transverse energy contained in a cone of radius $R$
around a photon of $E_T^\gamma$, and  where $n$, $R_0$ and $\epsilon_\gamma$ are parameters
of the algorithm.
Here the $(1- \cos R)$ term suppresses the collinear singularity present as $R\rightarrow 0$, but arbitrarily soft radiation is allowed inside the
cone $R_0$ in order to preserve the cancellation of infrared poles in the calculation.

The granularity of an experimental calorimeter is such that 
this smooth cone isolation can never be directly replicated in experimental analyses, thereby introducing an unwelcome disconnect between theoretical
calculations and the data.
However, the parameters appearing in the above isolation definition, $ \e_\g$ and $n$, are arbitrary and this allows them
to be tuned to replicate the features of a full calculation including fragmentation,
Eq.~(\ref{eq:fullcalc}).
Such a study was first performed at
NLO in~\cite{Bern:2011pa}, finding that the values $\e_\g = 0.025$ and $n=2$ result in
good agreement between the full calculation and the smooth cone
result to within a few percent.
A similar study was undertaken in the context of di-photon production at NNLO in~\cite{Campbell:2016yrh}, for
which the parameters $n=2$ and $\e_\g = 0.1$ were found to agree well with the fragmentation calculation.

The differences between the two types of
isolation criteria were further studied recently at NLO in~\cite{Campbell:2017dqk},
finding a small ($\sim2\%)$ correction that is independent of $E_T^\gamma$
over a range of values similar to the one studied in this paper. This correction is the same approximate
size as the missing higher-order uncertainty associated to the NNLO calculation. A full quantitative description of these theoretical uncertainties is beyond the scope of this work, so we
therefore do not account for the uncertainty due to the choice of isolation algorithm.
In this work, we adopt the smooth cone isolation, Eq.~(\ref{eq:smoothcone}),
with parameters
$n=2\,, \e_\g = 0.1$ and $R_0 = 0.4$, and motivated
by the above studies we assume that the residual
uncertainties due to the choice of isolation prescription are negligible
and should not have a bearing on the PDF fits that we perform.  

In order to account for the impact of Sudakov
effects induced by virtual loops of heavy electroweak gauge bosons,
we include the resummation of the electroweak Sudakov logarithms at
leading-logarithmic (LL) accuracy.
Following the procedure in~\cite{Becher:2013zua,Becher:2015yea}, we set
the QED coupling constant to be $\alpha_{\rm em}(m_Z) = 1/127.9$ in the calculation.

The
electroweak effects may be accounted for by an overall
rescaling of the cross-section of the form
\be
 \sigma^{\left(\text{NNLO QCD + LL EW}\right)} = [1 + \Delta_V ^{\rm ew}(E_T^\g,s)] \times \sigma^{\left(\text{NNLO QCD}\right)} \,,
\ee
where the LL electroweak correction $\Delta_V ^{\rm ew}(E_T^\g,s)$ is given
by~\cite{Becher:2013zua,Becher:2015yea}
\begin{equation}
  \label{eq:ewcorr}
 \Delta_V ^{\rm ew}(E_T^\g,s) = \frac{1.72 -21.68  E_T^\g +12.16 (E_T^\g)^2 -3.05 (E_T^\g)^3}{1 - 2.3355 \cdot 10^{-2} y + 1.2310 \cdot 10^{-3} y^2 }\,,
\end{equation}
with $y \equiv (\sqrt{s}-7)/7$, and $\sqrt{s}$ being
the hadronic center-of-mass energy expressed in TeV.

Since Eq.~(\ref{eq:ewcorr})
is only valid for $E_T ^\g \gtrsim M_Z,M_W$, we include in the fit
only data such that $E_T ^\gamma > 65$ GeV. %
This way, we can consistently use NNLO QCD and LL EW theory
for all the data bins in the fit. 
In the central rapidity bin, this extra cut has the additional advantage of minimising the contribution from fragmentation photons, which are not included
in our analysis, as the size of the fragmentation component decreases with $E_T^\g$.
After applying this kinematic cut, 
we have $N_{\rm dat}=63$ datapoints to include in the fit.
In addition, after removing the data from the most 
forward rapidity bin due to the poor $\chi^2$, as mentioned
in Sect.~\ref{sec:expts}, we are left with $N_{\rm dat}=49$ datapoints. 

Concerning the calculation of the NNLO QCD cross-sections,
we start by computing
theoretical predictions at NLO accuracy using \texttt{MCFM}~\cite{Boughezal:2016wmq} interfaced with \texttt{LHAPDF6}~\cite{Buckley:2014ana}
and {\tt APPLgrid} using the NNPDF3.1 NNLO PDF set~\cite{Ball:2017nwa} with the dynamical renormalization $(\mu_R)$ and factorization $(\mu_F)$ scales set equal to $E_T ^\g$.
The output of this calculation is a fast NLO interpolation grid, as required
for the inclusion of these measurements in a PDF fit.
Subsequently, 
the NNLO QCD corrections from Ref.~\cite{Campbell:2016lzl} are included in the form of bin-by-bin $K$-factors, defined as:
\begin{equation}
  \label{eq:kfactors}
  K \equiv  \frac{\d \sigma^\text{NNLO}}{\d E_T ^\g \d \eta^\g}({\rm NNLO~PDFs})
  \bigg/  \frac{\d \sigma^\text{NLO}}{\d E_T ^\g \d \eta^\g}({\rm NNLO~PDFs}) \,,
\end{equation}
so that only the perturbative order of the partonic cross-section
is varied, but the PDFs are kept the same.

The technical details regarding this NNLO QCD
computation can be found in the original
publications~\cite{Campbell:2016lzl,Campbell:2017dqk}, and we refer the interested reader to these works for more detail.
The only non-trivial change from these works is the manner in which the slicing variable $\tau_1^{\rm{cut}}$ is defined. In the original calculations this is set at a fixed value for the entire phase space, $\tau_1^{\rm{cut}}=0.08$ GeV.
In the present analysis we instead use a dynamic cut that is determined by the photon transverse momentum, $\tau_1^{\rm{cut}} = 0.001 \times E_T^{\gamma}$.
We find that this improves the overall performance of the computation, particularly in the determination of the NNLO corrections at high $E_T^{\gamma}$.  

\begin{figure}[t]
\centering
  \includegraphics[scale=0.73]{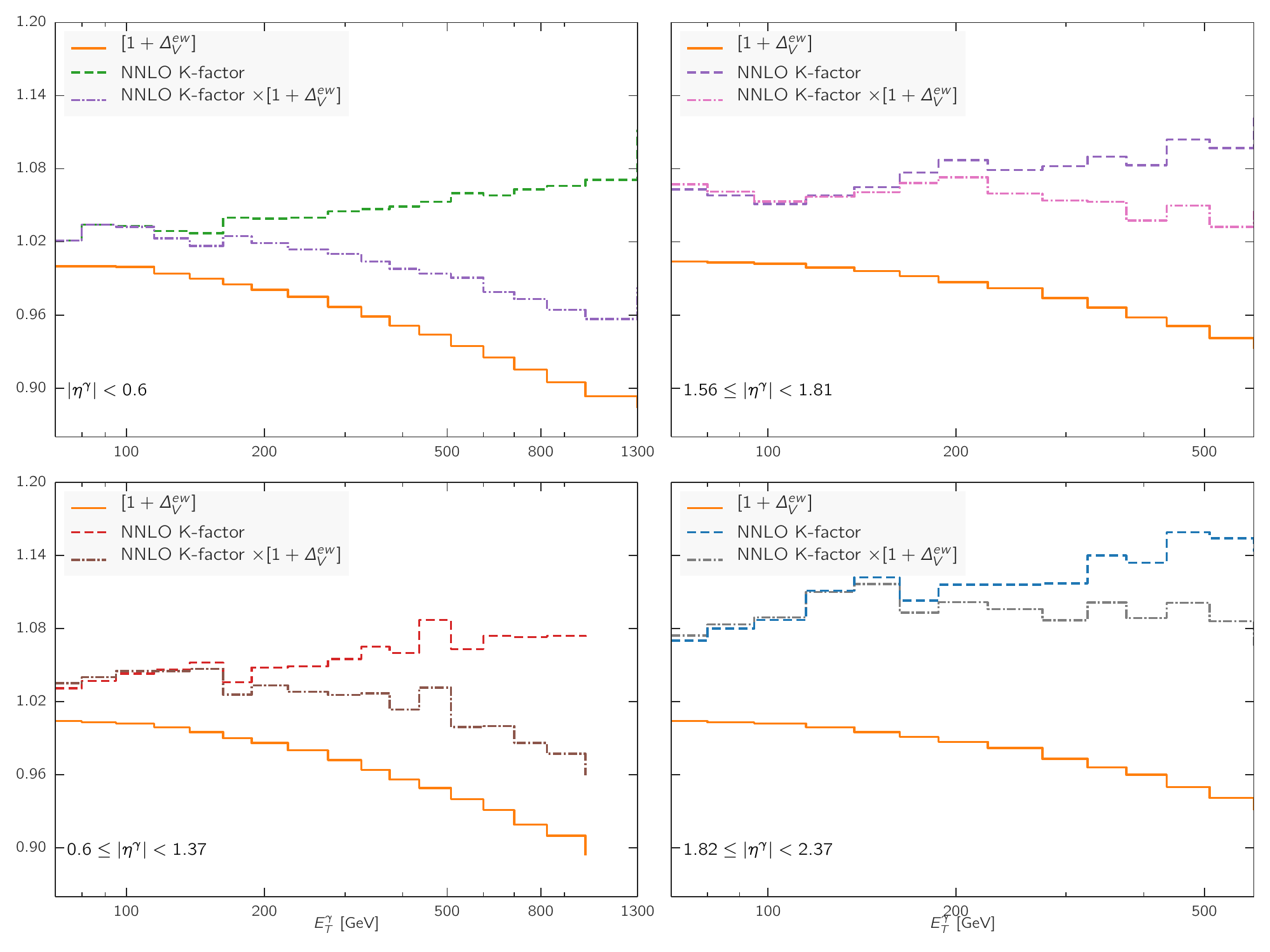} 
  \caption{\small The NNLO QCD $K$-factor, Eq.~(\ref{eq:kfactors}),
    and the LL electroweak correction $[1 + \Delta_V ^{\rm ew}(E_T^\g,s)]$,
Eq.~(\ref{eq:ewcorr}), in the four rapidity bins
of the ATLAS 8 TeV measurement.
We also show the results of its multiplicative combination, which indicates
the overall correction applied to the NLO QCD cross-section.
  }
\label{fig:kfactors}
\end{figure}

We show in Fig.~\ref{fig:kfactors} the NNLO QCD $K$-factors, Eq.~(\ref{eq:kfactors}),
 as well as the LL electroweak correction $[1 + \Delta_V ^{\rm ew}(E_T^\g,s)]$,
Eq.~(\ref{eq:ewcorr}), in the four rapidity bins
of the ATLAS 8 TeV measurement~\cite{Aad:2016xcr}.
Both corrections become more important in the high $E_T^\g$ regions, where
they deviate significantly from 1.
We also show in  Fig.~\ref{fig:kfactors} the results of the
multiplicative combination of the NNLO QCD $K$-factor
and of the LL EW effects, which represents
the overall correction applied to the NLO QCD cross-section.
We can observe that there is a partial cancellation between the two
higher-order effects, since each pulls the NLO cross-section
in an opposite direction.

\label{sec:APPLgrids}
As mentioned above, a fast interpolation of the NLO QCD
calculation, required for the subsequent PDF fit,
is constructed by interfacing \texttt{APPLgrid} with {\tt MCFM}.
These fast grids  may be used to compute the  cross-sections for any PDF set
other than the one used in the original calculation with a very small
calculational overhead.
Specifically, we have used {\tt MCFM} v6.8 interfaced with the \texttt{MCFM/APPLgrid} bridge code and with the \texttt{HOPPET}~\cite{Salam:2008qg}
PDF evolution program.
Note that in this respect {\tt MCFM} v6.8 had to be patched
  to reproduce the results of v8.0, which in turn
  correspond to the results of Ref.~\cite{Campbell:2016lzl},
  as we verified explicitly.

\begin{figure}[t]
\centering
      \includegraphics[scale=1.2]{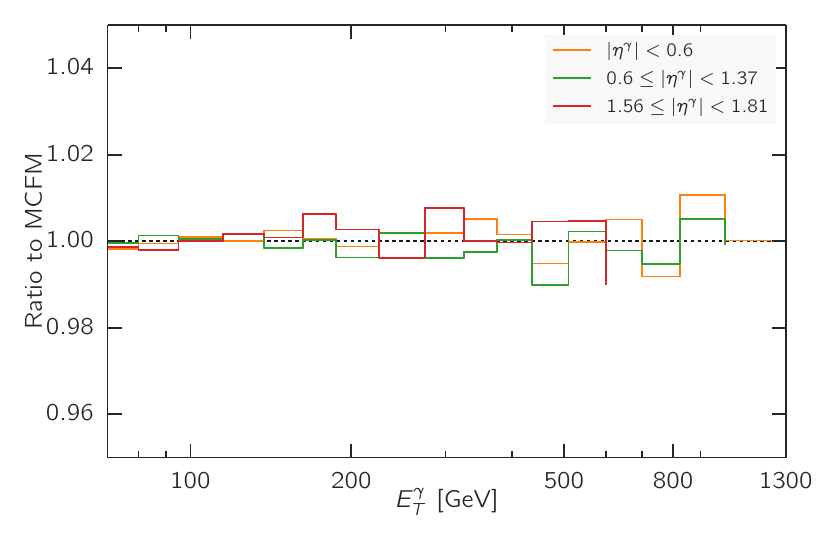} 
      \caption{\small The ratio  of the \texttt{APPLgrid} computations of the NLO QCD
        cross-section to the corresponding \texttt{MCFM} v6.8 result for
        the kinematics of the first three rapidity bins of the ATLAS
        8 TeV measurement, using in both cases the NNPDF3.1 set as input.}
\label{fig:applgrid_benchmark}
\end{figure}

In order to obtain sufficiently high numerical precision, we ran \texttt{MCFM} in 10 batches with different random seeds and combined the resulting grids using the \texttt{applgrid-combine} script.
This \texttt{MCFM/APPLgrid} computation was successfully
benchmarked with the NLO code of Ref.~\cite{Campbell:2016lzl},
finding excellent agreement.
In Fig.~\ref{fig:applgrid_benchmark} we plot the ratio of the
\texttt{APPLgrid} computations of the NLO QCD
        cross-section to the corresponding \texttt{MCFM} v6.8 result for
        the kinematics of the first three rapidity bins of the ATLAS
        8 TeV measurement, using in both cases the NNPDF3.1 set as input.
We find good agreement between the two methods within the uncertainties from the finite MC
integration statistics, which are typically at the
permille level.\footnote{These fast NLO {\tt APPLgrids}, matching the selection cuts and the
binning of the ATLAS 8 and 13 TeV direct photon production
measurements, are available from the authors.}

\section{Results}
\label{sec:results}

In this section we present the main results of this work, namely
the impact of the ATLAS direct photon production data at $\sqrt{s}=8$ TeV
on the NNPDF3.1 global analysis.
NNPDF3.1 is the most up-to-date NNPDF release, including a wealth of new
Tevatron and LHC datasets
from processes such as Drell-Yan and $t\bar{t}$ pair-production and
the transverse momentum of $Z$  bosons.
In contrast to previous fits, NNPDF3.1 independently parameterizes  the charm content
of the proton, eliminating any possible bias related to the assumption that the
charm PDF is generated perturbatively~\cite{Ball:2016neh}.

In this context, an important difference of the
present work as compared to the study of Ref.~\cite{dEnterria:2012kvo} is that
the latter was based on the NNPDF2.1 fit, where the information on
the gluon PDF was limited.
This is not the case in NNPDF3.1, where the gluon PDF is already
reasonably well constrained at medium and small-$x$ from
the combination of jet, $t\bar{t}$, and $Z$ $p_T$ data, and therefore
we expect the impact of the direct photon data on the gluon to be moderate.

Here we will also study the impact of the direct photon
production data on  fits based
on reduced datasets, in particular the NNPDF3.1 no-LHC data and collider-only fits.
We also compare other global PDF sets to the direct
photon measurements, specifically MMHT14,
CT14, and ABMP16.
The corresponding  comparisons with the ATLAS 13 TeV 
measurements as well as with the  13/8 cross-section ratio
will then be presented in the next section.

\subsection{Comparison to the experimental data}
\label{sec:data-theory}

\begin{figure}[t]
\centering
  \includegraphics[scale=0.72]{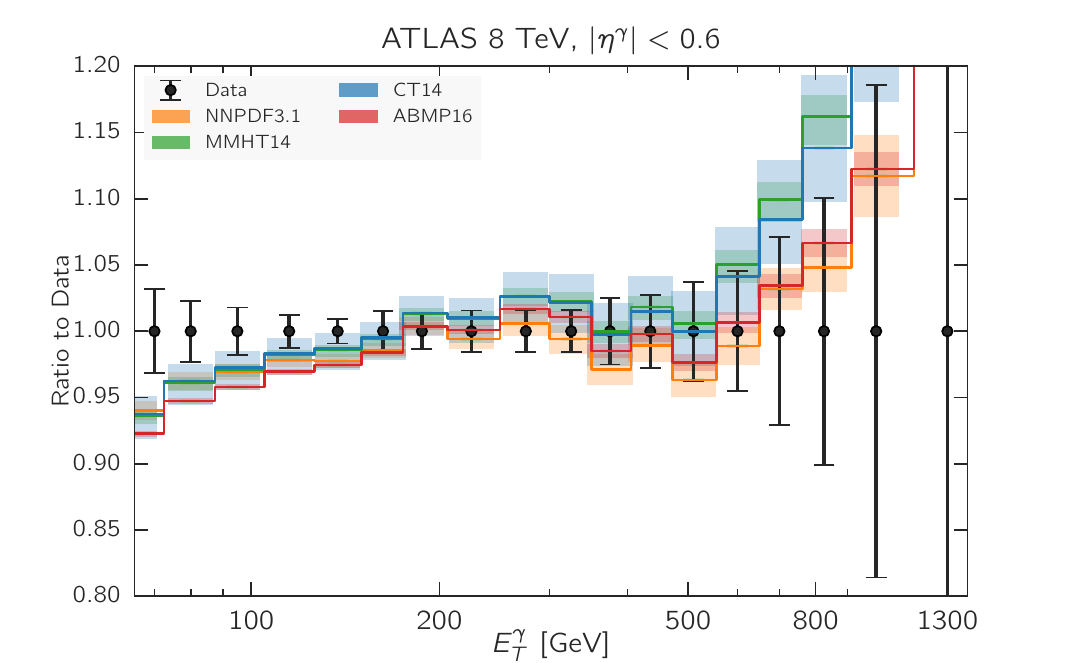} 
  \includegraphics[scale=0.72]{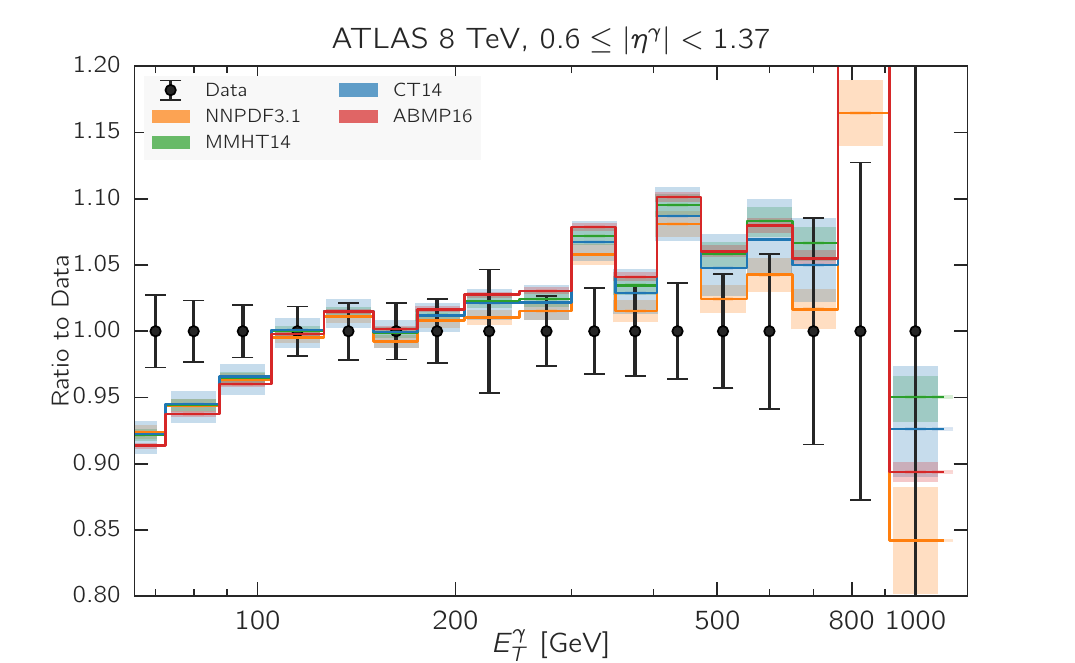} \\
  \includegraphics[scale=0.72]{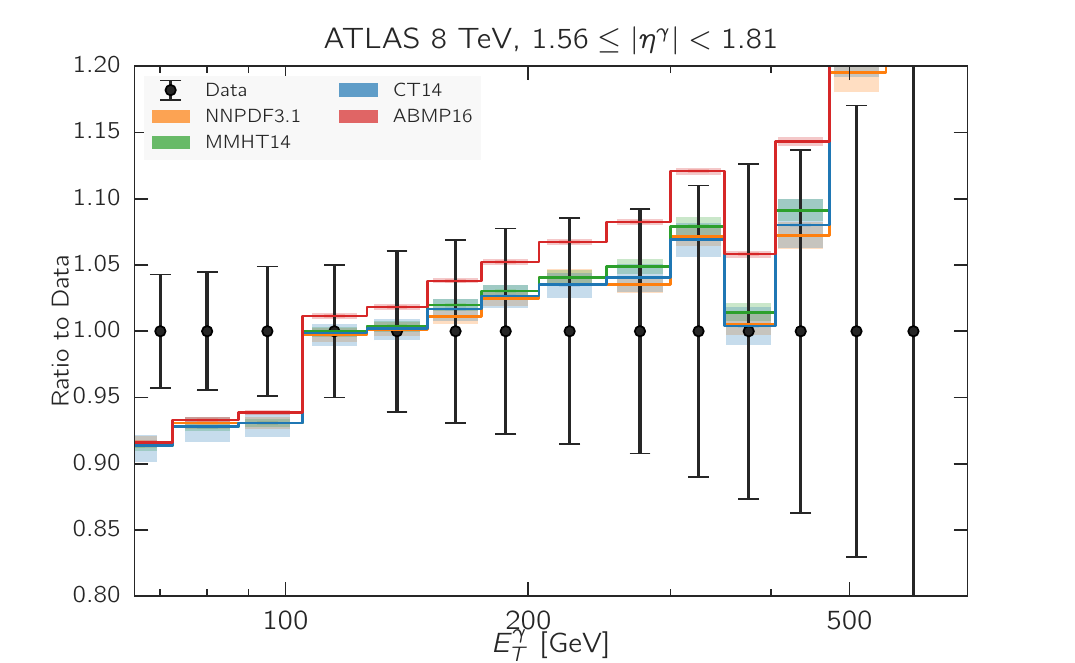}
    \includegraphics[scale=0.72]{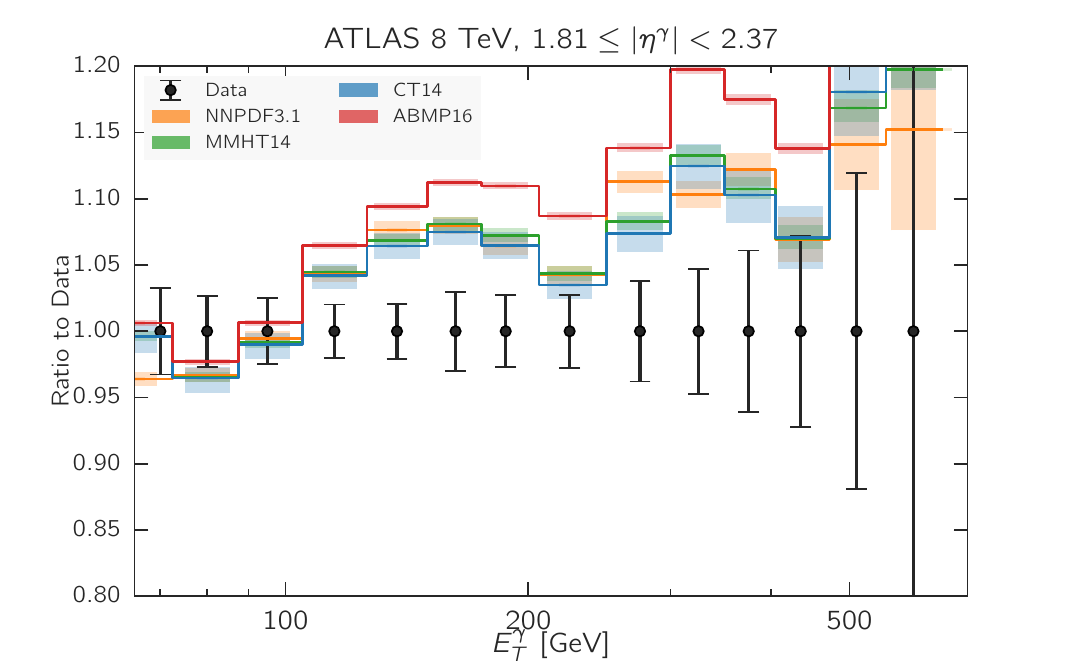}  
    \caption{\small Comparison between the theoretical
      predictions for direct photon production data
      computed with different PDF sets and
      the ATLAS 8 TeV data, normalized
      to the central value of the former.
      The experimental statistical and systematic uncertainties
       have been added in quadrature.
        The error bands for the theory predictions include only the PDF uncertainties.}
\label{fig:data_theory_ratios}
\end{figure}

To begin with,
using the NNLO QCD theory supplemented with LL electroweak corrections
described in Sect.~\ref{sec:theory},
we have computed the differential cross-sections for the $E_T^{\gamma}$
distributions of the ATLAS 8 TeV measurement for different PDF sets.
In all cases we  use their default value for the strong coupling
constant;
for NNPDF3.1, MMHT14 and CT14 this is $\as(m_Z) = 0.118$ and for ABMP16, $\as(m_Z) = 0.1147$.
In Fig.~\ref{fig:data_theory_ratios} we show the
comparison of these theoretical
      predictions normalized
      to the central value of the ATLAS measurements, where
      the error bars on the experimental data 
      are the sum in quadrature of the statistical and systematic uncertainties,
      while the error bands for the theory predictions include only
        the PDF uncertainties.

From the comparisons of Fig.~\ref{fig:data_theory_ratios} we see that
across the first three rapidity bins, the various sets are in good agreement; in particular in the 3rd bin, NNPDF, CT14 and MMHT14 are very close to each other.
We also
find that the NNPDF3.1 and ABMP16 sets lead to a better description of the high $E_T^\g$ region in the central bin.
On the other hand,
one can clearly observe in the most forward bin a large disagreement between theory and data, in particular for ABMP16.

These trends are further examined in Table~\ref{table:chi2_globalfits}, where we compare the total $\chi^2 /N_\text{dat}$ for the different PDF sets.
We note that in this $\chi^2$ computation the
\emph{experimental} definition of the covariance matrix is used~\cite{Ball:2012wy}, as opposed to the $t_0$ definition~\cite{Ball:2009qv} which is only used during the fitting.
From Table~\ref{table:chi2_globalfits} we see that
none of the four PDF sets manage to describe the most forward
rapidity bin in a satisfactory way.
We have verified that this is still the case even when this bin
is included in the PDF fit.
We show in Appendix~\ref{sec:systematicbreakdown} that even upon the 
inclusion of the covariance matrix, the poor description of the $\chi^2$ still exists in the 4th rapidity bin. We therefore exclude this
4th rapidity bin from
the analysis.

Then in Table~\ref{table:chi2_globalfits_nlo} we show the same $\chi^2$
comparison but now using only NLO QCD theory
and without the LL electroweak corrections.
In this case we find that
the $\chi^2$ in all bins is rather poor.
Interestingly, the most forward rapidity bin
exhibits a slight improvement in the $\chi^2$ values, which however, remain large. 
As the most forward rapidity bin corresponds to
  the small-$x$ region of one  of the incoming partons,
  it would be interesting to verify if the theoretical
  description of this bin would be improved
  by including NLL$x$ resummation of direct photon production,
  similar as what was done in~\cite{Ball:2017otu,Abdolmaleki:2018jln} for the
  HERA data.
For the three rapidity bins used in the fit, one finds
a dramatic improvement in the description of the data upon including the
higher-order QCD and EW effects.
This comparison highlights the phenomenological
importance of the recent NNLO QCD calculation, and why only 
now we can robustly include the direct photon measurements into the global PDF fits.
 
\begin{table}[h!]\centering
  \renewcommand{\arraystretch}{1.20}
\begin{tabular}{  c | c |c | c | c|c| c }
\multicolumn{7}{c}{$\chi^2 /N_\text{dat}$}   \\ \toprule
& 1st bin & 2nd bin & 3rd bin& 4th bin& Total &\multicolumn{1}{|p{3cm}}{\centering Total excluding \\ 4th bin}\\ \midrule
NNPDF3.1	&  0.81 	&  1.61 	& 0.89 	&  1.97	&  1.83	& 1.12  \\
MMHT14		& 1.94	& 2.49 	&1.02	&  2.19	&  2.31	&1.89  \\
CT14		& 1.63	& 2.18 	& 0.96	&  2.02 	&  2.01	&1.65  \\ 
ABMP16		& 1.38	&  2.70 	& 1.27	& 7.78	& 3.50  	& 1.91  \\ \bottomrule
\end{tabular}
\vspace{0.3cm}
\caption{\small The $\chi^2 /N_\text{dat}$ values of the 8 TeV ATLAS
  using NNLO QCD theory supplemented with LL electroweak corrections for
  different PDF sets.
  We provide the results for
  the four individual rapidity bins in Eq.~(\ref{eq:rapidities}),
  as well as their sum with and without the most forward bin.
}\label{table:chi2_globalfits}
\end{table}

\begin{table}[h!]\centering
  \renewcommand{\arraystretch}{1.20}
\begin{tabular}{  c | c |c | c | c|c| c }
\multicolumn{7}{c}{$\chi^2 /N_\text{dat}$}   \\ \toprule
& 1st bin & 2nd bin & 3rd bin& 4th bin& Total &\multicolumn{1}{|p{3cm}}{\centering Total excluding \\ 4th bin}\\ \midrule
NNPDF3.1	& 1.55  	&  2.35 	&  1.44	& 1.83 	&  1.69	& 1.71  \\
MMHT14		& 3.37	&3.43  	&1.57	&2.08  	&  2.73	&2.87  \\
CT14		&2.91 	&  3.03	& 1.51	& 1.99  	& 2.51 	& 2.57 \\ 
ABMP16		& 2.48	&   4.19	& 2.03	& 3.19	&  2.59 	&   2.53\\ \bottomrule
\end{tabular}
\vspace{0.3cm}
\caption{\small Same as in Table~\ref{table:chi2_globalfits} but with only NLO QCD theory and 
without LL electroweak corrections.
}\label{table:chi2_globalfits_nlo}
\end{table}

\subsection{Impact on the global fit}
\label{sec:globalfit}

In the following, we denote by  NNPDF3.1+ATLAS$\g$  the results of the fit obtained
by adding the ATLAS 8 TeV direct photon production cross-sections
to the NNPDF3.1 NNLO global analysis.
In Table~\ref{table:chi2_NNPDF_comp} we compare
the resulting values of $\chi^2/N_{\rm dat}$ for each of the three
rapidity bins included in the fit as well as for their
total.
We find that the inclusion of the ATLAS direct photon data
 improves the agreement between the theoretical
predictions and the experimental measurements, with
the total $\chi^2 /N_\text{dat}$ decreasing from 1.12 down
to 0.96.
This improvement is particularly marked in the second rapidity bin,
where $\chi^2 /N_\text{dat}$ is reduced from 1.61 to 1.37.

\begin{table}[h!]\centering
  \renewcommand{\arraystretch}{1.20}
{\begin{tabular}{  c | c |c | c | c|c| c }
 & \multicolumn{4}{c|}{$\chi^2 /N_\text{dat}$}   \\ \toprule
& 1st bin & 2nd bin & 3rd bin& Total \\ \midrule
NNPDF3.1	&  0.81 	&  1.61 	& 0.89	& 1.12  \\
NNPDF31+ATLAS$\g$		& 0.66	& 1.37 	&0.82	&  0.96 \\ \bottomrule\end{tabular}}
\vspace{0.3cm}
\caption{\small The $\chi^2 /N_\text{dat}$ values for
  the 8 TeV ATLAS data for the NNPDF3.1 and NNPDF3.1+ATLAS$\g$ fits,
both for three
rapidity bins included in the fit and for their total.
}\label{table:chi2_NNPDF_comp}
\end{table}

These results suggest that the ATLAS photon measurements seem to be consistent with the rest
of the datasets in NNPDF3.1.
In order to further investigate this issue, and
to determine if the ATLAS photon measurements are in tension
with some of the other datasets included in the fit,
In Table~\ref{table:chi2tab_31-nlo-nnlo-30} we provide the breakdown
of the $\chi^2/N_{\rm dat}$  values for the individual datasets,
comparing the results from the 
NNPDF3.1 and  NNPDF3.1+ATLAS$\g$ sets.

\begin{table}[h!]
\begin{center}
  \scriptsize
      \renewcommand{\arraystretch}{1.20}
\begin{tabular}{l|c|c}
Dataset  &  NNPDF3.1 & NNPDF3.1+ATLAS$\g$ \\
\toprule
NMC    &    1.30     &   1.28    \\    
SLAC    &    0.75      &   0.75     \\    
BCDMS    &    1.21      &   1.22    \\    
CHORUS    &    1.11      &   1.11    \\    
NuTeV dimuon    &    0.82      &   0.81    \\    
\hline
HERA I+II inclusive    &    1.16      &   1.16     \\      
HERA $\sigma_c^{\rm NC}$    &    1.45      &   1.45     \\    
HERA $F_2^b$    &    1.11      &   1.10  \\    
\midrule
DY E866 $\sigma^d_{\rm DY}/\sigma^p_{\rm DY}$ &    0.41     &   0.46       \\    
DY E886 $\sigma^p$    &    1.43     &   1.41    \\    
DY E605  $\sigma^p$   &    1.21     &   1.21   \\    
\hline
CDF $Z$ rap    &    1.48     &   1.49     \\    
CDF Run II $k_t$ jets    &    0.87      &   0.86    \\    
\hline
D0 $Z$ rap    &    0.60      &   0.60    \\    
D0 $W\to e\nu$  asy   &    2.70      &   2.73    \\    
D0 $W\to \mu\nu$  asy    &    1.56      &   1.56    \\    
\midrule
ATLAS total   &    {\bf 1.09}      &   {\bf 1.07}    \\    
ATLAS $W,Z$ 7 TeV 2010    &    0.96     &   0.96     \\    
ATLAS high-mass DY 7 TeV    &    1.54     &   1.61    \\    
ATLAS low-mass DY 2011    &    0.90      &   0.91     \\    
ATLAS $W,Z$ 7 TeV 2011    &    2.14     &   2.05     \\    
ATLAS jets 2010 7 TeV     &    0.94      &   0.92   \\    
ATLAS jets 2.76 TeV     &    1.03      &   1.01     \\    
ATLAS jets 2011 7 TeV     &    1.07     &   1.07   \\    
ATLAS $Z$ $p_T$ 8 TeV $(p_T^{ll},M_{ll})$       &    0.93    &   0.93  \\    
ATLAS $Z$ $p_T$ 8 TeV $(p_T^{ll},y_{ll})$    &    0.94      &   0.88   \\    
ATLAS $\sigma_{tt}^{\rm tot}$     &    0.86      &   1.09   \\    
ATLAS $t\bar{t}$ rap    &    1.45     &   1.39    \\
\hline
CMS total   &   {\bf  1.06}    &   {\bf 1.04 }   \\    
CMS $W$ asy 840 pb     &    0.78     &   0.78   \\    
CMS $W$ asy 4.7 fb     &    1.75     &   1.76    \\    
CMS Drell-Yan 2D 2011    &    1.27     &   1.29    \\    
CMS $W$ rap 8 TeV    &    1.01      &   1.06   \\    
CMS jets 7 TeV 2011     &    0.84      &   0.82    \\    
CMS jets 2.76 TeV    &    1.03      &   1.00     \\    
CMS $Z$ $p_T$ 8 TeV $(p_T^{ll},M_{ll})$    &    1.32    &   1.33   \\    
CMS $\sigma_{tt}^{\rm tot}$     &    0.20      &   0.24     \\    
CMS $t\bar{t}$ rap    &    0.94      &   0.93   \\
\hline 
LHCb total   &  {\bf  1.47}      &   {\bf 1.42}    \\
LHCb $Z$ 940 pb    &    1.49      &   1.49     \\    
LHCb $Z\to ee$ 2 fb    &    1.14      &   1.16    \\    
LHCb $W,Z \to \mu$ 7 TeV    &    1.76      &   1.69    \\    
LHCb $W,Z \to \mu$ 8 TeV    &    1.37     &   1.30     \\
\midrule
{\bf Total dataset}   &  {\bf  1.148}      & {\bf 1.146 }   \\    
\bottomrule
\end{tabular}
\vspace{0.3cm}
\caption{\small
\label{table:chi2tab_31-nlo-nnlo-30}
 \small The values of $\chi^2/N_{\rm dat}$ for
 all the datasets included in the present analysis,
 comparing the results from the 
 NNPDF3.1 and  NNPDF3.1+ATLAS$\g$ fits.
 The corresponding values for the
 ATLAS 8 TeV direct photon data are reported in
 Table~\ref{table:chi2_NNPDF_comp}.
}
\end{center}
\end{table}

\begin{table}[h!]
\begin{center}
  \footnotesize
      \renewcommand{\arraystretch}{1.20}
\begin{tabular}{l|c|c|c}
  &  NNPDF3.1 & NNPDF3.1+ATLAS$\g$ \\
  \toprule
Fixed-target lepton DIS& 1.207 & 1.203 \\
Fixed-target neutrino DIS&1.081 & 1.087 \\
  HERA  &1.166& 1.169\\
  \midrule
Fixed-target Drell-Yan&1.241 & 1.242\\
  Collider Drell-Yan  &1.356 &1.346 \\
  Top-quark pair production &1.065&1.049  \\
Inclusive jets& 0.939& 0.915  \\
$Z$ $p_T$ &0.997& 0.980 \\
\midrule
{\bf Total dataset}   &  {\bf  1.148}      & {\bf 1.146 }   \\    
\bottomrule
\end{tabular}
\vspace{0.3cm}
\caption{\small
  \label{table:chi2tab_31-nnlo-grouped}
  Same as Table~\ref{table:chi2tab_31-nlo-nnlo-30} now with
  individual experiments  grouped into families of processes.
}
\end{center}
\end{table}


We observe that the overall fit quality upon inclusion of the photon data is unchanged
within statistical fluctuations.
In addition, we find that the direct photon
data does not appear to exhibit any tensions with existing datasets.
In particular, there are no tensions with other datasets which constrain the gluon, such as
top-quark pair and inclusive jets production and the $Z$ transverse momentum distributions.
This stability is further highlighted by the comparison in
Table~\ref{table:chi2tab_31-nnlo-grouped}, where we have grouped datasets together in
families of related processes.
We find that the largest improvements in the values of the $\chi^2/N_{\rm dat}$
indeed correspond to those processes with sensitivity to the gluon PDF.
We can thus conclude that the constraints on the gluon from direct photon
production are consistent with those of the rest of the datasets
in NNPDF3.1.

\begin{figure}[t]
\centering
\includegraphics[scale=0.47]{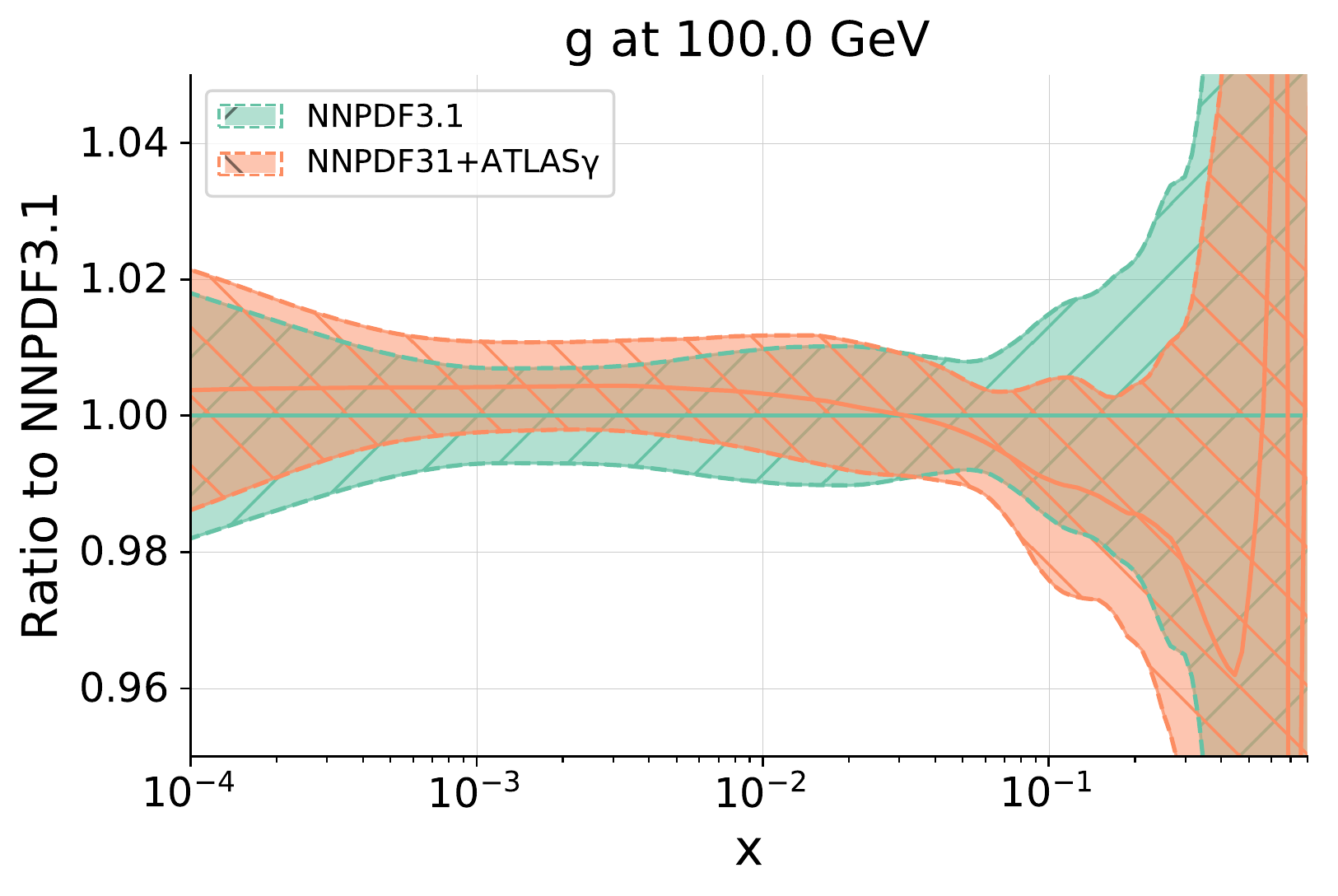}
 \includegraphics[scale=0.47]{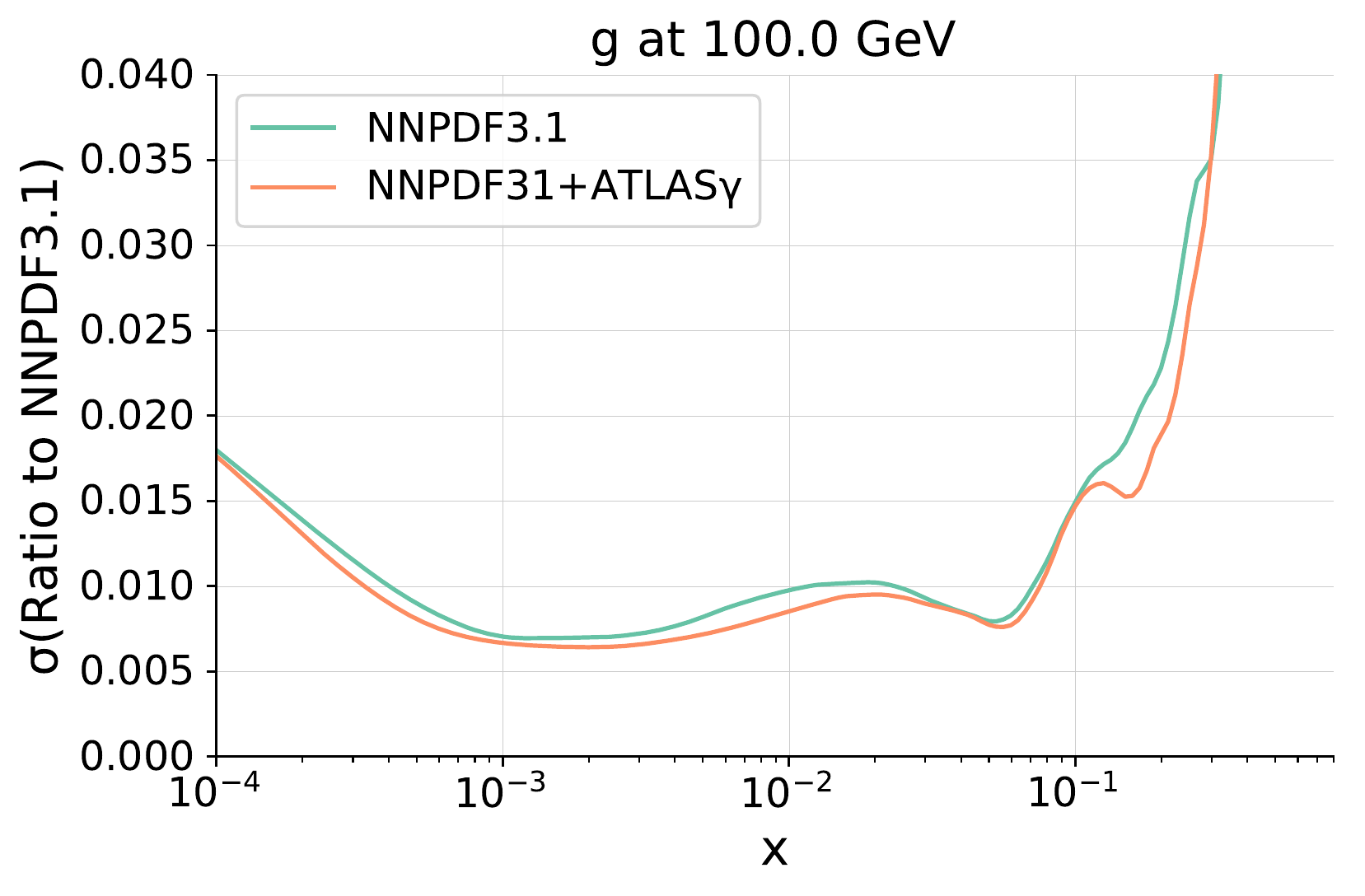}
 \caption{\small Left: comparison of the gluon PDF at $Q = 100$ GeV
   between the NNPDF3.1 and NNPDF3.1+ATLAS$\g$ fits, normalized
   to the central value of the former.
   Right: the corresponding relative one-sigma PDF uncertainties in each case.
   \label{fig:gluon_ratio}}
\end{figure}

In order to quantify the impact of the ATLAS direct photon data into the PDFs, in
Fig.~\ref{fig:gluon_ratio}
we show the comparison of the gluon PDF at $Q = 100$ GeV
   between the NNPDF3.1 and NNPDF3.1+ATLAS$\g$ fits, normalized
   to the central value of the former.
   In the same figure, we also compare the corresponding relative one-sigma PDF uncertainties in 
   both cases.
   We find two main implications of adding the photon data into NNPDF3.1.
   The first one is a moderate reduction of the gluon PDF
   uncertainties in the region $10^{-3}\lsim x \lsim 0.4$,
   which is consistent with the kinematic coverage spanned by the ATLAS measurements
   shown in Fig.~\ref{fig:kinematic_range}.

   The second is a downward shift of the gluon central value in the large-$x$ region,
   by an amount of up to two thirds of the PDF uncertainty.
   For instance at $x\simeq 0.4$ the gluon in NNPDF3.1+ATLAS$\g$ is about
   4\% smaller than in NNPDF3.1.
   Interestingly, the same trend was observed when adding top-quark pair differential
   distributions to NNPDF3.0~\cite{Czakon:2016olj}.
   The overall consistency of the ATLAS
   direct photon data with the NNPDF3.1 dataset is highlighted by the fact that
   in the whole range of $x$ the two fits are consistent within uncertainties.
   
\begin{figure}[h!]
\centering
\includegraphics[scale=0.47]{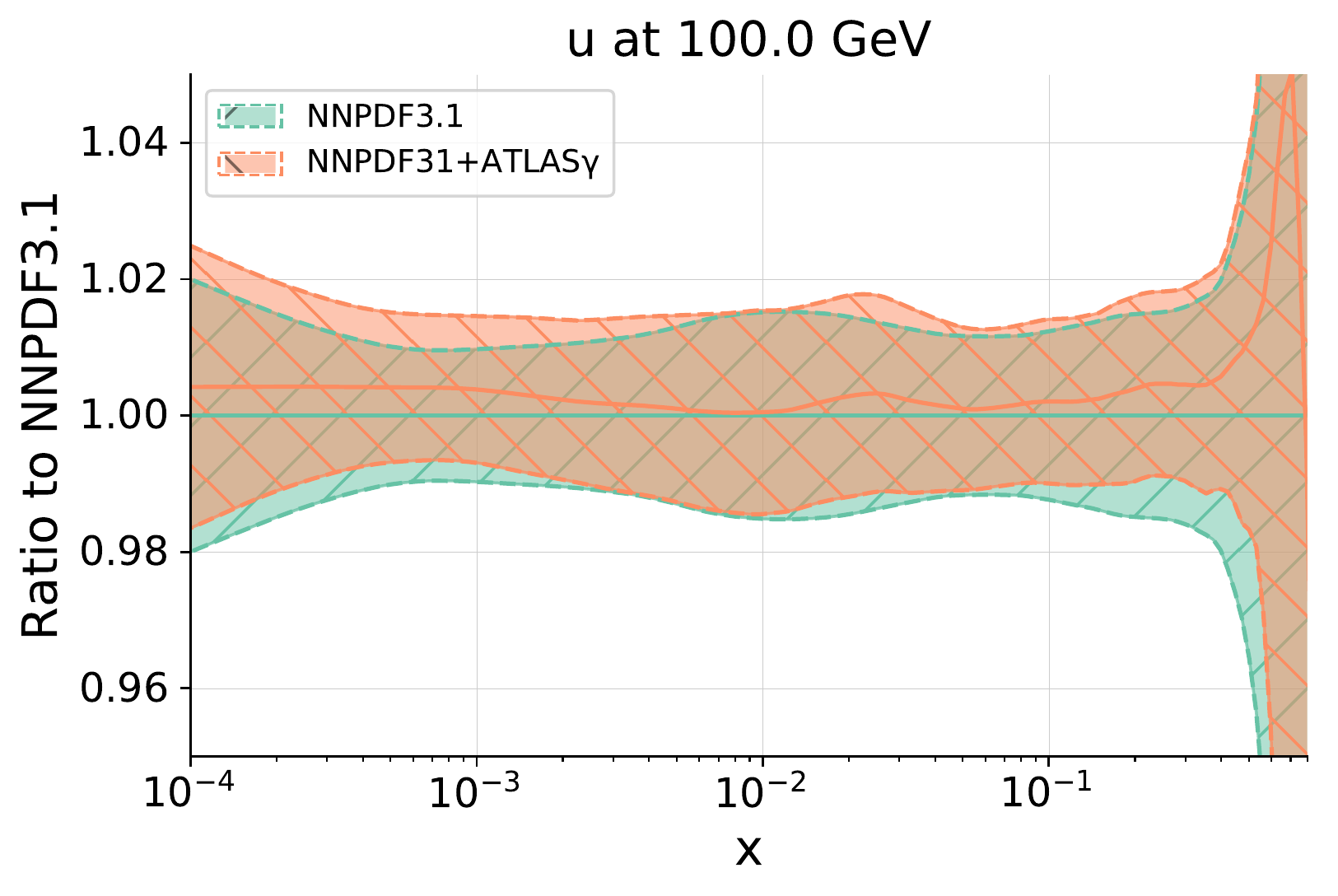}
 \includegraphics[scale=0.47]{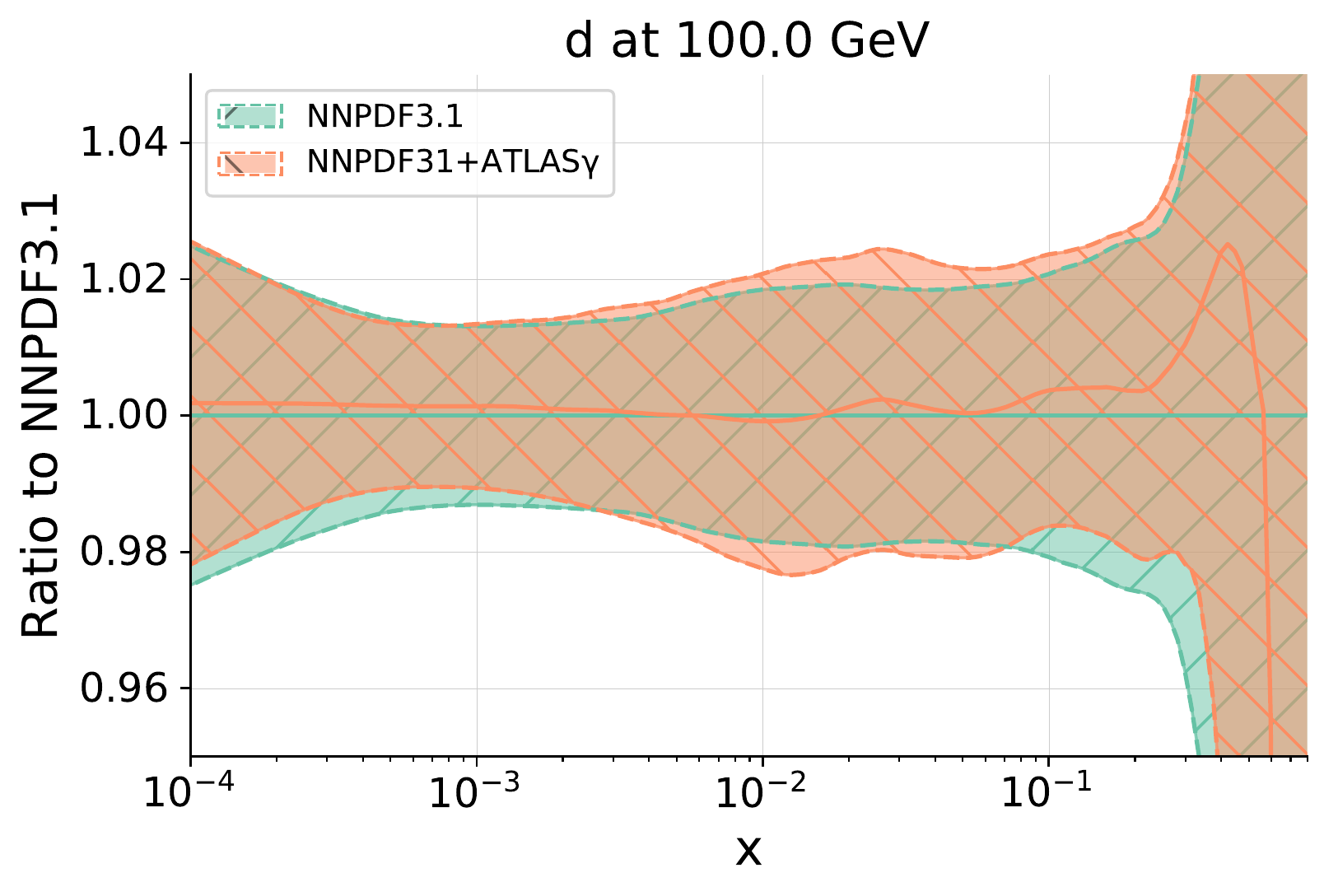}
\includegraphics[scale=0.47]{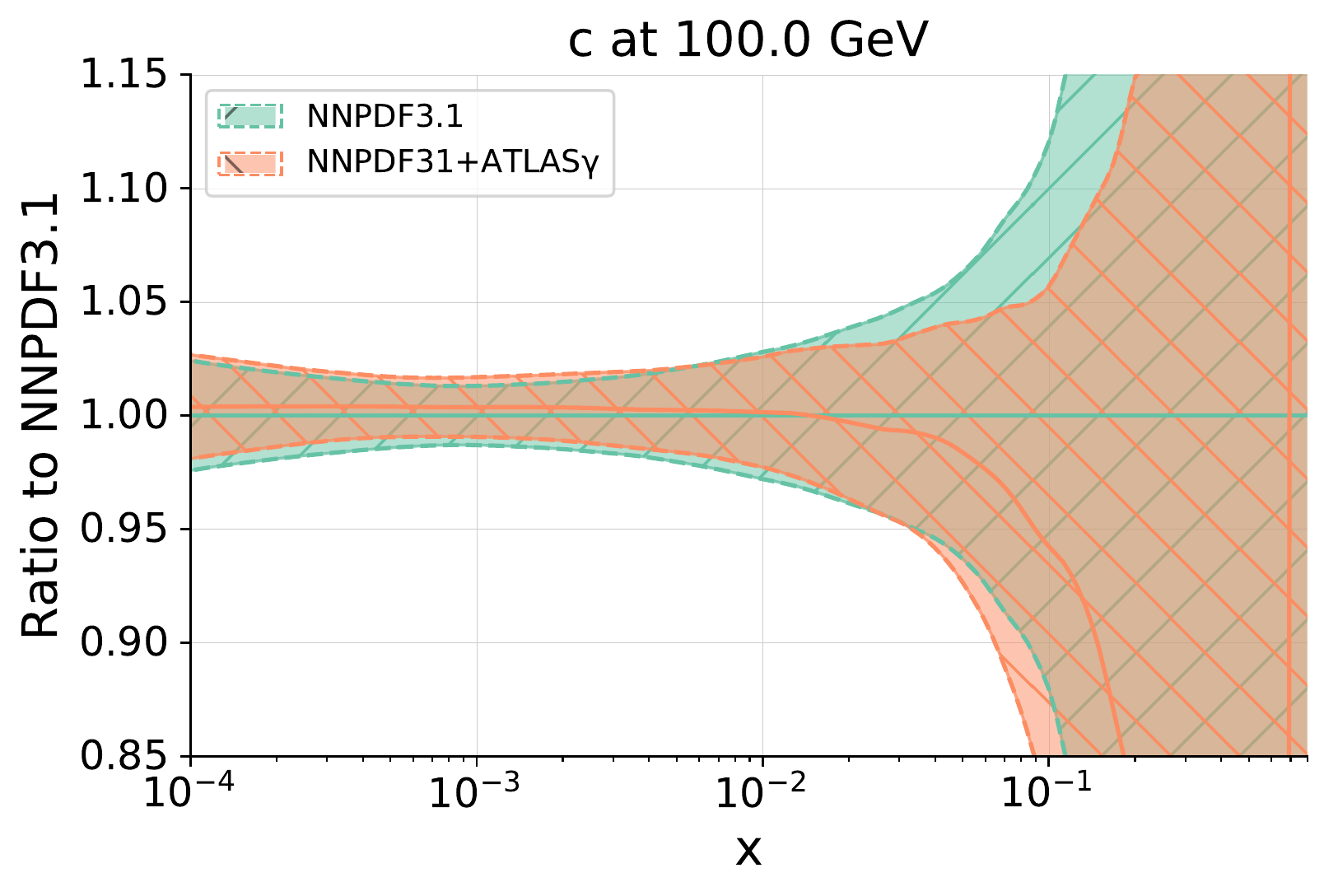}
 \includegraphics[scale=0.47]{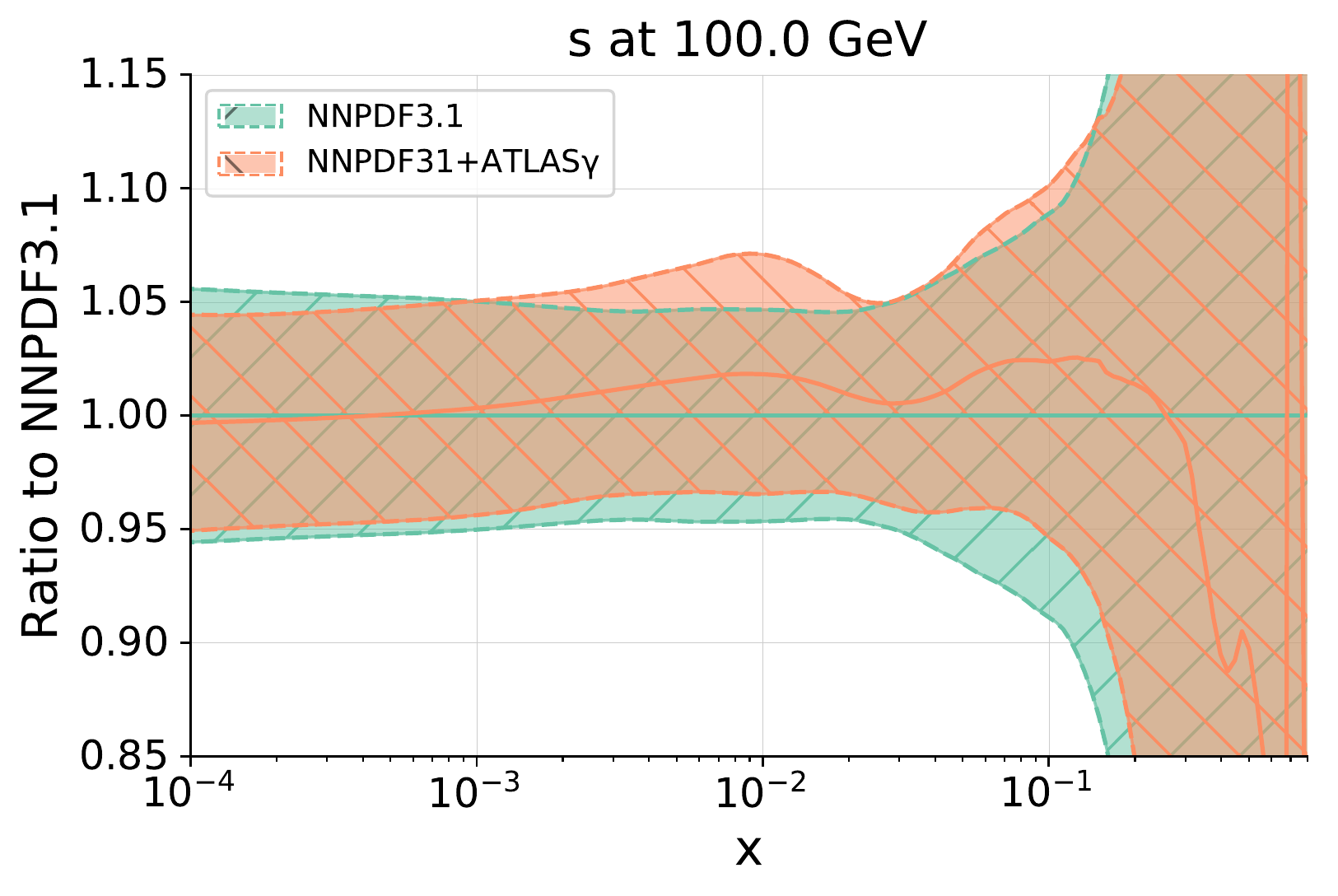}
 \caption{\small Comparison of the quark PDFs at $Q = 100$ GeV
   between the NNPDF3.1 and NNPDF3.1+ATLAS$\g$ fits, normalized
   to the central value of the former.
   \label{fig:quark_ratio}}
\end{figure}

In addition to the impact of the photon data on the gluon, it is important to determine 
if the new data is consistent with the quark PDFs.
In Fig.~\ref{fig:quark_ratio} we show the comparison of the quark PDFs at $Q = 100$ GeV
between the NNPDF3.1 and NNPDF3.1+ATLAS$\g$ fits. 
We find only rather small changes upon the addition of the photon data, both
in terms of central values and of uncertainties,
The exception is the charm PDF, which decreases in uncertainty across the full $x$ range, partly due
to its relation to the gluon via perturbative evolution.
We therefore conclude that the ATLAS data does not introduce tensions with the quark PDFs, and furthermore
does not strongly impact the size of their respective uncertainties. 

\begin{figure}[t]
\centering
  \includegraphics[scale=0.72]{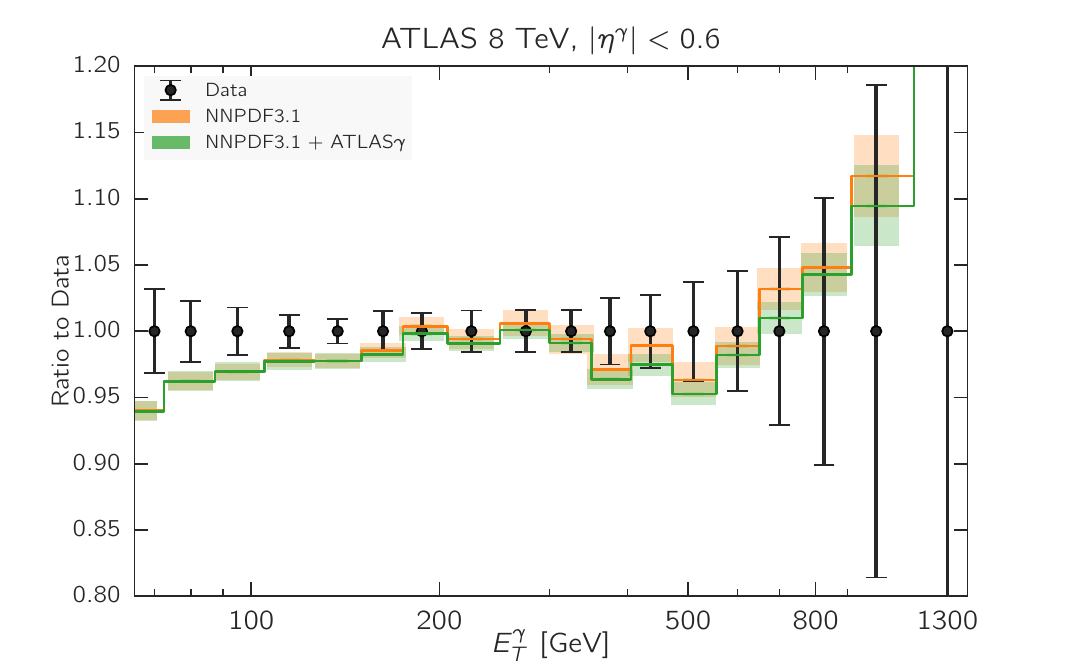} 
  \includegraphics[scale=0.72]{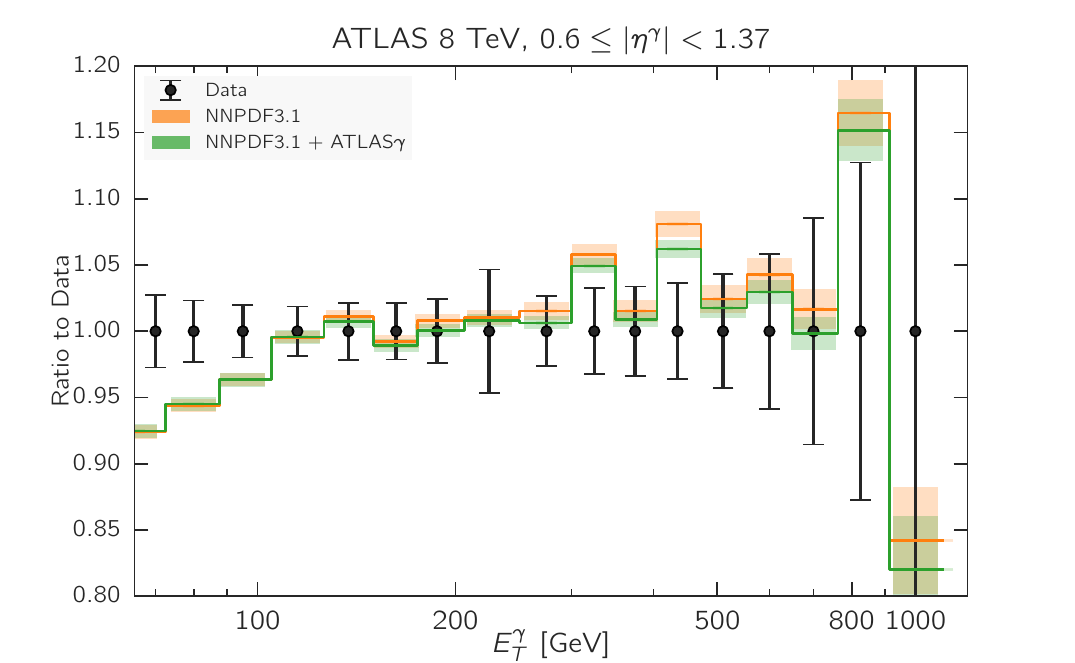} \\
  \includegraphics[scale=0.72]{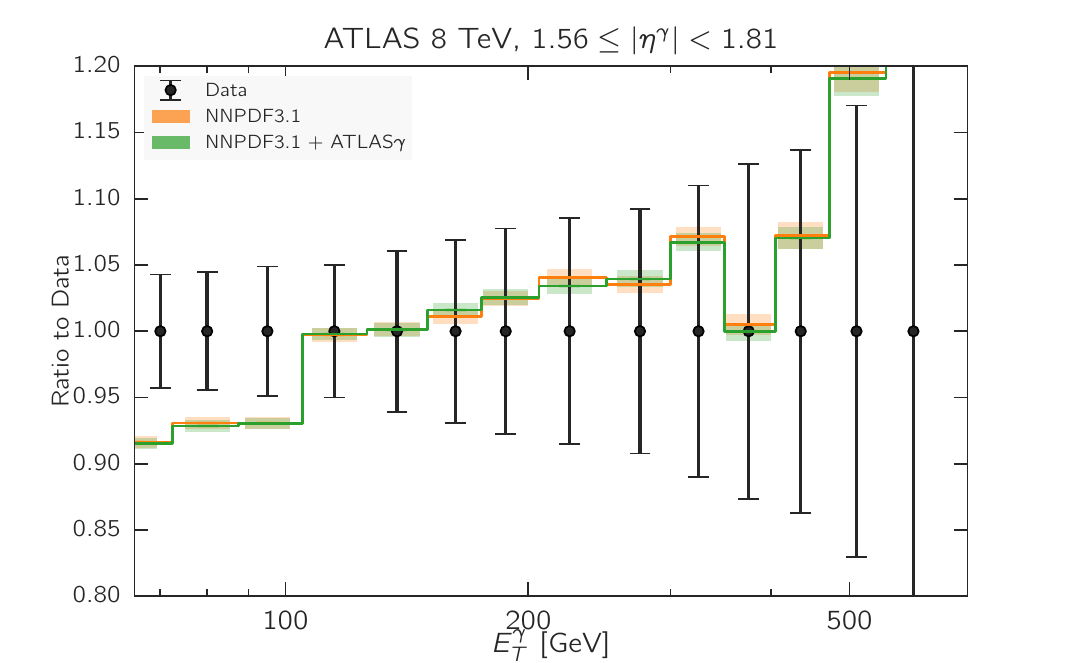} 
  \caption{\small Same as Fig.~\ref{fig:data_theory_ratios}, now
    comparing the NNPDF3.1 and NNPDF3.1+ATLAS$\g$ sets
    for the three rapidity bins of the ATLAS 8 TeV data included in the fit.}
\label{fig:data_theory_ratios_direct}
\end{figure}

Finally, in Fig.~\ref{fig:data_theory_ratios_direct} we show
the same comparison between theory predictions and experimental
data as in Fig.~\ref{fig:data_theory_ratios} now for the NNPDF3.1 and NNPDF3.1+ATLAS$\g$ sets
for the three rapidity bins of the ATLAS 8 TeV data included in the fit.
We can see how in this case the predictions obtained with
NNPDF3.1+ATLAS$\g$ as an input move closer to the central values
of the experimental data as compared to the NNPDF3.1 baseline, although
by a small amount.
These findings are consistent with the corresponding variations
at the 
PDF level discussed in Figs.~\ref{fig:gluon_ratio} and~\ref{fig:quark_ratio}.

\subsection{Impact on fits based on reduced datasets}
\label{sec:reduced_fits}

In addition to assessing the impact of the ATLAS direct photon production data when added to
the NNPDF3.1 dataset, we have also studied its impact on fits using reduced
datasets, specifically the NNPDF3.1 collider-only fit and no-LHC fits.
The former excludes all DIS and Drell-Yan fixed-target data, with the motivation that
collider observables might be cleaner and under better theoretical control, while
the latter excludes all LHC measurements for specific applications such as 
in searches for BSM physics.

In the two cases,  we find a good overall agreement between theory and data,
as indicated in Table~\ref{table:chi2_NNPDF_alternative_priors}.
For the  fit without LHC data, the total $\chi^2/N_{\rm dat}$ is reduced
from 1.49 to 1.00.
Recall that in this fit the constraints on the medium and large-$x$ gluon are much looser,
basically coming only from the Tevatron jet data, and thus one expects the impact of the ATLAS
direct photon data to be more significant.
For the collider only fit, the total $\chi^2/N_{\rm dat}$ is already very good to begin
with, 0.94, and is further reduced to $\chi^2/N_{\rm dat}=0.87$ upon
the addition of the photon data.
This moderate improvement is consistent with the fact that the bulk of the gluon-sensitive datasets
in NNPDF3.1 are already included in the collider-only dataset.
Another interesting result from Table~\ref{table:chi2_NNPDF_alternative_priors} is that
in all cases an improved description of the three rapidity bins is obtained.

\begin{table}[t]
  \centering
  \renewcommand{\arraystretch}{1.20}
  \small
{\begin{tabular}{  c | c |c | c | c}
PDF set& \multicolumn{4}{c}{$\chi^2 /N_\text{dat}$ } \\ \toprule
 & 1st bin & 2nd bin & 3rd bin & Total  \\ \midrule
NNPDF3.1 no LHC data  &1.26  & 2.07 & 0.96 &1.49    \\
NNPDF3.1 no LHC data+ATLAS$\g$ & 0.66 & 1.39 &0.84 & 1.00  \\
\midrule
 NNPDF3.1 collider only  & 0.89 & 1.29 & 0.68  &  0.94 \\
 NNPDF3.1 collider only+ATLAS$\g$ & 0.92 & 1.15 & 0.64 & 0.87  \\ \bottomrule
\end{tabular}}
\vspace{0.3cm}
\caption{\small Same as Table~\ref{table:chi2_NNPDF_comp},
  now for the NNPDF3.1 fits based on reduced datasets.
}\label{table:chi2_NNPDF_alternative_priors}
\end{table}

\begin{figure}[t]
\centering
\includegraphics[scale=0.47]{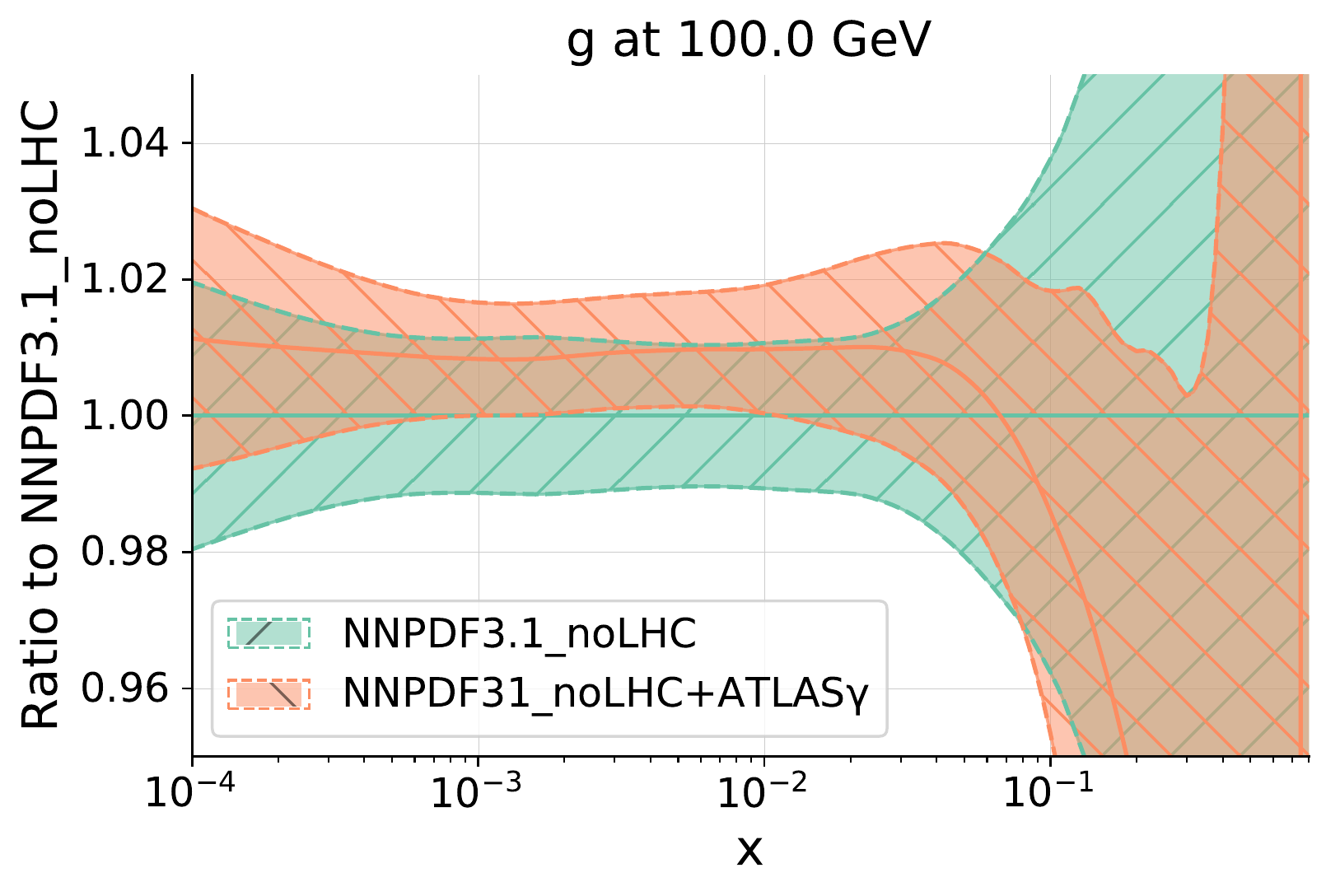}
\includegraphics[scale=0.47]{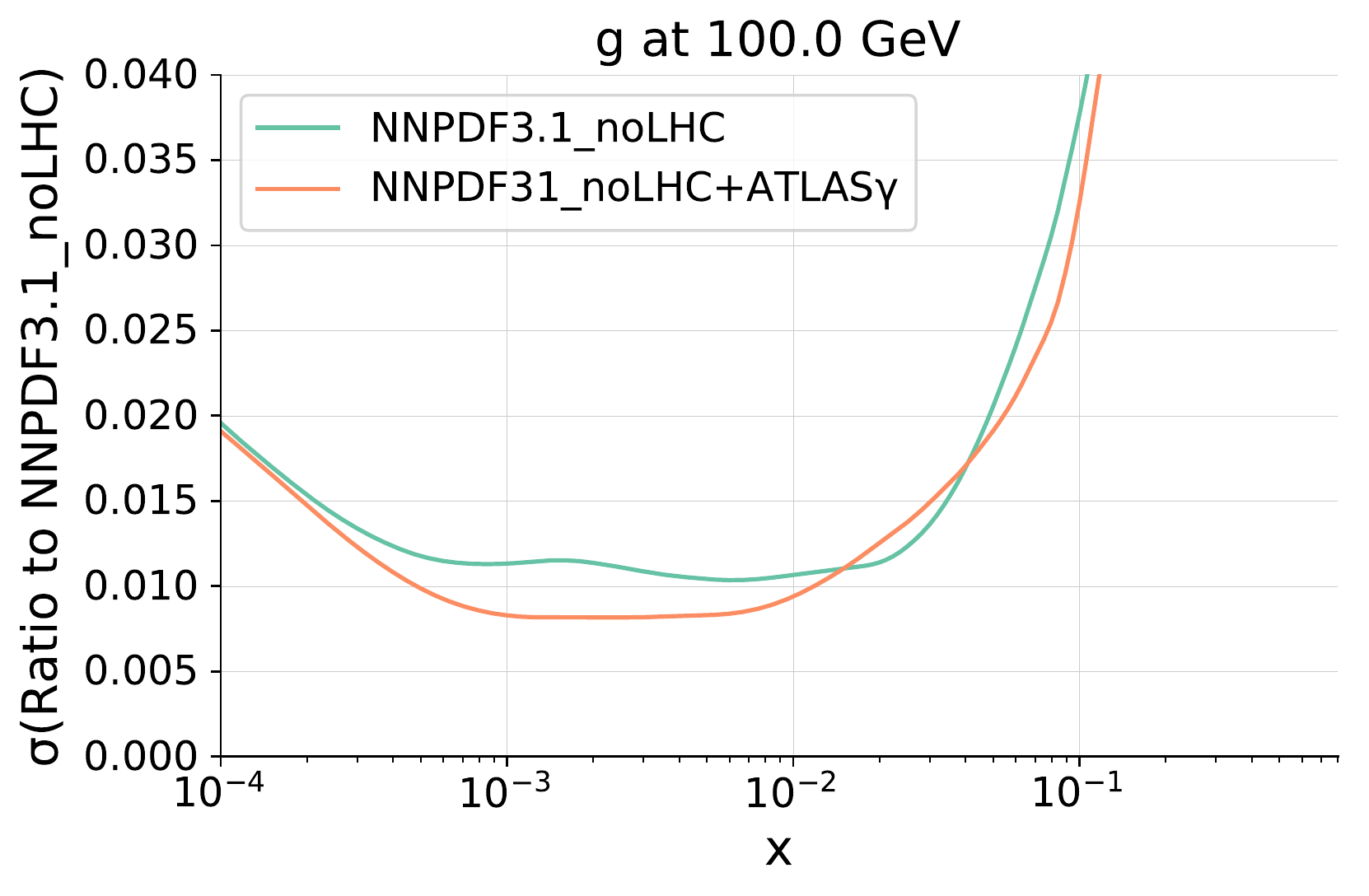}
\includegraphics[scale=0.47]{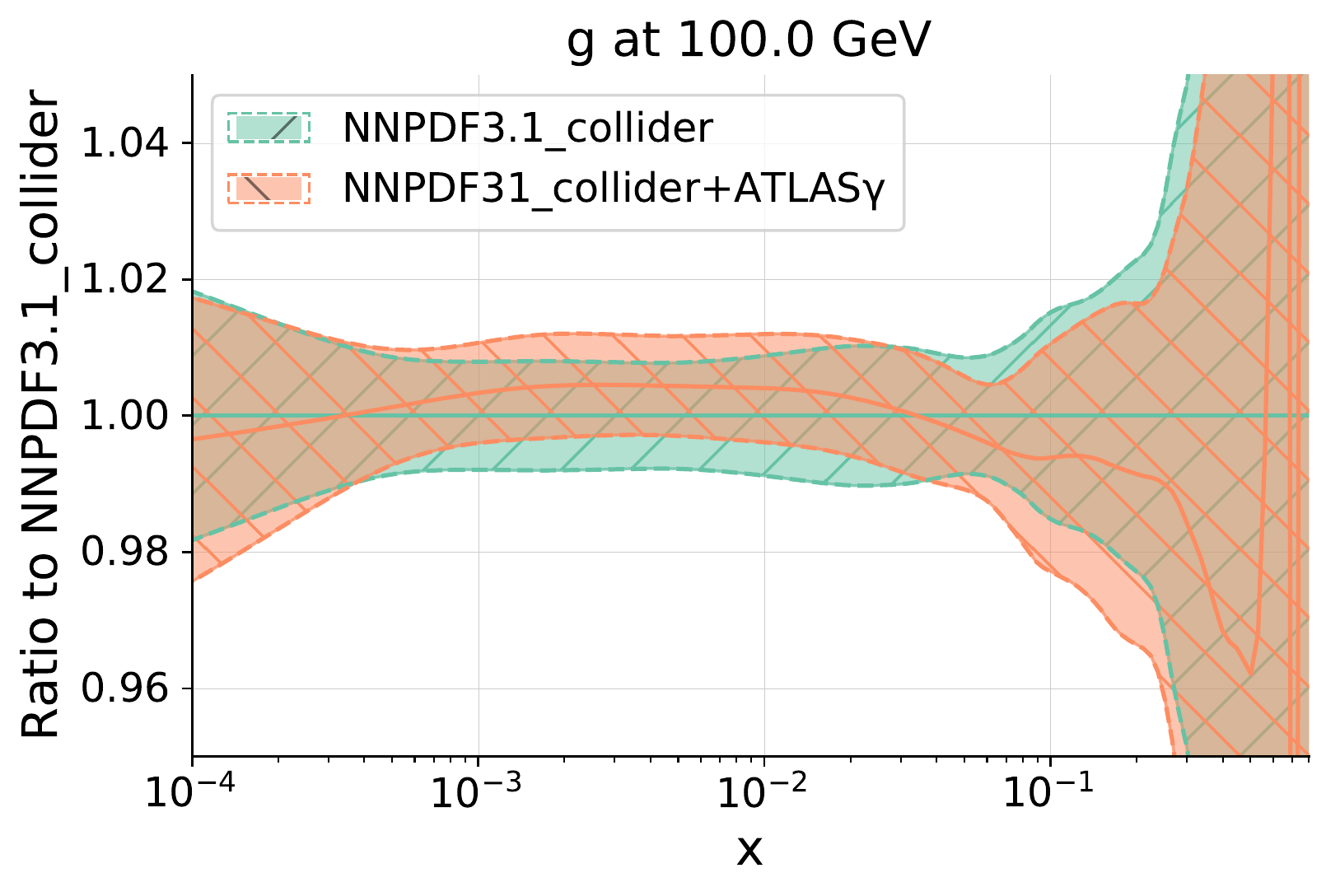}
\includegraphics[scale=0.47]{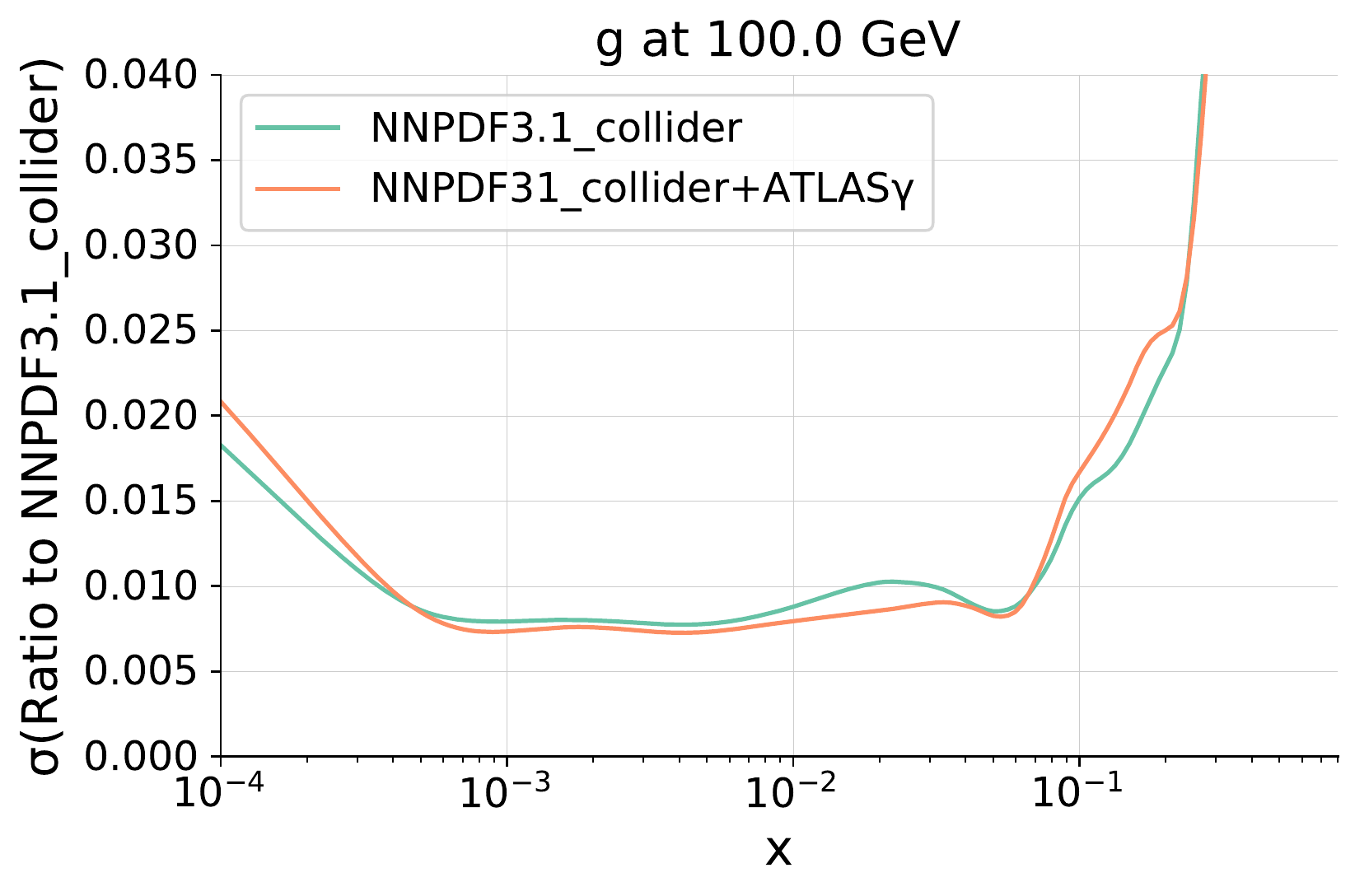}
\caption{\small Same as Fig.~\ref{fig:gluon_ratio} for the NNPDF3.1 no-LHC
  (upper) and collider-only (lower plots) fits. \label{fig:collider_err}}
\end{figure}

Next in Fig.~\ref{fig:collider_err} we show
the same comparisons as in Fig.~\ref{fig:gluon_ratio} but now
for the NNPDF3.1 no-LHC and collider-only fits.
In the case of the no-LHC fit, we find that the impact of adding the ATLAS photon
data is larger than in the global fit, both in terms of the shift
in the gluon central values and in the reduction of its PDF
uncertainties.
This is consistent with the fact that $g(x,Q^2)$ is less constrained in the
no-LHC fit than in the global fit.
The trend in central values is the same for the collider-only fits:
moderate enhancement at medium $x$ followed
by a suppression at large $x$.
The impact of the ATLAS photon data is also moderate at the level
of PDF uncertainties in the collider-only fit.

To summarize, the results of this study
demonstrate that the qualitative impact of the ATLAS 8 TeV direct photon production
data on the gluon PDF in the fits based on reduced datasets is consistent
with that of the global analysis.
In all cases, we find an improvement in the quantitative description
of the ATLAS data, as shown in Table~\ref{table:chi2_NNPDF_alternative_priors}.
Interestingly, we also find that the direct photon data prefer
a softer gluon at large $x$ irrespective of the input dataset used, a 
trend that is similar to the one induced by the top-quark pair differential cross-sections.

\section{Direct photon production at 13 TeV}
\label{sec:13tev}

In this section we present the comparison between state-of-the-art
theoretical predictions and experimental data for the
recent ATLAS measurements of direct photon production at 13 TeV~\cite{Aaboud:2017cbm}.
The motivation is two-fold.
On the one hand, we want to verify whether or not we
can quantitatively describe direct photon production at 13 TeV,
and in particular understand if the disagreement found for the most
forward bin at 8 TeV (see Sect.~\ref{sec:data-theory}) is also present
at a higher center-of-mass energy.
On the other hand, we aim to provide predictions for direct
photon production at 13 TeV that include the constraints from
the same process at 8 TeV: we will do this by using
the NNPDF3.1+ATLAS$\g$ fit constructed in
the previous section.

To begin with, in Table~\ref{table:chi2_globalfits_13tev}
we provide the $\chi^2/N_{\rm dat}$ values for different
NNLO PDF sets to the ATLAS 13 TeV measurements using
the theory settings described in Sect.~\ref{sec:theory}.
We also include here the predictions using the NNPDF3.1+ATLAS$\g$ set,
which accounts for the constraints of the 8 TeV photon measurements.
We find that the different PDF sets provide an equally satisfactory description
of this dataset, with the total $\chi^2/N_{\rm dat}\simeq 1$ in all cases.
In particular, we find an excellent description of the most forward
rapidity bin (with the exception perhaps of ABMP16), in contrast to what
was found at 8 TeV.
One should note, however, that this measurement is based on a relatively small
integrated luminosity, $\mathcal{L}_{\rm int}=3.2$ fb$^{-1}$,
and therefore its uncertainties are larger than for the 8 TeV case,
explaining the reduced discrimination power.

\begin{table}[t]\centering
  \renewcommand{\arraystretch}{1.20}
\begin{tabular}{  c | c |c | c | c|c }
    PDF set& \multicolumn{5}{c}{$\chi^2 /N_\text{dat}$}
    \\\toprule
& 1st bin & 2nd bin & 3rd bin& 4th bin& Total \\ \midrule
NNPDF3.1			&0.68   	& 0.53   	& 0.28  	&0.47  	&  0.65 \\
NNPDF3.1+ATLAS$\g$     & 0.70	&0.49	&0.30	&0.46	& 0.65 \\
\midrule
MMHT14				&0.81 	&0.73  	&0.27 	&  0.45	& 0.70 \\
CT14				& 0.75	&0.65  	& 0.28  &  0.41	&0.64  \\ 
ABMP16				& 0.82	&   0.89	& 0.20 	&  1.56 	&1.05  \\
\bottomrule
\end{tabular}
\vspace{0.3cm}
\caption{\small
  Same as Table.~\ref{table:chi2_globalfits}
for the ATLAS 13 TeV direct photon production measurements.
}\label{table:chi2_globalfits_13tev}
\end{table}

\begin{figure}[t]
\centering
  \includegraphics[scale=0.72]{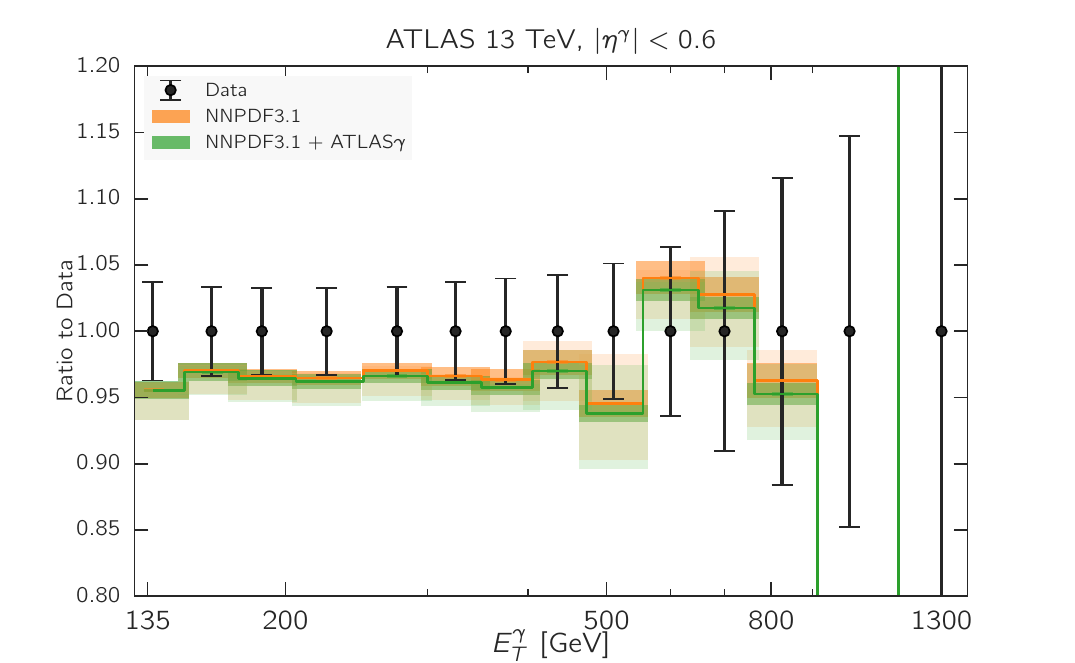} 
  \includegraphics[scale=0.72]{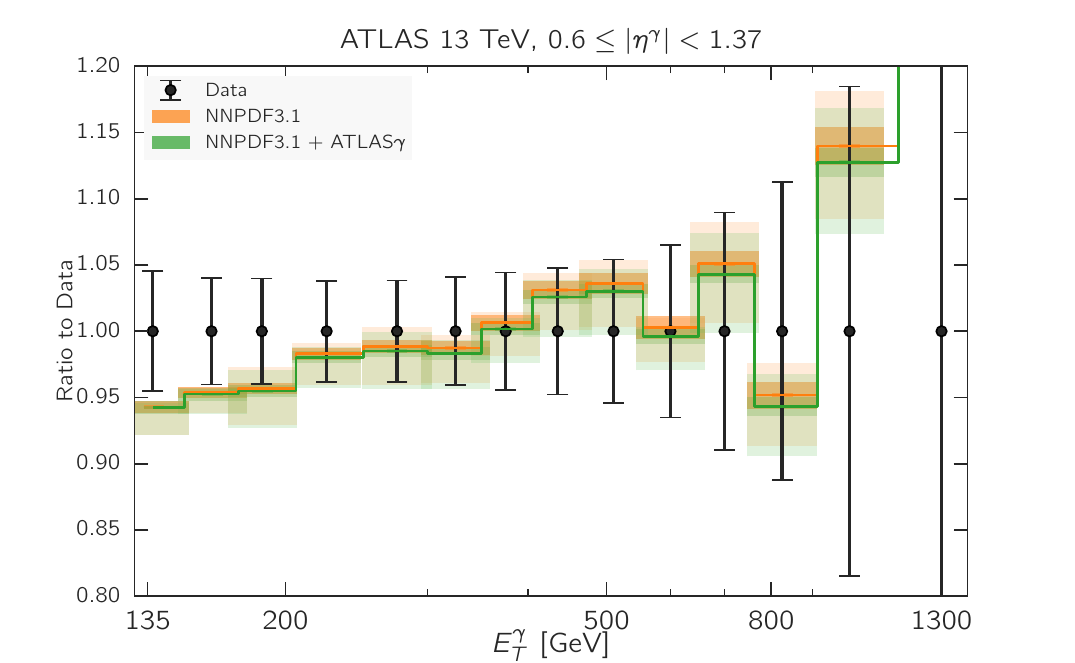} \\
  \includegraphics[scale=0.72]{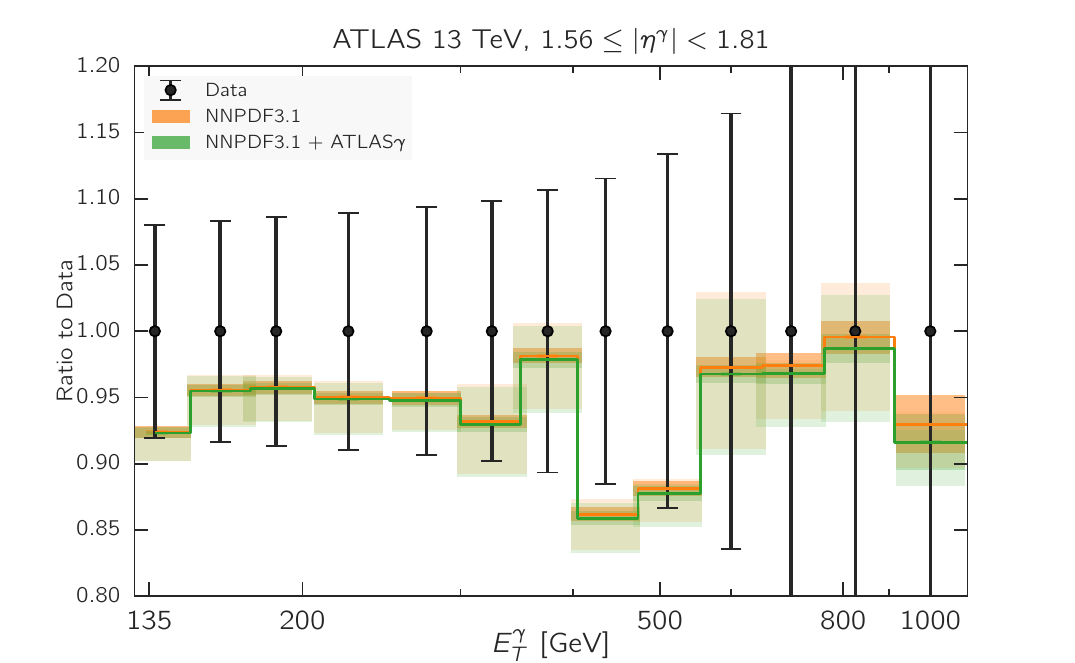}
  \includegraphics[scale=0.72]{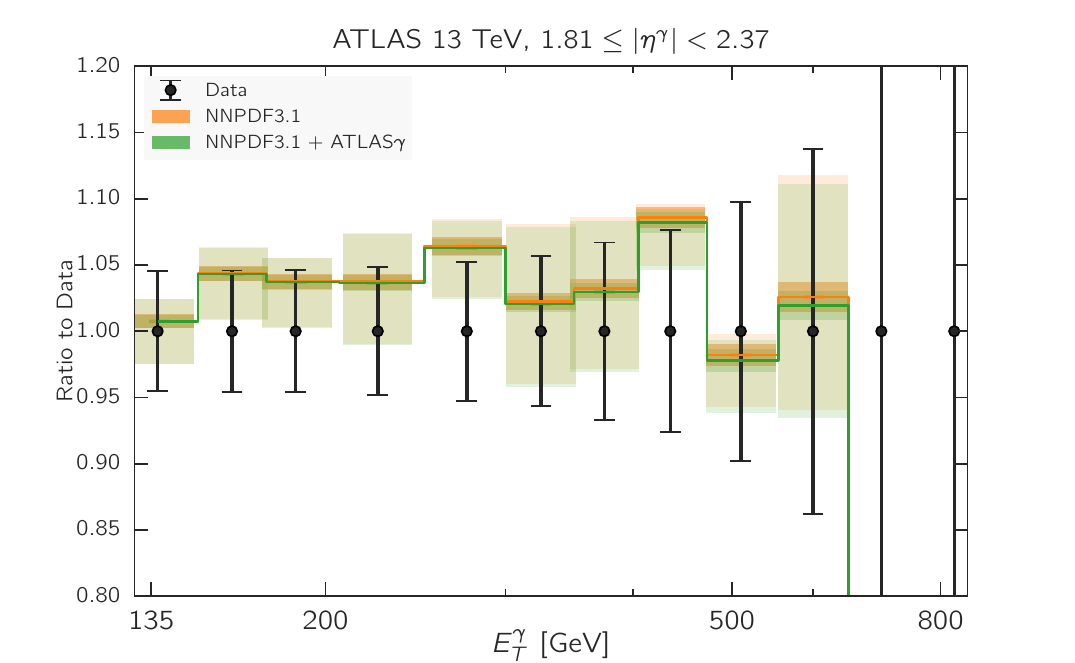}  
  \caption{\small  Same as Fig.~\ref{fig:data_theory_ratios_direct}
    for the ATLAS 13 TeV direct photon measurements.
    In addition to the PDF uncertainties shown in the previous cases (darker bands),
    here we also include the scale uncertainties associated to the NNLO QCD calculation (lighter bands).
     }
\label{fig:data_theory_ratios_13tev}
\end{figure}

As can be seen
from Table~\ref{table:chi2_globalfits_13tev},  the differences
in the values of $\chi^2$ between NNPDF3.1 and NNPDF3.1+ATLAS$\gamma$ are small.
This may be further observed in Fig.~\ref{fig:data_theory_ratios_13tev},
where we compare the theory predictions for the 13 TeV data with both
NNPDF3.1 and NNPDF3.1+ATLAS$\g$.
 In addition to the PDF uncertainties shown in the previous cases (darker bands),
 here we also include the scale uncertainties associated with the NNLO QCD calculation (lighter bands),
 as discussed below.
The two PDF sets are in good agreement with each other
and the limited statistics of the measurement do not allow us to discriminate
among them.
This can also be seen from the fact that
the experimental uncertainties
are significantly larger than the differences between the two
theoretical predictions.
It is also interesting to take a closer look
at the most forward rapidity bin of the 13 TeV measurement,
which in the 8 TeV case had to be excluded from the fit.
Here instead we find reasonably good agreement between theory and data, although again,
there are larger experimental errors in this bin and therefore one cannot conclude
that the description of the 13 TeV data is better than at 8 TeV. 

As mentioned above, we also indicate in Fig.~\ref{fig:data_theory_ratios_13tev} the 
scale uncertainties associated with the NNLO QCD calculation (shown as the lighter error bands)
in addition to the standard PDF uncertainties.
These scale uncertainties have
been estimated using the standard practice of independently varying the renormalization $\mu_R$ and factorization
$\mu_F$ scales by a factor of two.
For the majority of $E_T^\g$ bins, the scale uncertainty is $\mathcal{O}(5\%)$, reaching a
maximum of $\mathcal{O}(10\%)$ in the most forward rapidity bin at high $E_T^\g$.
At NLO,
we find the typical size of the scale uncertainty to be approximately double that of the NNLO one,
thus compounding the requirement to have the NNLO predictions in order to adequately
describe the direct photon data.

One of the main differences that arises in the comparison between data and theory
at 8 TeV and 13 TeV, as we discussed, is that the most forward rapidity bin is poorly described
in the former case, while it is reasonably well described in the latter.
A possible way forward to understand the origin of this discrepancy
is to take ratios of the cross-section measurements at the two centre-of-mass energies.
Such ratios are useful since many theoretical and experimental systematic
uncertainties cancel out~\cite{Mangano:2012mh}, allowing us to elucidate
possible issues arising for individual center-of-mass energies.
With this motivation, we have constructed the following ratio
\be
\label{eq:ratiodef}
R_{13/8}(E_T^\gamma,\eta^\gamma) \equiv \frac{
  d\sigma(E_T^\gamma,\eta^\gamma)
}{
  dE_T^\gamma d\eta^\gamma
}\Bigg|_{13~{\rm TeV}}\Bigg/
\frac{
  d\sigma(E_T^\gamma,\eta^\gamma)
}{
  dE_T^\gamma d\eta^\gamma
  }\Bigg|_{8~{\rm TeV}} \, ,
\ee
for those bins where both $E_T^\g$ and $\eta^\g$ overlap between the
two center of mass energies, corresponding to a total
of 47 bins.
Since the experimental covariance matrix is not available
at 13 TeV and the description of the 4th rapidity bin is poor at 8 TeV both with and without the covariance matrix, the uncertainty
on the ratio Eq.~(\ref{eq:ratiodef}) is obtained by adding
in quadrature the total experimental errors in the numerator
and the denominator.
For the theoretical calculation of Eq.~(\ref{eq:ratiodef}), the correlation
between the PDF uncertainties at 8 TeV and 13 TeV is accounted for.

\begin{figure}[t]
\centering
  \includegraphics[scale=0.71]{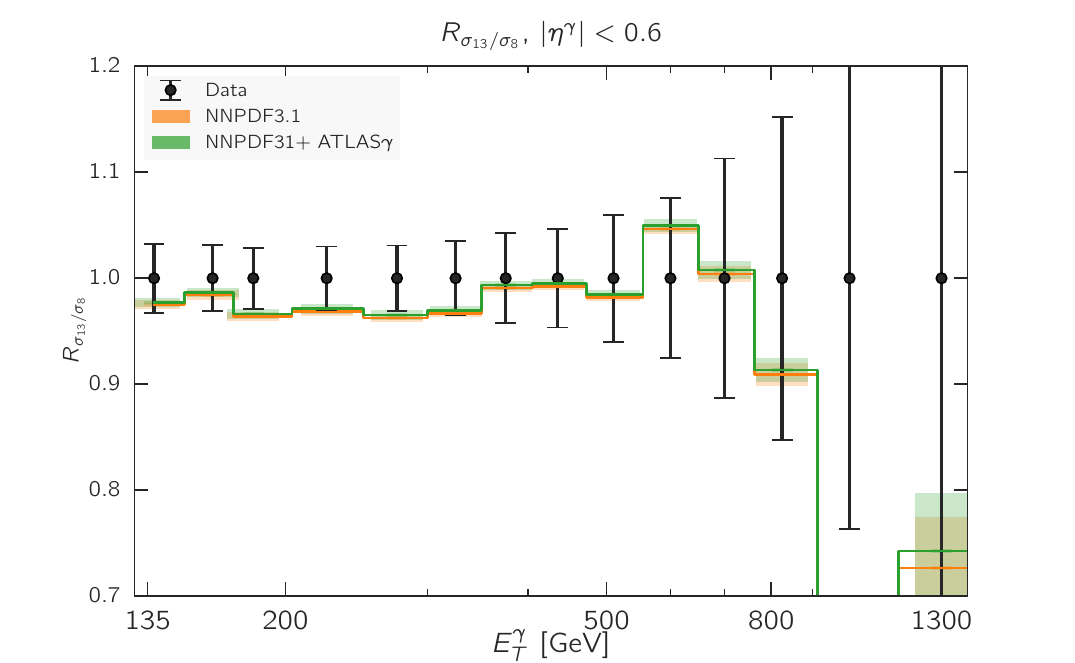} 
  \includegraphics[scale=0.71]{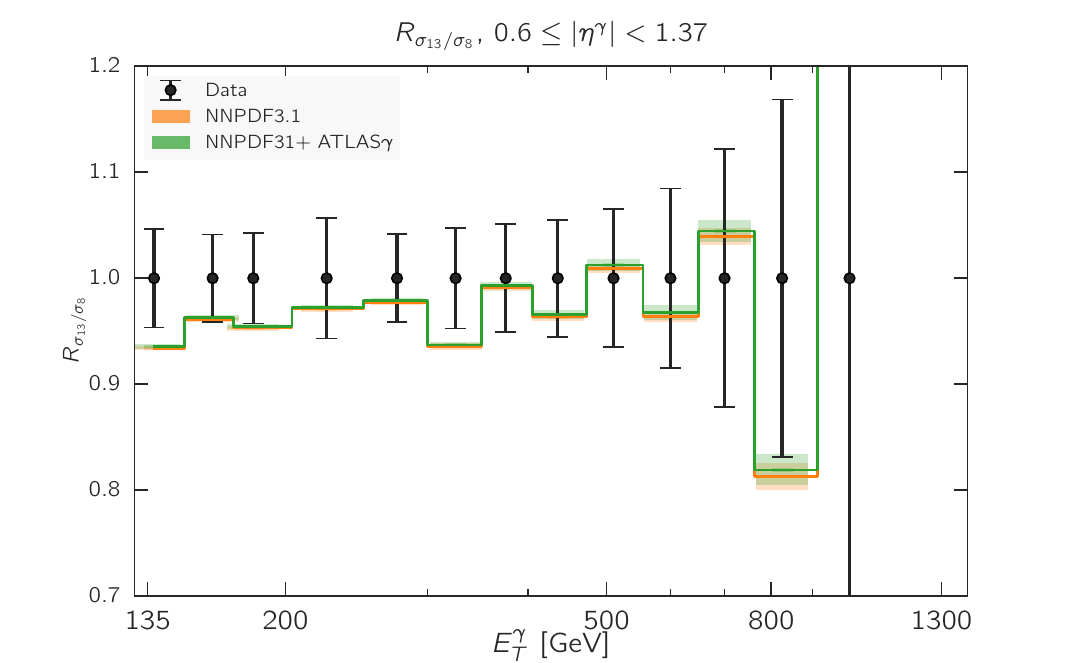} \\
  \includegraphics[scale=0.71]{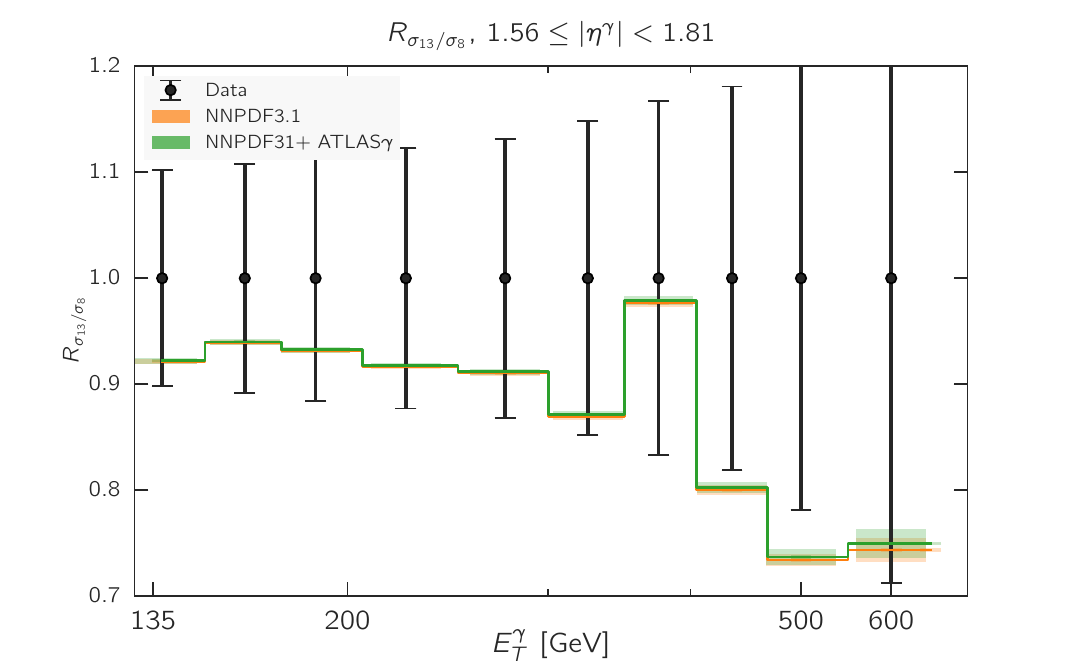}
    \includegraphics[scale=0.71]{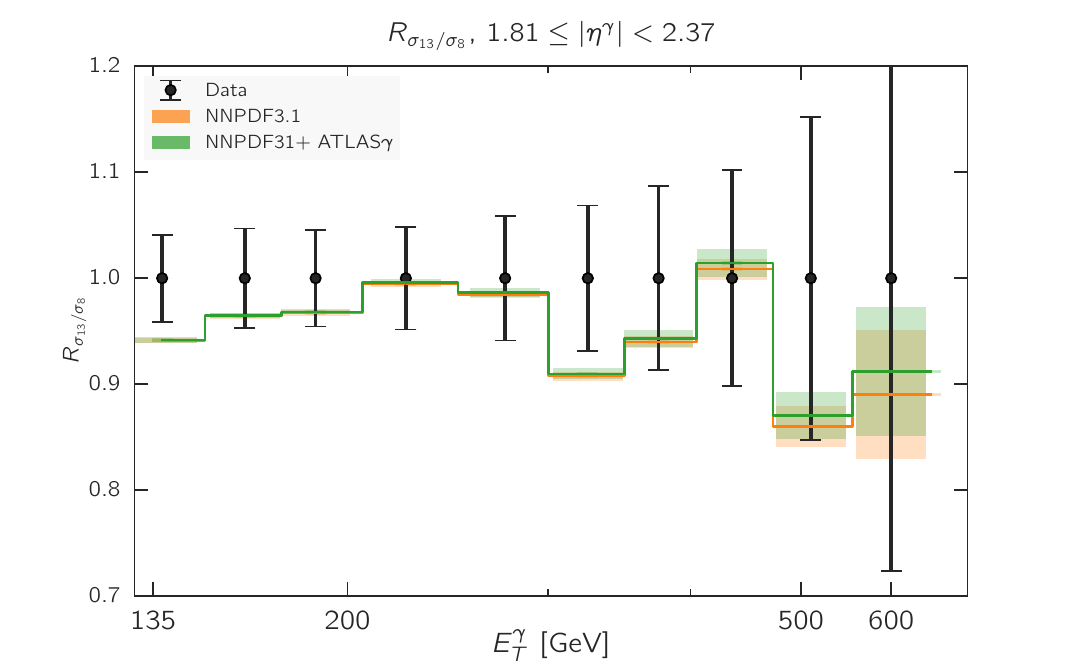}  
    \caption{\small Comparison between the experimental
      measurements of the $R_{13/8}(E_T^\gamma,\eta^\gamma)$ ratio
      and the corresponding theoretical calculations using NNPDF3.1 and NNPDF3.1+ATLAS$\g$,
      normalized to the central experimental value.
      The theory band includes only the contribution from the
      PDF uncertainties.
    }
\label{fig:data_theory_813tev}
\end{figure}

In Fig.~\ref{fig:data_theory_813tev}
we show a  comparison between the experimental
measurements of the $R_{13/8}(E_T^\gamma,\eta^\gamma)$ ratio, Eq.~(\ref{eq:ratiodef}),
with the corresponding calculations using the NNPDF3.1 and NNPDF3.1+ATLAS$\g$ sets,
normalized to the central value of the experimental data.
Here the theoretical uncertainty band includes only the contribution from the
PDF uncertainties.
From this comparison we find that there is good agreement
between data and theory for all the bins,
including for the most forward rapidity bin which
was problematic at 8 TeV.
The results of Fig.~\ref{fig:data_theory_813tev} suggest
that the underlying reason for the disagreement at 8 TeV in
the most forward bin, either an inadequacy of the theory
calculation or some issue with the experimental measurement,
is a common effect between the two center of mass energies
which mostly cancels out when computing their ratio.
      
\begin{figure}[t]
  \centering
  \includegraphics[scale=0.72]{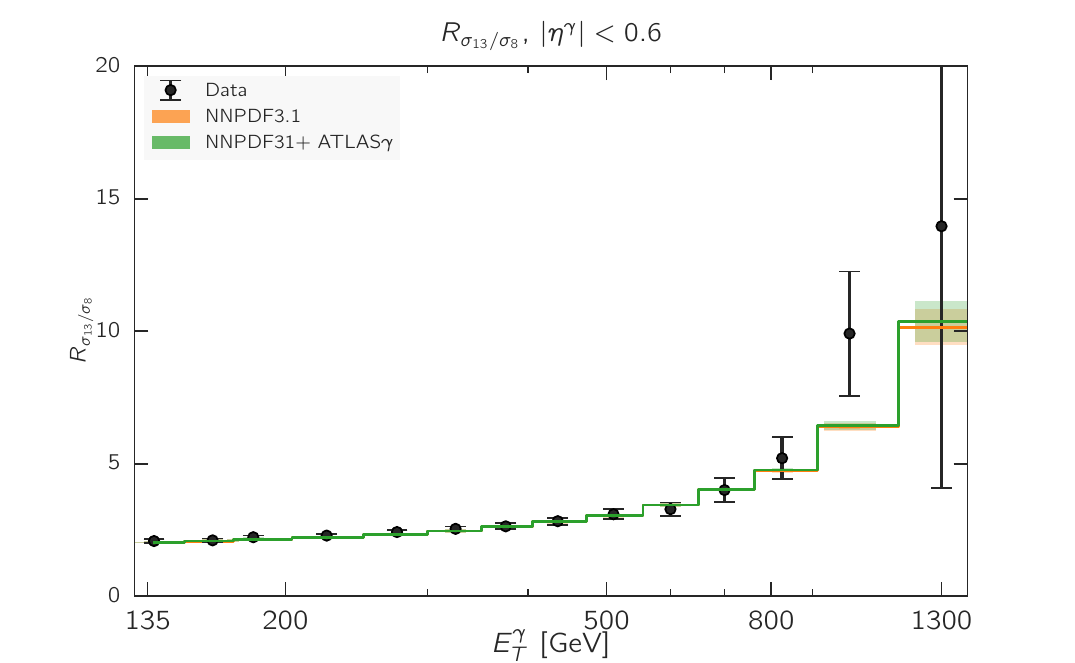} 
\includegraphics[scale=0.72]{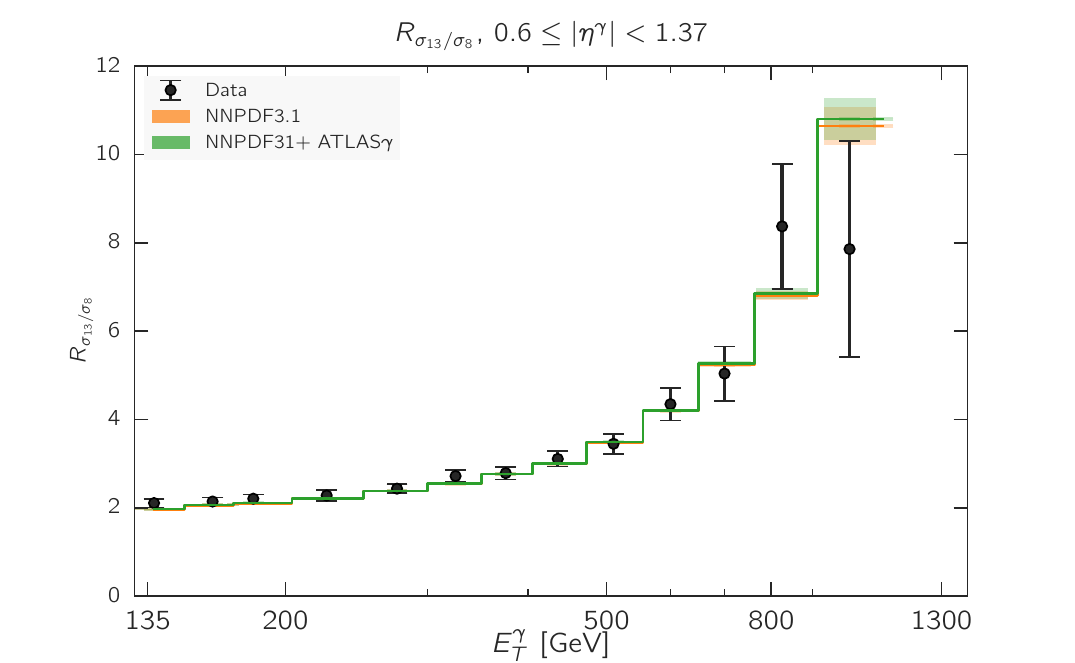} \\
\includegraphics[scale=0.72]{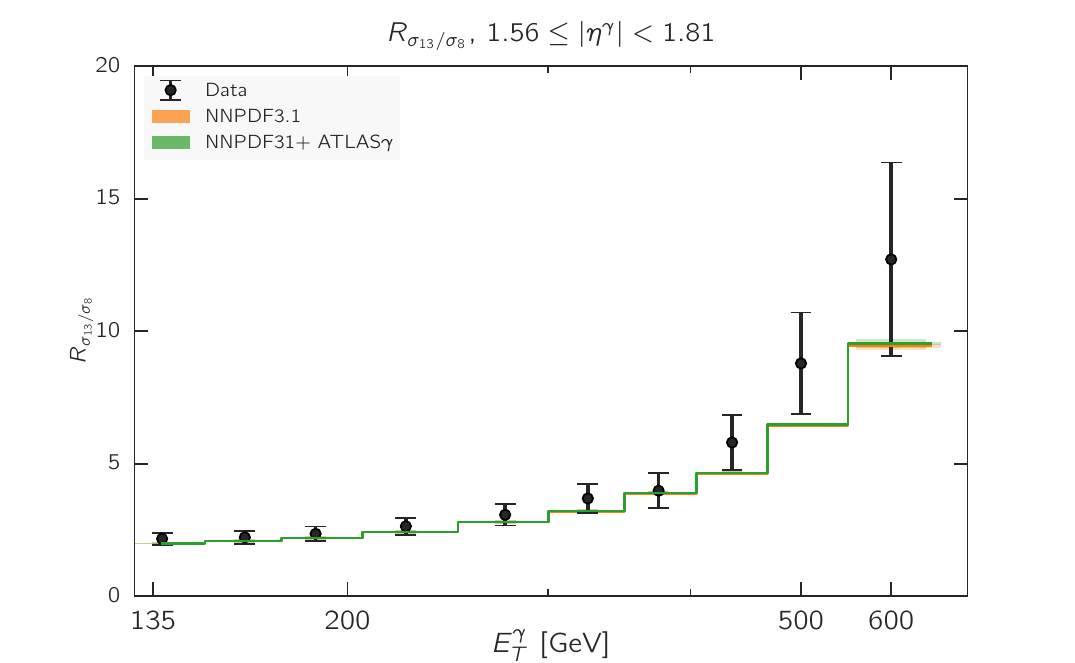}
\includegraphics[scale=0.72]{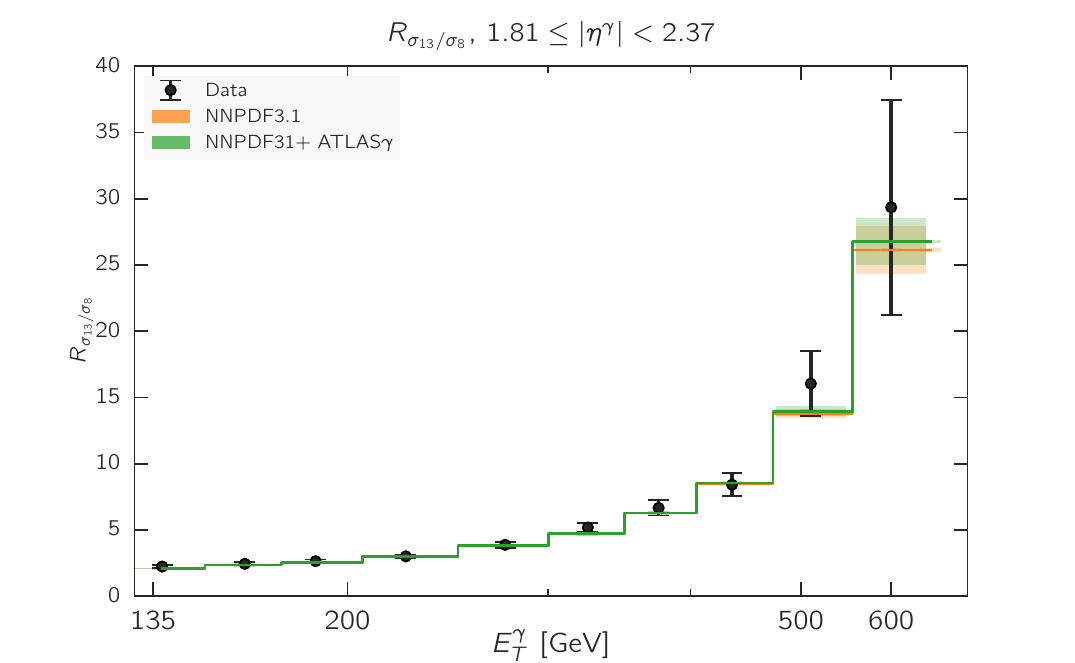}  
    \caption{\small Same as in Fig.~\ref{fig:data_theory_813tev} without normalizing to the
      experimental data.
    }
\label{fig:data_theory_813tev_unnorm}
\end{figure}

In order to further understand how the cross-section ratio Eq.~(\ref{eq:ratiodef})
behaves as a function of $E_T^\g$ and $\eta^\g$,
in Fig.~\ref{fig:data_theory_813tev_unnorm} we show the same comparison as in Fig.~\ref{fig:data_theory_813tev}
but this time without normalizing to the experimental data.
We can see that there is excellent agreement between theory and data in all rapidity bins, both at low and
high values of $E_T^\g$; moreover, 
we can also observe that the trend in the data-theory agreement is consistent across all the rapidity bins.
These results therefore compound the argument that there is some inadequacy in either
the theory or the experimental analysis for the most forward bin at $\sqrt{s}=8$ TeV.
      
In order to substantiate this point, in Table~\ref{table:chi2_globalfits_ratio}
we provide the $\chi^2/N_{\rm dat}$ values for the ratio of cross-sections between
13 TeV and 8 TeV,
Eq.~(\ref{eq:ratiodef}), with different input PDFs.
As we can see from this comparison, all PDF sets are in agreement
with the cross-section ratios, both for the total dataset and
for the individual rapidity bins.
In particular, we find that the description of the cross-section ratio $R_{13/8}$
in the most forward bin is the best with NNPDF3.1+ATLAS$\g$.
Table~\ref{table:chi2_globalfits_ratio} provides further evidence that the origin
of the disagreement between theory and data at 8 TeV in this bin is common
with the 13 TeV case, since it mostly cancels in the ratio.

\begin{table}[t]\centering
  \renewcommand{\arraystretch}{1.20}
\begin{tabular}{  c | c |c | c | c|c }
    PDF set& \multicolumn{5}{c}{$\chi^2 /N_\text{dat}$}
    \\\toprule
& 1st bin & 2nd bin & 3rd bin& 4th bin& Total \\ \midrule
NNPDF3.1			&  0.66  	& 0.75  	&  0.79 	& 0.45   &0.68    \\
NNPDF3.1+ATLAS$\g$     &  0.58	& 0.72	&0.77 	&0.41   & 0.64     \\
\midrule
MMHT14				&  0.96	&  0.85 	& 0.83   &  0.50   &  0.82     \\
CT14				& 0.90	&0.80   	& 0.80     &   0.52  & 0.79 \\ 
ABMP16				& 0.84	&  0.90  	& 0.95     &  0.69    & 0.87    \\
\bottomrule
\end{tabular}
\vspace{0.3cm}
\caption{\small
  Same as Table~\ref{table:chi2_globalfits}
  for the ratio of cross-sections between 13 TeV and 8 TeV,
  Eq.~(\ref{eq:ratiodef}).
}\label{table:chi2_globalfits_ratio}
\end{table}

\section{Summary}
\label{sec:conc}

The quantitative understanding of the detailed features of photon production at the LHC is
of crucial importance for a wide range of analyses, from searches for Higgs decays and BSM resonances
to precision Standard Model measurements.
In this work, we have revisited the possibility of using direct photon production 
from the LHC to constrain the parton distribution functions of the proton
within a global QCD fit.
By using state-of-the-art NNLO QCD calculations combined with LL electroweak corrections,
we have quantified the impact of the ATLAS 8 TeV photon production
data on the gluon PDF from the NNPDF3.1 global analysis.

Our results indicate that the LHC direct photon production data
leads to both
a moderate reduction of the gluon uncertainties at medium-$x$
and a preference for a somewhat
softer central value at large-$x$.
These effects are more marked when the direct photon data is added on top
of fits based on reduced datasets, in particular the NNPDF3.1 no-LHC fit. 
We have also demonstrated that including both NNLO QCD and LL electroweak corrections is
required in order to achieve a quantitative agreement with
the experimental data for the entire kinematic range in $E_T^\g$ and $\eta^\g$.
Moreover, we find that the constraints from the direct photon data are consistent
with those of other gluon-sensitive measurements included in NNPDF3.1 such as the
$Z$ $p_T$, inclusive jets, and $t\bar{t}$ differential distributions.

Here we have also provided theoretical predictions for the ATLAS measurements
of direct photon production at 13 TeV as well as for the ratio of cross-sections
between 13 TeV and 8 TeV.
In this case, we find that  due to the relatively
small integrated luminosity used for the 13 TeV measurement,
its discrimination power is rather limited.
It would therefore be important to repeat the 13 TeV analysis using the full
integrated luminosity of Run II, in order to complement the information provided
by the 8 TeV data.
In this respect, it is essential that the experimental collaborations
make public the covariance matrices of their measurements,
else their lack of availability
limits the physics output that can be extracted from their own data.

Our results demonstrate that there is no reason, neither in principle
nor in practice, for excluding collider direct photon data
from a global PDF analysis.
Indeed, the most precise LHC measurements available agree well with state-of-the-art
theoretical predictions, and the latter  can be included in global PDF analyses using fast interpolation tables.
The information provided by the ATLAS 8 TeV direct photon measurements
turns out to be consistent with the constraints provided by other gluon-sensitive datasets included in
NNPDF3.1, and leads to a moderate
reduction of the gluon uncertainties.
For these reasons, collider direct photon production should be rightfully restored to its
well-deserved position as a full member of the global PDF analysis toolbox.\\

\vspace{0.2cm}

The main output of this work, the  NNPDF3.1+ATLAS$\g$ NNLO fit,
is available in the {\tt LHAPDF6} format~\cite{Buckley:2014ana}
from the NNPDF collaboration webpage
\begin{center}
\url{http://nnpdf.mi.infn.it/for-users/unpolarized-pdf-sets/}
\end{center}
with the file name \texttt{NNPDF31\_nnlo\_as\_0118\_directphoton}.
In addition,
the fast NLO tables
computed using {\tt MCFM} and {\tt APPLgrid} for the ATLAS 8 TeV and 13 TeV direct
photon measurements produced in this work, together with the corresponding
NNLO/NLO $K$-factors Eq.~(\ref{eq:kfactors}),
are also publicly available from the same website.

\subsection*{Acknowledgments}
We thank Ulla Blumenschein,  Ana Cueto G\'omez,  Jan Kretzschmar,
Claudia Glasman, Matthias Schott, and
Juan Terr\'on Cuadrado for information
and discussions concerning the ATLAS direct photon measurements.
J.~R. and E.~S. are grateful to our colleagues of the NNPDF collaboration for their support and useful discussions.
E.~S. thanks Rhorry Gauld for providing useful scripts and Alberto Guffanti
for help with the production of {\tt APPLgrids}.
J.~R. and E.~S. are supported by the ERC Starting Grant ``PDF4BSM''.
The work of J.~R. is also supported by the Dutch Organization for Scientific Research
(NWO).
C.W. is supported by a National Science Foundation CAREER award PHY-1652066.
This manuscript has been authored by Fermi Research Alliance, LLC under Contract No. DE-AC02-07CH11359 with the
U.S. Department of Energy, Office of Science, Office of High Energy Physics.

\appendix
\section{The impact of correlations in the systematic uncertainties}
\label{sec:systematicbreakdown}

The baseline results of this work, presented
in Sect.~\ref{sec:results}, have been obtained by adding
in quadrature the statistical and experimental uncertainties,
except for the luminosity uncertainty which is treated as fully
correlated among all the data bins.
As mentioned in Sect.~\ref{sec:expts}, we only realised
after the completion of this work 
that
the full breakdown of the experimental systematic uncertainties
corresponding to the ATLAS 8 TeV measurement
had just been posted in HepData.
The same breakdown of the
experimental systematic uncertainties, however, is not yet available
for the 13 TeV data.
In this appendix, we study the impact of accounting for
the effects of the correlations among the systematic
uncertainties of the 8 TeV ATLAS data
both at the level of parton distributions
and of the values of the $\chi^2$.

We will consider here two scenarios for the correlation model.
In the first one, all sources of systematic uncertainty are fully correlated
among the data bins.
In the second one, a subset of sources of systematic errors will
be considered as uncorrelated between data bins.
Specifically, using the notation used in the corresponding {\tt HepData} entry\footnote{\url{https://hepdata.net/record/ins1457605}},
the following sources are treated as point-to-point uncorrelated: {\tt sysPhotonID},
{\tt sysPhotonIsolation}, {\tt sysBackgroundID}, {\tt sysBackgroundIsolation}
{\tt sysEnergyResolution}.
The reason for these two choices is that, initially, the ATLAS analyzers
recommended to treat all errors as fully correlated, but at a later stage
they provided us with an updated recommendation for their correlation model based
on the decorrelation of some systematic sources.

\begin{figure}[h!]
\centering
\includegraphics[scale=0.47]{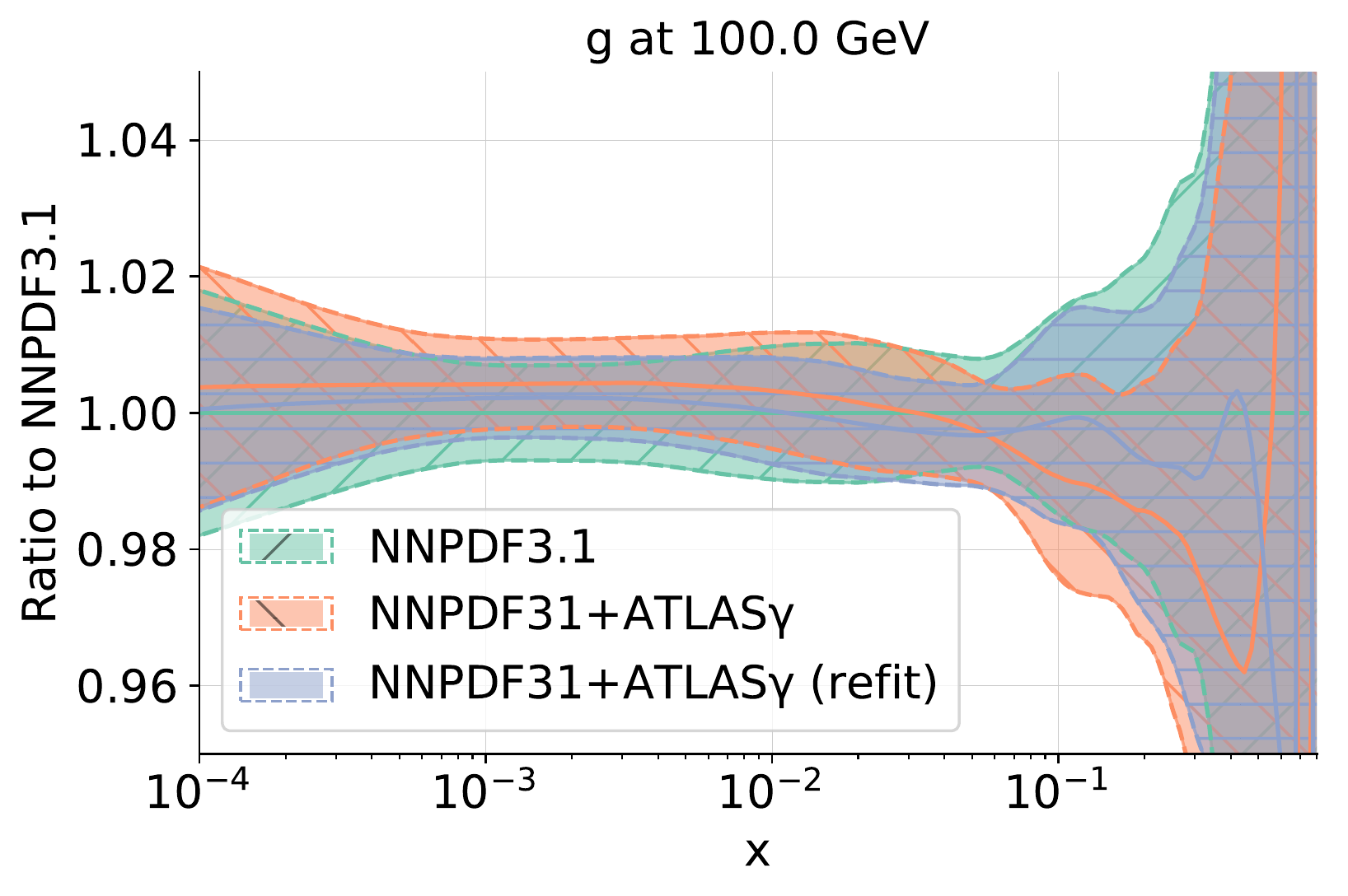}
 \includegraphics[scale=0.47]{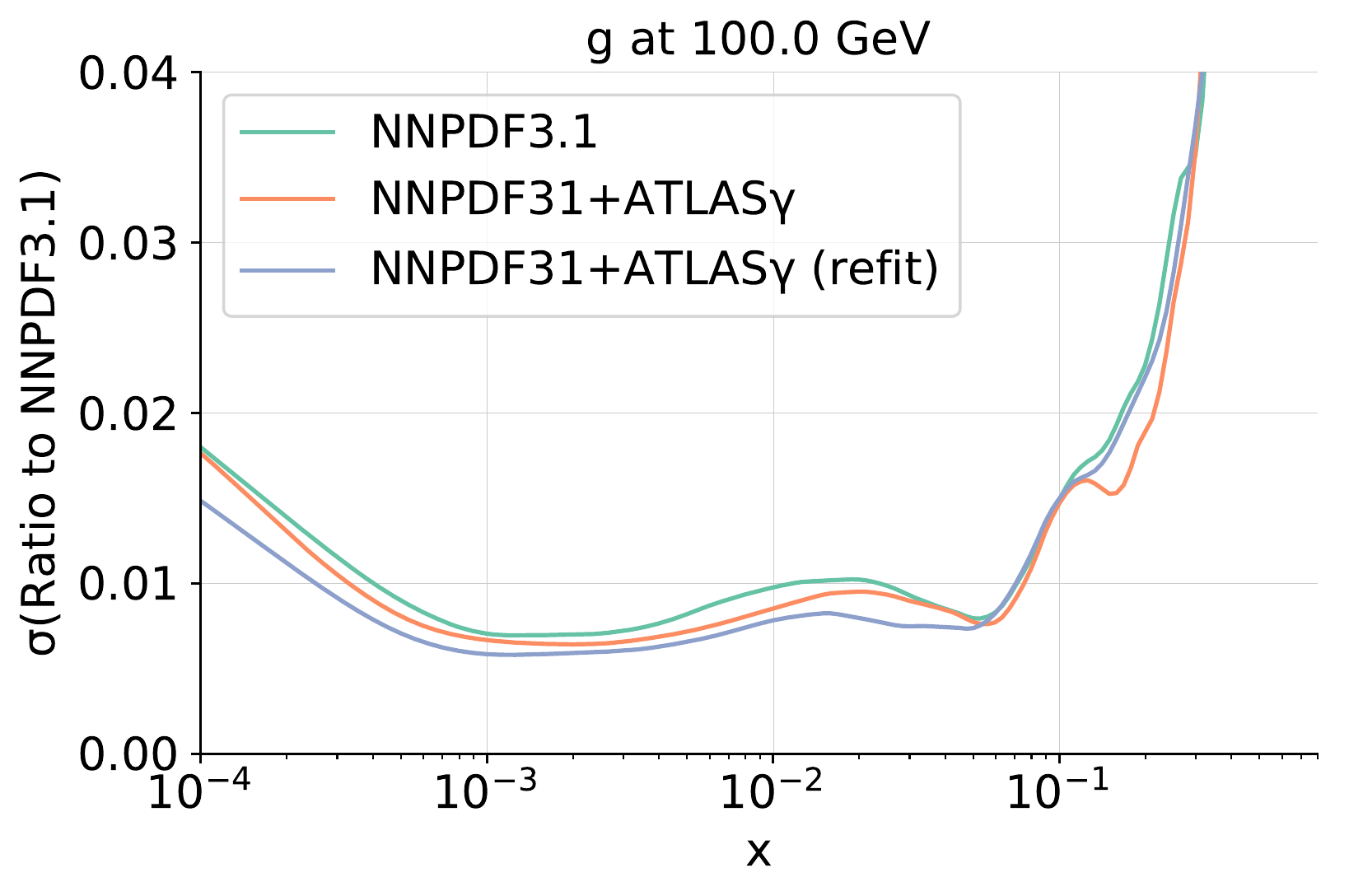}
 \caption{\small Same as Fig.~\ref{fig:gluon_ratio},
   now adding to the comparison the results of the fit including
   the correlations between the experimental
   systematic uncertainties (labelled as
   ``refit'').
   \label{fig:gluon_ratio_refit}}
\end{figure}

Let us first of all discuss the impact of
the experimental correlations at the PDF level.
We will show results with the first correlation model, where all sources of
systematic uncertainties are fully correlated bin-to-bin.
In Fig.~\ref{fig:gluon_ratio_refit} we show
the same comparison as in Fig.~\ref{fig:gluon_ratio},
 now adding the results of the fit including
 the correlations between the experimental
 systematic uncertainties (labelled as ``refit'').
 Reassuringly, the results are reasonably similar as compared
 to the fit where these correlations are neglected, though
 there are some small differences.
 In particular, concerning the PDF central values, the shift
 as compared to NNPDF3.1 is reduced once the correlations
 are accounted for.
 At the level of uncertainties, we also find a similar pattern as the
 one shown in Fig.~\ref{fig:gluon_ratio},
 with the main differences being that the PDF error reduction is a bit
 more marked for $x\lsim 0.05$ and is somewhat less important
for larger values of $x$.
All in all, the qualitative impact of the ATLAS 8 TeV direct photon data on the
PDFs is similar irrespective of whether or not one includes the information
on correlations in the $\chi^2$ definition.

Turning now to the comparison at the fit quality level, in
Table~\ref{table:chi2_NNPDF_comp_refit} we show
the values of the $\chi^2$ per data point including
  the information on the correlation of systematic
  uncertainties.
  We display the results for three fits: the baseline NNPDF3.1,
  the NNPDF3.1+ATLAS$\gamma$ fit presented in Sect.~\ref{sec:results},
  and the corresponding fit using the correlated $\chi^2$
  for the minimisation (labelled as ``refit'') and shown in
  Fig.~\ref{fig:gluon_ratio_refit}.
  Note that the first two of these fits have been obtained
  from the minimisation of $\chi^2$ which does not include
  the information on the correlations of experimental systematic uncertainties.
  The consistent values of the $\chi^2/N_{\rm dat}$ for
  these two fits were reported in Table~\ref{table:chi2_NNPDF_comp}, where
  excellent agreement with the data was found.

\begin{table}[h!]\centering
  \renewcommand{\arraystretch}{1.20}
{\begin{tabular}{  c | c |c | c | c|c| c }
 & \multicolumn{4}{c|}{$\chi^2 /N_\text{dat}$}   \\ \toprule
& 1st bin & 2nd bin & 3rd bin  & Total \\ \midrule
NNPDF3.1	&  1.52 	& 3.95  	& 2.57  &   3.03   \\
NNPDF31+ATLAS$\g$	& 1.40	&  3.90	& 2.53  & 3.02  \\ 
NNPDF31+ATLAS$\g$ (refit)	& 1.51	&  3.93	& 2.57  &  3.00   \\ \bottomrule\end{tabular}}
\vspace{0.3cm}
\caption{\small The values of the $\chi^2$ per data point including
  the information on the correlation of systematic
  uncertainties.
  We show the results for three fits: the baseline NNPDF3.1,
  the NNPDF3.1+ATLAS$\gamma$ fit presented in Sect.~\ref{sec:results},
  and the corresponding fit using the correlated $\chi^2$
  for the minimisation (labelled as ``refit'') and shown in
  Fig.~\ref{fig:gluon_ratio_refit}.
  Note that the first two of these fits have been obtained
  from the minimisation of $\chi^2$ which does not include
  the information on the correlations of systematic uncertainties.
}\label{table:chi2_NNPDF_comp_refit}
\end{table}

From the comparison between the results of
Tables~\ref{table:chi2_NNPDF_comp_refit}
and~\ref{table:chi2_NNPDF_comp} we find that the fit quality
is poorer once the information on correlated systematics
is accounted for.
In particular one gets a total $\chi^2/N_{\rm dat}\simeq 3$
for the NNPDF3.1+ATLAS$\gamma$ fit as compared to $\chi^2/N_{\rm dat}\simeq 1$
in the corresponding case where these correlations are neglected.
The description of the first rapidity bin is still satisfactory, but
not that of the second and third rapidity bins.
While the origin of these poor $\chi^2$ values is still
not understood, similar issues have been reported
in the case of the ATLAS inclusive jet measurements by different groups
(see {\it e.g.}~\cite{Harland-Lang:2017ytb,Ball:2017nwa}).
Here we have tried to vary the nominal correlation model, for instance by
neglecting the correlations between different rapidity bins, but this
does not modify the numbers in Table~\ref{table:chi2_NNPDF_comp_refit}
in any significant way.

We have also verified that the most forward rapidity bin is still
very poorly described once the full information on experimental
correlations is taken into account.
Using NNPDF3.1 as input, one finds $\chi^2/N_{\rm dat}\simeq 5.2$
for the fourth rapidity bin of the ATLAS 8 TeV measurement, a number
which is mostly unchanged when this bin is included in the fit.
This result provides further evidence in support of our decision to exclude
this bin from our baseline fits.

Let us now turn to discuss the case of the second correlation
model mentioned above, namely the one where a selected number of sources
of systematic error are treated as uncorrelated.
In Table~\ref{table:chi2_NNPDF_comp_refit2} we show the same
comparison as in Table~\ref{table:chi2_NNPDF_comp_refit} where
now the numbers
  in the second row have been computed using the partial decorrelation
  model for the experimental covariance matrix.
  We can observe a marked improvement as compared to the case
  where all systematic errors are treated as fully correlated among
data bins, although the data/theory agreement is still not ideal.
This result
suggests that a further study of the correlation model of this measurement
might further improve the numerical agreement between theory and data,
for example in the case of a partial decorrelation of some of the other systematic sources.
Such a study, which might be advantageous also for other LHC
measurements such as jet production, is however beyond the scope
of this paper.

\begin{table}[h!]\centering
  \renewcommand{\arraystretch}{1.20}
{\begin{tabular}{  c | c |c | c | c|c| c }
 & \multicolumn{4}{c|}{$\chi^2 /N_\text{dat}$}   \\ \toprule
& 1st bin & 2nd bin & 3rd bin  & Total \\ \midrule
   NNPDF3.1	&  1.52 	& 3.95  	& 2.57  &   3.03    \\
NNPDF3.1 (part. decorr.)	& 1.09	&  2.64	& 1.88  &  1.98   \\
    \bottomrule\end{tabular}}
\vspace{0.3cm}
\caption{\small
  Same as Table~\ref{table:chi2_NNPDF_comp_refit}, where now the numbers
  in the second row have been computed using the partial decorrelation
  model for the experimental covariance matrix.
  See text for more details.
}\label{table:chi2_NNPDF_comp_refit2}
\end{table}

To summarise, we find that once the information on the experimental
correlated systematics is included in the $\chi^2$ definition, the differences
at the PDF level are rather moderate and consistent with our baseline results,
but that the numerical values of the $\chi^2$ are higher.
We also find that these $\chi^2$ values are rather sensitive to the
underlying correlation model, in particular to whether some specific sources
of systematic errors are correlated or not between data bins.
These poor $\chi^2$ values deserve further investigation
in the future in order to elucidate their underlying
origin.
It is in any case reassuring that the impact of the ATLAS direct photon data
at the PDF level is mostly unaffected by this, validating
the results presented in Sects.~\ref{sec:results} and~\ref{sec:13tev}
of this work.

\section{Reweighting study}
\label{sec:Reweightingres}
\label{sec:reweighting_intro}

An alternative strategy to quantify the impact of the ATLAS direct
photon production measurements on the NNPDF3.1 global analysis
is the
Bayesian reweighting procedure~\cite{Ball:2010gb,Ball:2011gg}.
This technique
 allows one to determine the effects of a new dataset onto a
Monte Carlo
PDF set without refitting, and is therefore
less computationally intensive.
The only required inputs are
the values of the figure of merit $\chi^2_k$ to the
new dataset computed using
each of the $N_{\rm rep}$ replicas of the prior PDF set.

Given a dataset with $n_{\rm dat}$ data points,
the Bayesian reweighting procedure assigns a weight  $w_k$
to the $k$-th Monte Carlo replica
given by
\begin{equation}
  \label{eq:rwformula}
  w_k = \frac{(\chi_k^2)^{\frac{1}{2}(n_{\rm dat}-1)}e^{-\frac{1}{2}\chi_k^2}}{\frac{1}{N_{\rm rep}}\sum_{k=1}^{N_{\rm rep}} (\chi_k^2)^{\frac{1}{2}(n_{\rm dat}-1)}e^{-\frac{1}{2}\chi_k^2}} \,,\quad
  k=1,\ldots,N_{\rm rep}\, ,
\end{equation}
so that replicas that lead to theoretical predictions in disagreement
with the new dataset (and that thus lead to larger $\chi^2_k$)
receive a small weight and are thus effectively discarded.
The reweighting procedure also defines the
effective number of replicas, $N_{\text{eff}}$, given by the Shannon entropy: 
\begin{equation} \label{eq:effective_replicas}
  N_{\text{eff}} = \text{exp} \bigg\{\frac{1}{N_{\rm rep}}\sum_{k=1}^{
    N_{\rm rep}} w_k \log(N_{\rm rep}/w_k) \bigg\} \,,
\end{equation}
which allow us to  quantify how strongly the new data restricts the prior PDF set by how many replicas are left.
The interpretation of Eq.~(\ref{eq:effective_replicas})
is that the smaller the ratio $N_{\rm eff}/N_{\rm rep}$, the higher the amount
of new information that is being added by this specific
experiment into the fit.

Using identical experimental data, kinematic cuts, and theoretical settings
adopted to produce the NNPDF3.1+ATLAS$\g$ set described in Sect.~\ref{sec:results},
we have reweighted the NNPDF3.1 NNLO set with $N_{\rm rep}=100$ replicas
using Eq.~(\ref{eq:rwformula}).
The resulting reweighted PDF set is denoted by NNPDF3.1+ATLAS$\g$(rw).
This reweighted set has been subsequently unweighted to a reduced number of replicas equal to the effective number of replicas $\widetilde{N}_{\rm rep}=N_{\rm eff}$, which
can then be directly compared with the results of Sect.~\ref{sec:results}.
For the total ATLAS
direct photon production 8 TeV dataset, we find that  $N_\text{eff} = 91$.
In other words, $N_\text{eff}/N_{\rm rep}=0.91$, reflecting the moderate impact
of the direct photon data on the NNPDF3.1 analysis

In Table~\ref{table:chi2_neff} we provide the same comparison
as in Table~\ref{table:chi2_NNPDF_comp},
  now adding as well the $\chi^2/N_{\rm dat}$ values
  corresponding to 
  NNPDF3.1+ATLAS$\g$(rw) set.
  We also indicate the value for $N_{\rm eff}$
  corresponding to the total dataset.
  Note that
  we evaluate the reweighted $\chi^2$ in each of the
  rapidity bin using the
  weights $\omega_k$ computed from the full dataset.
  The overall agreement between the fitted the reweighted
  versions of NNPDF3.1+ATLAS$\g$(rw)  is reasonably
  good, with residual differences traced back to the moderate
  loss of information involved in the reweighting.
  From Table~\ref{table:chi2_neff} we observe that the value of
  $\chi^2/N_{\rm dat}$ for the ATLAS direct photon data is
  1.12 before the fit, reduced to 0.96 in the fit and 1.03 after reweighting,
  so both methods
  yield consistent results
  taking into account the statistical fluctuations
  of the $\chi^2$ itself.
  
 \begin{table}[t]\centering
   \small
         \renewcommand{\arraystretch}{1.25}
{\begin{tabular}{ c | c |c | c | c  }
    & \multicolumn{3}{c|}{$\chi^2 /N_\text{dat}$}  & $N_\text{eff}$  \\
    \toprule
& NNPDF3.1 & NNPDF3.1+ATLAS$\g$ (fit) &NNPDF3.1+ATLAS$\g$ (rw) & \\ \midrule
1st bin 		& 0.81	& 0.66 & 0.71 & -  \\  
2nd bin	& 1.61	&1.37  & 1.53 &  -   \\
3rd bin	& 0.89	&0.82 & 0.88 & -  \\ \midrule
  Total	&1.12 &	0.96 & 1.03 & 91 \\\bottomrule
\end{tabular}}
\vspace{0.3cm}
\caption{\small Same as
  Table~\ref{table:chi2_NNPDF_comp},
  now with the $\chi^2/N_{\rm dat}$ values
  corresponding to the reweighted
  NNPDF3.1+ATLAS$\g$ set.
  We also provide the value of $N_{\rm eff}$
  corresponding to the total dataset.
}\label{table:chi2_neff}
\end{table}

\begin{figure}[t]
\centering
  \includegraphics[scale=0.47]{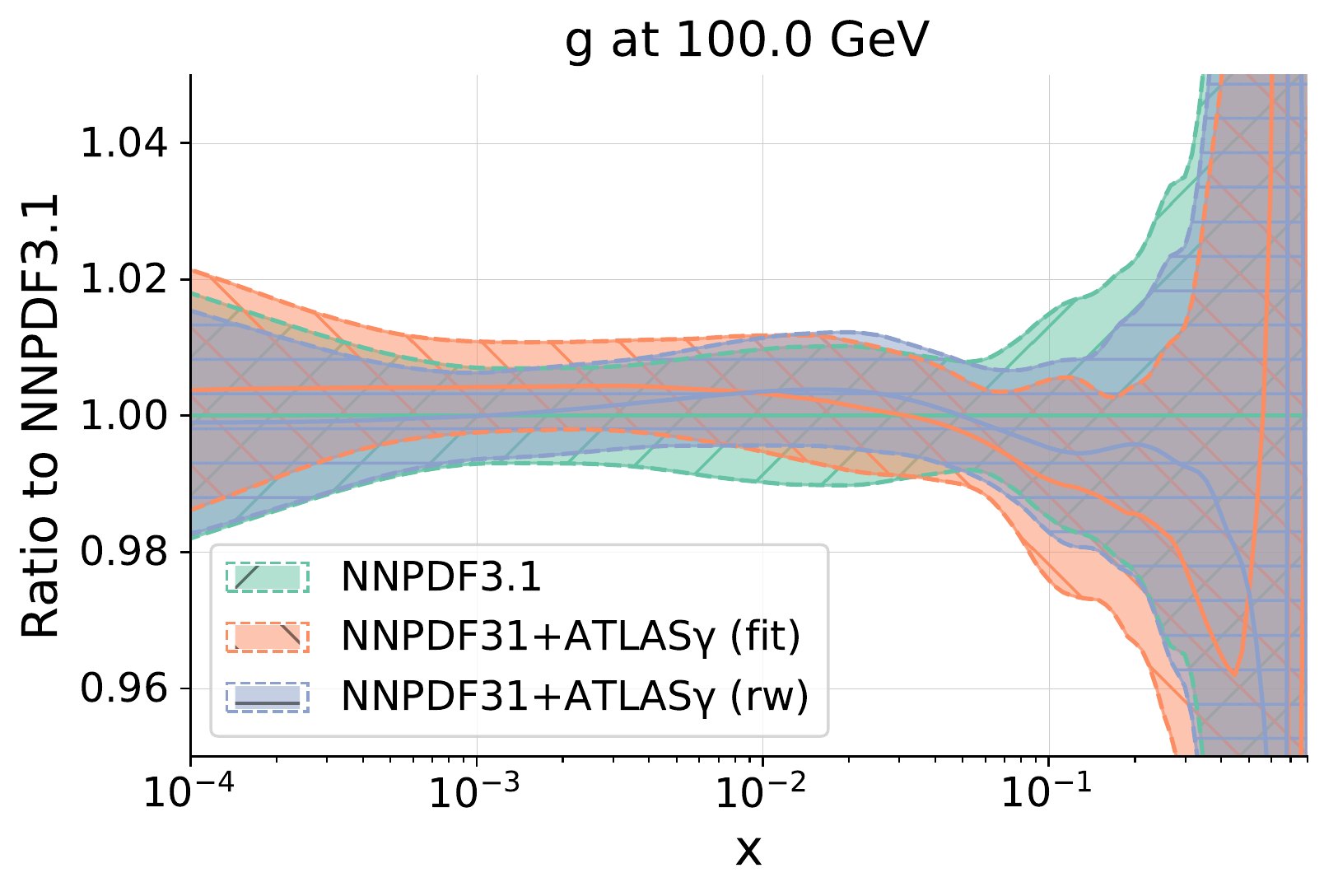} 
  \includegraphics[scale=0.47]{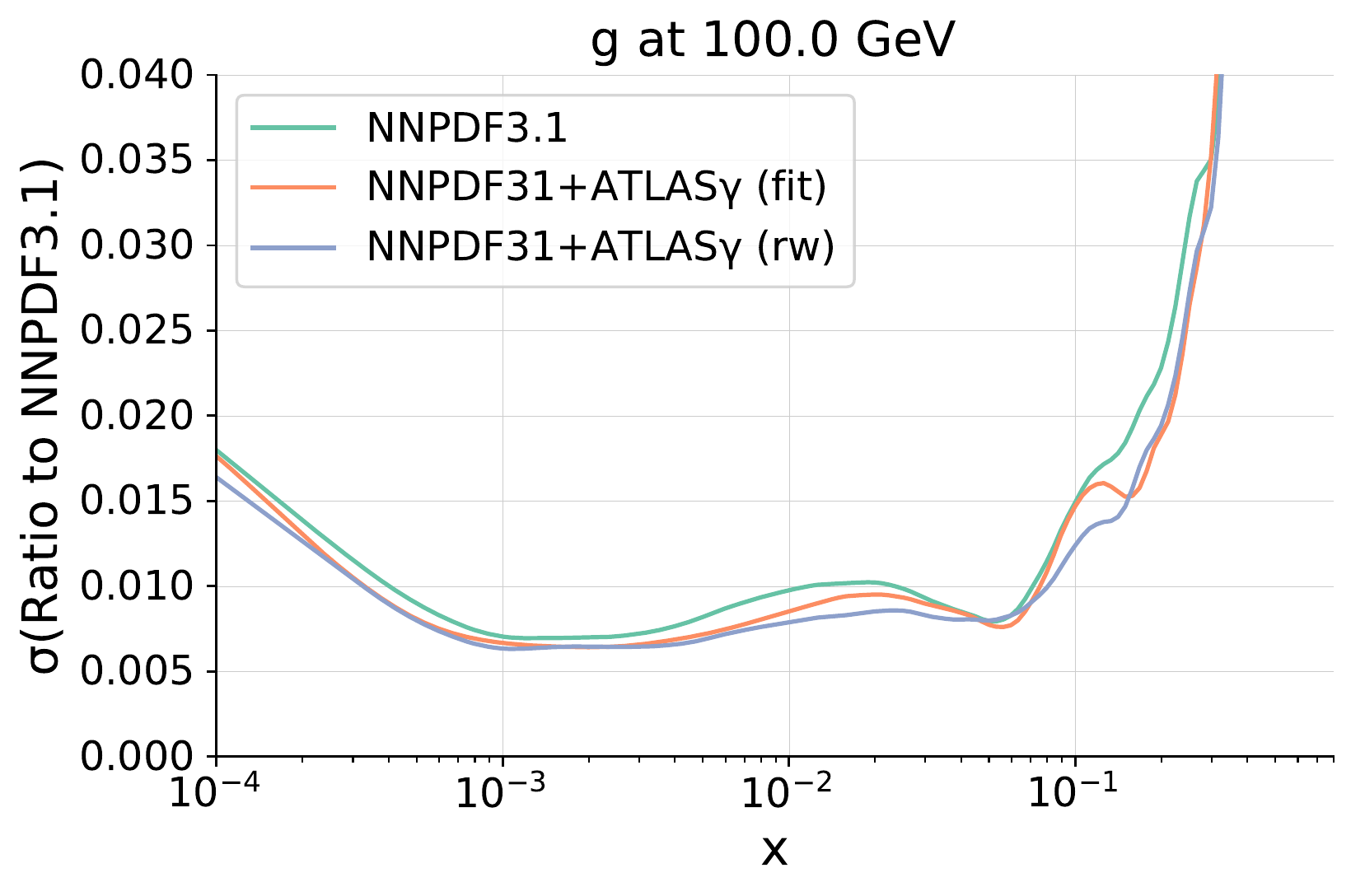} 
  \caption{\small Same as Fig.~\ref{fig:gluon_ratio}
    now adding to the comparison the results of
    the NNPDF3.1+ATLAS$\g$(rw) set obtained by applying
    the Bayesian reweighting procedure.
  }
\label{fig:reweighting_gluon_comp}
\end{figure}

 Concerning the comparison between fitting
 and reweighting at the PDF level, in Fig.~\ref{fig:reweighting_gluon_comp}
 we show an updated version of Fig.~\ref{fig:gluon_ratio} now
 adding also the results of
 the NNPDF3.1+ATLAS$\g$(rw) set.
 We see that the results of both gluons are close to each other, and
 in particular that both the fitted and reweighted versions of 
 NNPDF3.1+ATLAS$\g$(rw) exhibit the clear preference for a somewhat
 softer gluon at large $x$.
 In terms of the gluon PDF uncertainty, also here the results of
 the fitted and reweighted sets are reasonably similar.
 Note that in particular in the large-$x$ region, $x\gsim 0.2$,
 the reduction of the PDF uncertainties as compared
 to the baseline determined using the two methods is identical.
 
To summarize, the fitted and the reweighted versions of the NNPDF3.1+ATLAS$\g$ PDF
set are in good agreement with each other.
This agreement could be further
improved if a larger prior would have been used, specifically
the NNPDF3.1 NNLO set with $N_{\rm eff}=1000$ replicas.
This is however not required here, since our goal was limited
to providing a 
validation of the qualitative features of the
fit results by means of an independent technique.


\bibliography{prompt_photon}

\providecommand{\href}[2]{#2}\begingroup\raggedright\begin{thebibliography}{10}

\bibitem{Forte:2013wc}
S.~Forte and G.~Watt, \emph{{Progress in the Determination of the Partonic
  Structure of the Proton}},
  \href{https://doi.org/10.1146/annurev-nucl-102212-170607}{\emph{Ann.Rev.Nucl.Part.Sci.}
  {\bfseries 63} (2013) 291},
  [\href{https://arxiv.org/abs/1301.6754}{{\ttfamily 1301.6754}}].

\bibitem{Butterworth:2015oua}
J.~Butterworth et~al., \emph{{PDF4LHC recommendations for LHC Run II}},
  \href{https://doi.org/10.1088/0954-3899/43/2/023001}{\emph{J. Phys.}
  {\bfseries G43} (2016) 023001},
  [\href{https://arxiv.org/abs/1510.03865}{{\ttfamily 1510.03865}}].

\bibitem{Rojo:2015acz}
J.~Rojo et~al., \emph{{The PDF4LHC report on PDFs and LHC data: Results from
  Run I and preparation for Run II}},
  \href{https://doi.org/10.1088/0954-3899/42/10/103103}{\emph{J. Phys.}
  {\bfseries G42} (2015) 103103},
  [\href{https://arxiv.org/abs/1507.00556}{{\ttfamily 1507.00556}}].

\bibitem{Gao:2017yyd}
J.~Gao, L.~Harland-Lang and J.~Rojo, \emph{{The Structure of the Proton in the
  LHC Precision Era}},  \href{https://arxiv.org/abs/1709.04922}{{\ttfamily
  1709.04922}}.

\bibitem{Rojo:2014kta}
J.~Rojo, \emph{{Constraints on parton distributions and the strong coupling
  from LHC jet data}},
  \href{https://doi.org/10.1142/S0217751X15460057}{\emph{Int. J. Mod. Phys.}
  {\bfseries A30} (2015) 1546005},
  [\href{https://arxiv.org/abs/1410.7728}{{\ttfamily 1410.7728}}].

\bibitem{Czakon:2016olj}
M.~Czakon, N.~P. Hartland, A.~Mitov, E.~R. Nocera and J.~Rojo, \emph{{Pinning
  down the large-x gluon with NNLO top-quark pair differential distributions}},
  \href{https://doi.org/10.1007/JHEP04(2017)044}{\emph{JHEP} {\bfseries 04}
  (2017) 044}, [\href{https://arxiv.org/abs/1611.08609}{{\ttfamily
  1611.08609}}].

\bibitem{Boughezal:2017nla}
R.~Boughezal, A.~Guffanti, F.~Petriello and M.~Ubiali, \emph{{The impact of the
  LHC Z-boson transverse momentum data on PDF determinations}},
  \href{https://doi.org/10.1007/JHEP07(2017)130}{\emph{JHEP} {\bfseries 07}
  (2017) 130}, [\href{https://arxiv.org/abs/1705.00343}{{\ttfamily
  1705.00343}}].

\bibitem{Gauld:2016kpd}
R.~Gauld and J.~Rojo, \emph{{Precision determination of the small-$x$ gluon
  from charm production at LHCb}},
  \href{https://doi.org/10.1103/PhysRevLett.118.072001}{\emph{Phys. Rev. Lett.}
  {\bfseries 118} (2017) 072001},
  [\href{https://arxiv.org/abs/1610.09373}{{\ttfamily 1610.09373}}].

\bibitem{Gauld:2017rbf}
R.~Gauld, J.~Rojo and E.~Slade, \emph{{The small-x gluon from forward charm
  production: implications for a 100 TeV proton collider}},  in \emph{{25th
  International Workshop on Deep Inelastic Scattering and Related Topics (DIS
  2017) Birmingham, UK, April 3-7, 2017}}, 2017,
  \href{https://arxiv.org/abs/1705.04217}{{\ttfamily 1705.04217}}.

\bibitem{Ball:2017nwa}
{\scshape NNPDF} collaboration, R.~D. Ball et~al., \emph{{Parton distributions
  from high-precision collider data}},
  \href{https://doi.org/10.1140/epjc/s10052-017-5199-5}{\emph{Eur. Phys. J.}
  {\bfseries C77} (2017) 663},
  [\href{https://arxiv.org/abs/1706.00428}{{\ttfamily 1706.00428}}].

\bibitem{Nocera:2017zge}
E.~R. Nocera and M.~Ubiali, \emph{{Constraining the gluon PDF at large x with
  LHC data}},  2017, \href{https://arxiv.org/abs/1709.09690}{{\ttfamily
  1709.09690}},
  \href{http://inspirehep.net/record/1627625/files/arXiv:1709.09690.pdf}{http://inspirehep.net/record/1627625/files/arXiv:1709.09690.pdf}.

\bibitem{Ichou:2010wc}
R.~Ichou and D.~d'Enterria, \emph{{Sensitivity of isolated photon production at
  TeV hadron colliders to the gluon distribution in the proton}},
  \href{https://doi.org/10.1103/PhysRevD.82.014015}{\emph{Phys. Rev.}
  {\bfseries D82} (2010) 014015},
  [\href{https://arxiv.org/abs/1005.4529}{{\ttfamily 1005.4529}}].

\bibitem{dEnterria:2012kvo}
D.~d'Enterria and J.~Rojo, \emph{{Quantitative constraints on the gluon
  distribution function in the proton from collider isolated-photon data}},
  \href{https://doi.org/10.1016/j.nuclphysb.2012.03.003}{\emph{Nucl. Phys.}
  {\bfseries B860} (2012) 311--338},
  [\href{https://arxiv.org/abs/1202.1762}{{\ttfamily 1202.1762}}].

\bibitem{Bourhis:1997yus}
L.~Bourhis, M.~Fontannaz and J.~P. Guillet, \emph{{Quarks and gluon
  fragmentation functions into photons}},
  \href{https://doi.org/10.1007/s100520050158}{\emph{Eur. Phys. J.} {\bfseries
  C2} (1998) 529--537}, [\href{https://arxiv.org/abs/hep-ph/9704447}{{\ttfamily
  hep-ph/9704447}}].

\bibitem{dflm}
M.~Diemoz, F.~Ferroni, E.~Longo and G.~Martinelli, \emph{{Parton Densities from
  Deep Inelastic Scattering to Hadronic Processes at Super Collider Energies}},
  \href{https://doi.org/10.1007/BF01560387}{\emph{Z. Phys.} {\bfseries C39}
  (1988) 21}.

\bibitem{PhysRevD.42.798}
P.~N. Harriman, A.~D. Martin, W.~J. Stirling and R.~G. Roberts, \emph{Parton
  distributions extracted from data on deep-inelastic lepton scattering, prompt
  photon production, and the drell-yan process},
  \href{https://doi.org/10.1103/PhysRevD.42.798}{\emph{Phys. Rev. D} {\bfseries
  42} (Aug, 1990) 798--810}.

\bibitem{Morfin:1990ck}
J.~G. Morfin and W.-K. Tung, \emph{{Parton distributions from a global QCD
  analysis of deep inelastic scattering and lepton pair production}},
  \href{https://doi.org/10.1007/BF01412323}{\emph{Z. Phys.} {\bfseries C52}
  (1991) 13--30}.

\bibitem{Harland-Lang:2014zoa}
L.~A. Harland-Lang, A.~D. Martin, P.~Motylinski and R.~S. Thorne, \emph{{Parton
  distributions in the LHC era: MMHT 2014 PDFs}},
  \href{https://doi.org/10.1140/epjc/s10052-015-3397-6}{\emph{Eur. Phys. J.}
  {\bfseries C75} (2015) 204},
  [\href{https://arxiv.org/abs/1412.3989}{{\ttfamily 1412.3989}}].

\bibitem{Dulat:2015mca}
S.~Dulat, T.-J. Hou, J.~Gao, M.~Guzzi, J.~Huston, P.~Nadolsky et~al.,
  \emph{{New parton distribution functions from a global analysis of quantum
  chromodynamics}},
  \href{https://doi.org/10.1103/PhysRevD.93.033006}{\emph{Phys. Rev.}
  {\bfseries D93} (2016) 033006},
  [\href{https://arxiv.org/abs/1506.07443}{{\ttfamily 1506.07443}}].

\bibitem{Alekhin:2017kpj}
S.~Alekhin, J.~Bl{\"u}mlein, S.~Moch and R.~Placakyte, \emph{{Parton
  Distribution Functions, $\alpha_s$ and Heavy-Quark Masses for LHC Run II}},
  \href{https://doi.org/10.1103/PhysRevD.96.014011}{\emph{Phys. Rev.}
  {\bfseries D96} (2017) 014011},
  [\href{https://arxiv.org/abs/1701.05838}{{\ttfamily 1701.05838}}].

\bibitem{Campbell:2016lzl}
J.~M. Campbell, R.~K. Ellis and C.~Williams, \emph{{Direct Photon Production at
  Next to Next to Leading Order}},
  \href{https://doi.org/10.1103/PhysRevLett.118.222001}{\emph{Phys. Rev. Lett.}
  {\bfseries 118} (2017) 222001},
  [\href{https://arxiv.org/abs/1612.04333}{{\ttfamily 1612.04333}}].

\bibitem{Becher:2013zua}
T.~Becher and X.~Garcia~i Tormo, \emph{{Electroweak Sudakov effects in $W, Z$
  and $\gamma$ production at large transverse momentum}},
  \href{https://doi.org/10.1103/PhysRevD.88.013009}{\emph{Phys. Rev.}
  {\bfseries D88} (2013) 013009},
  [\href{https://arxiv.org/abs/1305.4202}{{\ttfamily 1305.4202}}].

\bibitem{Aad:2016xcr}
{\scshape ATLAS} collaboration, G.~Aad et~al., \emph{{Measurement of the
  inclusive isolated prompt photon cross section in pp collisions at $
  \sqrt{s}=8 $ TeV with the ATLAS detector}},
  \href{https://doi.org/10.1007/JHEP08(2016)005}{\emph{JHEP} {\bfseries 08}
  (2016) 005}, [\href{https://arxiv.org/abs/1605.03495}{{\ttfamily
  1605.03495}}].

\bibitem{Frixione:1998jh}
S.~Frixione, \emph{{Isolated photons in perturbative QCD}},
  \href{https://doi.org/10.1016/S0370-2693(98)00454-7}{\emph{Phys.Lett.}
  {\bfseries B429} (1998) 369},
  [\href{https://arxiv.org/abs/hep-ph/9801442}{{\ttfamily hep-ph/9801442}}].

\bibitem{Aaboud:2016tru}
{\scshape ATLAS} collaboration, M.~Aaboud et~al., \emph{{Search for resonances
  in diphoton events at $\sqrt{s}$=13 TeV with the ATLAS detector}},
  \href{https://doi.org/10.1007/JHEP09(2016)001}{\emph{JHEP} {\bfseries 09}
  (2016) 001}, [\href{https://arxiv.org/abs/1606.03833}{{\ttfamily
  1606.03833}}].

\bibitem{Aaboud:2017yyg}
{\scshape ATLAS} collaboration, M.~Aaboud et~al., \emph{{Search for new
  phenomena in high-mass diphoton final states using 37 fb$^{-1}$ of
  proton--proton collisions collected at $\sqrt{s}=13$ TeV with the ATLAS
  detector}}, {\emph{Submitted to: Phys. Lett.} (2017) },
  [\href{https://arxiv.org/abs/1707.04147}{{\ttfamily 1707.04147}}].

\bibitem{CMS_Zgamma}
\emph{Search for high-mass z resonances in proton--proton collisions at s=8 and
  13 tev using jet substructure techniques},
  \href{https://doi.org/https://doi.org/10.1016/j.physletb.2017.06.062}{\emph{Physics
  Letters B} (2017) }.

\bibitem{Aaboud:2017tcq}
{\scshape ATLAS} collaboration, M.~Aaboud et~al., \emph{{Study of $WW\gamma$
  and $WZ\gamma$ production in $pp$ collisions at $\sqrt{s} = 8$ TeV and search
  for anomalous quartic gauge couplings with the ATLAS experiment}},
  \href{https://doi.org/10.1140/epjc/s10052-017-5180-3}{\emph{Eur. Phys. J.}
  {\bfseries C77} (2017) 646},
  [\href{https://arxiv.org/abs/1707.05597}{{\ttfamily 1707.05597}}].

\bibitem{Khachatryan:2016vif}
{\scshape CMS} collaboration, V.~Khachatryan et~al., \emph{{Measurement of
  electroweak-induced production of W$\gamma$ with two jets in pp collisions at
  $ \sqrt{s}=8 $ TeV and constraints on anomalous quartic gauge couplings}},
  \href{https://doi.org/10.1007/JHEP06(2017)106}{\emph{JHEP} {\bfseries 06}
  (2017) 106}, [\href{https://arxiv.org/abs/1612.09256}{{\ttfamily
  1612.09256}}].

\bibitem{Khachatryan:2016yro}
{\scshape CMS} collaboration, V.~Khachatryan et~al., \emph{{Measurement of the
  $ \mathrm{ Z } \gamma \rightarrow \nu \bar{\nu} \gamma$ production cross
  section in pp collisions at $\sqrt{s}=$ 8 TeV and limits on anomalous $
  \mathrm{ ZZ } \gamma$ and $ \mathrm{Z} \gamma \gamma$ trilinear gauge boson
  couplings}},
  \href{https://doi.org/10.1016/j.physletb.2016.06.080}{\emph{Phys. Lett.}
  {\bfseries B760} (2016) 448--468},
  [\href{https://arxiv.org/abs/1602.07152}{{\ttfamily 1602.07152}}].

\bibitem{Aad:2015uqa}
{\scshape ATLAS} collaboration, G.~Aad et~al., \emph{{Evidence of
  $W\gamma\gamma$ Production in pp Collisions at $\sqrt{s} = 8$ TeV and Limits
  on Anomalous Quartic Gauge Couplings with the ATLAS Detector}},
  \href{https://doi.org/10.1103/PhysRevLett.115.031802}{\emph{Phys. Rev. Lett.}
  {\bfseries 115} (2015) 031802},
  [\href{https://arxiv.org/abs/1503.03243}{{\ttfamily 1503.03243}}].

\bibitem{Sirunyan:2017ewk}
{\scshape CMS} collaboration, A.~M. Sirunyan et~al., \emph{{Search for new
  physics in the monophoton final state in proton-proton collisions at sqrt(s)
  = 13 TeV}},  \href{https://arxiv.org/abs/1706.03794}{{\ttfamily 1706.03794}}.

\bibitem{Sirunyan:2017yse}
{\scshape CMS} collaboration, A.~M. Sirunyan et~al., \emph{{Search for
  supersymmetry in events with at least one photon, missing transverse
  momentum, and large transverse event activity in proton-proton collisions at
  $ \sqrt{s}=13 $ TeV}},
  \href{https://doi.org/10.1007/JHEP12(2017)142}{\emph{JHEP} {\bfseries 12}
  (2017) 142}, [\href{https://arxiv.org/abs/1707.06193}{{\ttfamily
  1707.06193}}].

\bibitem{Aaboud:2017uak}
{\scshape ATLAS} collaboration, M.~Aaboud et~al., \emph{{Search for dark matter
  in association with a Higgs boson decaying to two photons at $\sqrt{s}$= 13
  TeV with the ATLAS detector}},
  \href{https://arxiv.org/abs/1706.03948}{{\ttfamily 1706.03948}}.

\bibitem{Sirunyan:2017hnk}
{\scshape CMS} collaboration, A.~M. Sirunyan et~al., \emph{{Search for
  associated production of dark matter with a Higgs boson decaying to $
  \mathrm{b}\overline{\mathrm{b}} $ or $\gamma \gamma$ at $ \sqrt{s}=13$ TeV}},
  \href{https://doi.org/10.1007/JHEP10(2017)180}{\emph{JHEP} {\bfseries 10}
  (2017) 180}, [\href{https://arxiv.org/abs/1703.05236}{{\ttfamily
  1703.05236}}].

\bibitem{Ball:2010gb}
{\scshape The NNPDF} collaboration, R.~D. Ball et~al., \emph{{Reweighting
  NNPDFs: the W lepton asymmetry}},
  \href{https://doi.org/10.1016/j.nuclphysb.2011.03.017}{\emph{Nucl. Phys.}
  {\bfseries B849} (2011) 112--143},
  [\href{https://arxiv.org/abs/1012.0836}{{\ttfamily 1012.0836}}].

\bibitem{Ball:2011gg}
R.~D. Ball, V.~Bertone, F.~Cerutti, L.~Del~Debbio, S.~Forte et~al.,
  \emph{{Reweighting and Unweighting of Parton Distributions and the LHC W
  lepton asymmetry data}},
  \href{https://doi.org/10.1016/j.nuclphysb.2011.10.018}{\emph{Nucl.Phys.}
  {\bfseries B855} (2012) 608--638},
  [\href{https://arxiv.org/abs/1108.1758}{{\ttfamily 1108.1758}}].

\bibitem{Aaboud:2017cbm}
{\scshape ATLAS} collaboration, M.~Aaboud et~al., \emph{{Measurement of the
  cross section for inclusive isolated-photon production in $pp$ collisions at
  $\sqrt s=13$ TeV using the ATLAS detector}},
  \href{https://arxiv.org/abs/1701.06882}{{\ttfamily 1701.06882}}.

\bibitem{Acosta:2002ya}
{\scshape CDF} collaboration, D.~Acosta et~al., \emph{{Comparison of the
  isolated direct photon cross sections in $p\bar{p}$ collisions at $\sqrt{s}=$
  1.8 TeV and $\sqrt{s}=$ 0.63 TeV}},
  \href{https://doi.org/10.1103/PhysRevD.65.112003}{\emph{Phys. Rev.}
  {\bfseries D65} (2002) 112003},
  [\href{https://arxiv.org/abs/hep-ex/0201004}{{\ttfamily hep-ex/0201004}}].

\bibitem{Abazov:2001af}
{\scshape D0} collaboration, V.~M. Abazov et~al., \emph{{The ratio of the
  isolated photon cross sections at $\sqrt{s} = 630$ GeV and 1800 GeV}},
  \href{https://doi.org/10.1103/PhysRevLett.87.251805}{\emph{Phys. Rev. Lett.}
  {\bfseries 87} (2001) 251805},
  [\href{https://arxiv.org/abs/hep-ex/0106026}{{\ttfamily hep-ex/0106026}}].

\bibitem{Abe:1994rra}
{\scshape CDF} collaboration, F.~Abe et~al., \emph{{A Precision measurement of
  the prompt photon cross-section in $p\bar{p}$ collisions at $\sqrt{s} = 1.8$
  TeV}}, \href{https://doi.org/10.1103/PhysRevLett.73.2662}{\emph{Phys. Rev.
  Lett.} {\bfseries 73} (1994) 2662--2666}.

\bibitem{Abe:1993qb}
{\scshape CDF} collaboration, F.~Abe et~al., \emph{{A Prompt photon
  cross-section measurement in $\bar{p}p$ collisions at $\sqrt{s} = 1.8$ TeV}},
  \href{https://doi.org/10.1103/PhysRevD.48.2998}{\emph{Phys. Rev.} {\bfseries
  D48} (1993) 2998--3025}.

\bibitem{Abe:1992fd}
{\scshape CDF} collaboration, F.~Abe et~al., \emph{{Measurement of the isolated
  prompt photon cross-sections in $\bar{p}p$ collisions at $\sqrt{s} = 1.8$
  TeV}}, \href{https://doi.org/10.1103/PhysRevLett.68.2734}{\emph{Phys. Rev.
  Lett.} {\bfseries 68} (1992) 2734--2738}.

\bibitem{Abbott:1999kd}
{\scshape D0} collaboration, B.~Abbott et~al., \emph{{The isolated photon
  cross-section in $p\bar{p}$ collisions at $\sqrt{s} = 1.8$ TeV}},
  \href{https://doi.org/10.1103/PhysRevLett.84.2786}{\emph{Phys. Rev. Lett.}
  {\bfseries 84} (2000) 2786--2791},
  [\href{https://arxiv.org/abs/hep-ex/9912017}{{\ttfamily hep-ex/9912017}}].

\bibitem{Abachi:1996qz}
{\scshape D0} collaboration, S.~Abachi et~al., \emph{{Isolated photon
  cross-section in the central and forward rapidity regions in $p\bar{p}$
  collisions at $\sqrt{s} = 1.8$ TeV}},
  \href{https://doi.org/10.1103/PhysRevLett.77.5011}{\emph{Phys. Rev. Lett.}
  {\bfseries 77} (1996) 5011--5015},
  [\href{https://arxiv.org/abs/hep-ex/9603006}{{\ttfamily hep-ex/9603006}}].

\bibitem{Aaltonen:2013ama}
{\scshape CDF} collaboration, T.~Aaltonen et~al., \emph{{Measurement of the
  cross section for direct-photon production in association with a heavy quark
  in $p\bar{p}$ collisions at $\sqrt{s}$ = 1.96 TeV}},
  \href{https://doi.org/10.1103/PhysRevLett.111.042003}{\emph{Phys. Rev. Lett.}
  {\bfseries 111} (2013) 042003},
  [\href{https://arxiv.org/abs/1303.6136}{{\ttfamily 1303.6136}}].

\bibitem{Aaltonen:2009ty}
{\scshape CDF} collaboration, T.~Aaltonen et~al., \emph{{Measurement of the
  Inclusive Isolated Prompt Photon Cross Section in $p\bar{p}$ Collisions at
  $\sqrt{s}=1.96$~TeV using the CDF Detector}},
  \href{https://doi.org/10.1103/PhysRevD.80.111106}{\emph{Phys. Rev.}
  {\bfseries D80} (2009) 111106},
  [\href{https://arxiv.org/abs/0910.3623}{{\ttfamily 0910.3623}}].

\bibitem{Abazov:2005wc}
{\scshape D0} collaboration, V.~M. Abazov et~al., \emph{{Measurement of the
  isolated photon cross section in $p \bar{p}$ collisions at $\sqrt{s}$ = 1.96
  TeV}}, \href{https://doi.org/10.1016/j.physletb.2007.06.047,
  10.1016/j.physletb.2006.04.048}{\emph{Phys. Lett.} {\bfseries B639} (2006)
  151--158}, [\href{https://arxiv.org/abs/hep-ex/0511054}{{\ttfamily
  hep-ex/0511054}}].

\bibitem{Aad:2010sp}
{\scshape ATLAS} collaboration, G.~Aad et~al., \emph{{Measurement of the
  inclusive isolated prompt photon cross section in $pp$ collisions at
  $\sqrt{s}=7$ TeV with the ATLAS detector}},
  \href{https://doi.org/10.1103/PhysRevD.83.052005}{\emph{Phys. Rev.}
  {\bfseries D83} (2011) 052005},
  [\href{https://arxiv.org/abs/1012.4389}{{\ttfamily 1012.4389}}].

\bibitem{Aad:2011tw}
{\scshape ATLAS} collaboration, G.~Aad et~al., \emph{{Measurement of the
  inclusive isolated prompt photon cross-section in $pp$ collisions at
  $\sqrt{s}=$ 7 TeV using 35 pb$^{-1}$ of ATLAS data}},
  \href{https://doi.org/10.1016/j.physletb.2011.11.010}{\emph{Phys. Lett.}
  {\bfseries B706} (2011) 150--167},
  [\href{https://arxiv.org/abs/1108.0253}{{\ttfamily 1108.0253}}].

\bibitem{Aad:2013zba}
{\scshape ATLAS} collaboration, G.~Aad et~al., \emph{{Measurement of the
  inclusive isolated prompt photons cross section in pp collisions at
  $\sqrt{s}=7$TeV with the ATLAS detector using 4.6fb$^{-1}$}},
  \href{https://doi.org/10.1103/PhysRevD.89.052004}{\emph{Phys. Rev.}
  {\bfseries D89} (2014) 052004},
  [\href{https://arxiv.org/abs/1311.1440}{{\ttfamily 1311.1440}}].

\bibitem{Chatrchyan:2011ue}
{\scshape CMS} collaboration, S.~Chatrchyan et~al., \emph{{Measurement of the
  Differential Cross Section for Isolated Prompt Photon Production in pp
  Collisions at 7 TeV}},
  \href{https://doi.org/10.1103/PhysRevD.84.052011}{\emph{Phys.Rev.} {\bfseries
  D84} (2011) 052011}, [\href{https://arxiv.org/abs/1108.2044}{{\ttfamily
  1108.2044}}].

\bibitem{Khachatryan:2010fm}
{\scshape CMS} collaboration, V.~Khachatryan et~al., \emph{{Measurement of the
  Isolated Prompt Photon Production Cross Section in $pp$ Collisions at
  $\sqrt{s} = 7$~TeV}},
  \href{https://doi.org/10.1103/PhysRevLett.106.082001}{\emph{Phys. Rev. Lett.}
  {\bfseries 106} (2011) 082001},
  [\href{https://arxiv.org/abs/1012.0799}{{\ttfamily 1012.0799}}].

\bibitem{Carli:2010rw}
T.~Carli et~al., \emph{{A posteriori inclusion of parton density functions in
  NLO QCD final-state calculations at hadron colliders: The APPLGRID Project}},
  \href{https://doi.org/10.1140/epjc/s10052-010-1255-0}{\emph{Eur.Phys.J.}
  {\bfseries C66} (2010) 503},
  [\href{https://arxiv.org/abs/0911.2985}{{\ttfamily 0911.2985}}].

\bibitem{Bern:2011pa}
Z.~Bern, G.~Diana, L.~J. Dixon, F.~Febres~Cordero, S.~Hoche, H.~Ita et~al.,
  \emph{{Driving Missing Data at Next-to-Leading Order}},
  \href{https://doi.org/10.1103/PhysRevD.84.114002}{\emph{Phys. Rev.}
  {\bfseries D84} (2011) 114002},
  [\href{https://arxiv.org/abs/1106.1423}{{\ttfamily 1106.1423}}].

\bibitem{Campbell:2016yrh}
J.~M. Campbell, R.~K. Ellis, Y.~Li and C.~Williams, \emph{{Predictions for
  diphoton production at the LHC through NNLO in QCD}},
  \href{https://doi.org/10.1007/JHEP07(2016)148}{\emph{JHEP} {\bfseries 07}
  (2016) 148}, [\href{https://arxiv.org/abs/1603.02663}{{\ttfamily
  1603.02663}}].

\bibitem{Campbell:2017dqk}
J.~M. Campbell, R.~K. Ellis and C.~Williams, \emph{{Driving missing data at the
  LHC: NNLO predictions for the ratio of $\gamma+j$ and $Z+j$}},
  \href{https://doi.org/10.1103/PhysRevD.96.014037}{\emph{Phys. Rev.}
  {\bfseries D96} (2017) 014037},
  [\href{https://arxiv.org/abs/1703.10109}{{\ttfamily 1703.10109}}].

\bibitem{Becher:2015yea}
T.~Becher and X.~Garcia~i Tormo, \emph{{Addendum: Electroweak Sudakov effects
  in W, Z and gamma production at large transverse momentum}},
  \href{https://doi.org/10.1103/PhysRevD.92.073011}{\emph{Phys. Rev.}
  {\bfseries D92} (2015) 073011},
  [\href{https://arxiv.org/abs/1509.01961}{{\ttfamily 1509.01961}}].

\bibitem{Boughezal:2016wmq}
R.~Boughezal, J.~M. Campbell, R.~K. Ellis, C.~Focke, W.~Giele, X.~Liu et~al.,
  \emph{Color singlet production at {NNLO} in {MCFM}}, {\emph{Eur. Phys. J.}
  {\bfseries C77} (2017) 7},
  [\href{https://arxiv.org/abs/1605.08011}{{\ttfamily 1605.08011}}].

\bibitem{Buckley:2014ana}
A.~Buckley, J.~Ferrando, S.~Lloyd, K.~Nordstr{\"o}m, B.~Page et~al.,
  \emph{{LHAPDF6: parton density access in the LHC precision era}},
  \href{https://doi.org/10.1140/epjc/s10052-015-3318-8}{\emph{Eur.Phys.J.}
  {\bfseries C75} (2015) 132},
  [\href{https://arxiv.org/abs/1412.7420}{{\ttfamily 1412.7420}}].

\bibitem{Salam:2008qg}
G.~P. Salam and J.~Rojo, \emph{{A Higher Order Perturbative Parton Evolution
  Toolkit (HOPPET)}},
  \href{https://doi.org/10.1016/j.cpc.2008.08.010}{\emph{Comput. Phys. Commun.}
  {\bfseries 180} (2009) 120--156},
  [\href{https://arxiv.org/abs/0804.3755}{{\ttfamily 0804.3755}}].

\bibitem{Ball:2016neh}
{\scshape NNPDF} collaboration, R.~D. Ball, V.~Bertone, M.~Bonvini,
  S.~Carrazza, S.~Forte, A.~Guffanti et~al., \emph{{A Determination of the
  Charm Content of the Proton}},
  \href{https://doi.org/10.1140/epjc/s10052-016-4469-y}{\emph{Eur. Phys. J.}
  {\bfseries C76} (2016) 647},
  [\href{https://arxiv.org/abs/1605.06515}{{\ttfamily 1605.06515}}].

\bibitem{Ball:2012wy}
R.~D. Ball et~al., \emph{{Parton Distribution Benchmarking with LHC Data}},
  \href{https://doi.org/10.1007/JHEP04(2013)125}{\emph{JHEP} {\bfseries 04}
  (2013) 125}, [\href{https://arxiv.org/abs/1211.5142}{{\ttfamily 1211.5142}}].

\bibitem{Ball:2009qv}
{\scshape The NNPDF} collaboration, R.~D. Ball et~al., \emph{{Fitting Parton
  Distribution Data with Multiplicative Normalization Uncertainties}},
  \href{https://doi.org/10.1007/JHEP05(2010)075}{\emph{JHEP} {\bfseries 05}
  (2010) 075}, [\href{https://arxiv.org/abs/0912.2276}{{\ttfamily 0912.2276}}].

\bibitem{Ball:2017otu}
R.~D. Ball, V.~Bertone, M.~Bonvini, S.~Marzani, J.~Rojo and L.~Rottoli,
  \emph{{Parton distributions with small-x resummation: evidence for BFKL
  dynamics in HERA data}},
  \href{https://doi.org/10.1140/epjc/s10052-018-5774-4}{\emph{Eur. Phys. J.}
  {\bfseries C78} (2018) 321},
  [\href{https://arxiv.org/abs/1710.05935}{{\ttfamily 1710.05935}}].

\bibitem{Abdolmaleki:2018jln}
{\scshape xFitter Developers' Team} collaboration, H.~Abdolmaleki et~al.,
  \emph{{Impact of low-$x$ resummation on QCD analysis of HERA data}},
  \href{https://arxiv.org/abs/1802.00064}{{\ttfamily 1802.00064}}.

\bibitem{Mangano:2012mh}
M.~L. Mangano and J.~Rojo, \emph{{Cross Section Ratios between different CM
  energies at the LHC: opportunities for precision measurements and BSM
  sensitivity}}, \href{https://doi.org/10.1007/JHEP08(2012)010}{\emph{JHEP}
  {\bfseries 1208} (2012) 010},
  [\href{https://arxiv.org/abs/1206.3557}{{\ttfamily 1206.3557}}].

\bibitem{Harland-Lang:2017ytb}
L.~A. Harland-Lang, A.~D. Martin and R.~S. Thorne, \emph{{The Impact of LHC Jet
  Data on the MMHT PDF Fit at NNLO}},
  \href{https://doi.org/10.1140/epjc/s10052-018-5710-7}{\emph{Eur. Phys. J.}
  {\bfseries C78} (2018) 248},
  [\href{https://arxiv.org/abs/1711.05757}{{\ttfamily 1711.05757}}].

\end{thebibliography}\endgroup

\end{document}